\documentclass[aps,prx,10pt,amsmath,amssymb,floatfix,superscriptaddress,notitlepage,twocolumn]{revtex4-2}

\usepackage[colorlinks=true,linkcolor=blue]{hyperref}
\usepackage{graphicx}
\usepackage{times}
\usepackage{amssymb}
\usepackage{amsmath}
\usepackage{mathtools}
\usepackage{amsthm}

\usepackage{bm}         
\usepackage{color}
\usepackage[utf8]{inputenc}
\usepackage[T1]{fontenc}
\usepackage[normalem]{ulem}

\usepackage{xcolor}

\begin{document}

\title{Flux-tunable parity-protected qubit based on a single full-shell nanowire Josephson junction}

\author{G. Giavaras}\email{g.giavaras@gmail.com}
\affiliation{Instituto de Ciencia de Materiales de Madrid (ICMM),
Consejo Superior de Investigaciones Cient\'ificas (CSIC), Sor
Juana In\'es de la Cruz 3, 28049 Madrid, Spain}

\author{Rub\'en Seoane Souto}
\affiliation{Instituto de Ciencia de Materiales de Madrid (ICMM),
Consejo Superior de Investigaciones Cient\'ificas (CSIC), Sor
Juana In\'es de la Cruz 3, 28049 Madrid, Spain}

\author{Maria Jos\'e Calder\'on}
\affiliation{Instituto de Ciencia de Materiales de Madrid (ICMM),
Consejo Superior de Investigaciones Cient\'ificas (CSIC), Sor
Juana In\'es de la Cruz 3, 28049 Madrid, Spain}

\author{Ram\'on Aguado}\email{ramon.aguado@csic.es}
\affiliation{Instituto de Ciencia de Materiales de Madrid (ICMM),
Consejo Superior de Investigaciones Cient\'ificas (CSIC), Sor
Juana In\'es de la Cruz 3, 28049 Madrid, Spain}

\begin{abstract}
Leveraging the higher harmonics content of the Josephson potential
in a superconducting circuit offers a promising route in the
search for new qubits with increased protection against
decoherence. In this work, we demonstrate how the flux tunability
of a hybrid semiconductor-superconductor Josephson junction based
on a single full-shell nanowire enables this possibility. Near one
flux quantum, $\Phi\approx \Phi_0=h/2e$, we find that the qubit
system can be tuned from a gatemon regime to a parity-protected
regime with qubit eigenstates localized in phase space in the
$\varphi_0=0$ and $\pi$ minima of the Josephson potential ($\cos
2\varphi_0$). Estimates of qubit coherence and relaxation times
due to different noise sources are presented.
\end{abstract}

\maketitle

\section{Introduction}

The coupling of a physical qubit to its noisy environment
inevitably causes relaxation and dephasing, which is one of the
main obstacles in the practical development of a scalable quantum
computer~\cite{mohseni2025buildquantumsupercomputerscaling}. In
addition to quantum error correction~\cite{Google-error}, an
interesting approach is to engineer intrinsic noise protection by
designing the underlying qubit Hamiltonian and achieving
computational states that are largely decoupled from local noise
channels~\cite{DanonAPL2021}.

Among the various approaches for noise protection, a powerful idea
is to encode qubits in global degrees of freedom that are
insensitive to local fluctuations, which is the main idea behind
topological qubits~\cite{Sarma_NPJQ2015,Aguado:PT20}. Another
interesting approach is to encode qubits in local decoherence-free
subspaces, based on effective Hamiltonians with parity
symmetry~\cite{Douçot_2012,PRXQuantum.2.030101}. Although partial
noise protection is possible in superconducting circuits based on
a single quantum degree of freedom, full protection in both the
relaxation ($T_1$) and dephasing ($T_2$) channels is challenging,
and usually complex multimode circuits are
needed~\cite{PRXQuantum.2.010339,PhysRevA.87.052306}.

Given the above context, it is important to design simple circuits
to realize parity-protected
qubits~\cite{PRXQuantum.2.030101,Smith_NPJ2020,PRXQuantum.3.030329,PRXQuantum.3.030303,PhysRevX.12.021002}.
A basic advantage of the parity protection is that relaxation due
to charge induced noise can be suppressed leading to an enhanced
$T_1$ time. In hybrid superconductor-semiconductor Josephson
junctions important steps have been taken towards the realization
of parity-protected qubits by leveraging the higher harmonics
content of the Josephson potential~\footnote{We neglect Fourier
terms $E^{'}_{J,M}\sin(M\varphi_0)$ which are nonzero only when
both time-reversal and inversion symmetries are broken. They could
be incorporated in our analysis in such symmetry-broken
situations.}
\begin{equation}\label{fourier}
V_{J}(\varphi_0) = \sum_{M \geqslant 1} E_{J,M}\cos(M\varphi_0),
\end{equation}
with $\varphi_0$ being the superconducting phase difference and
$E_{J,M}$ the Fourier components. Here the key observation is that
control of $E_{J,M}$, for example using electrostatic gates, can
result in a double well (DW) Josephson potential $V_{J}(\varphi_0)
= E_{J,2} \cos(2\varphi_0)$, which is needed to define a
parity-protected qubit~\cite{Smith_NPJ2020, protected_exper,
PhysRevB.105.L180502}. Because the qubit eigenstates of the DW
Josephson potential are also eigenstates of the Cooper-pair parity
operator the charge induced noise in a superconducting qubit
circuit can be suppressed.

\begin{figure}
\label{cartoon} \centering
\includegraphics[width=8.5cm]{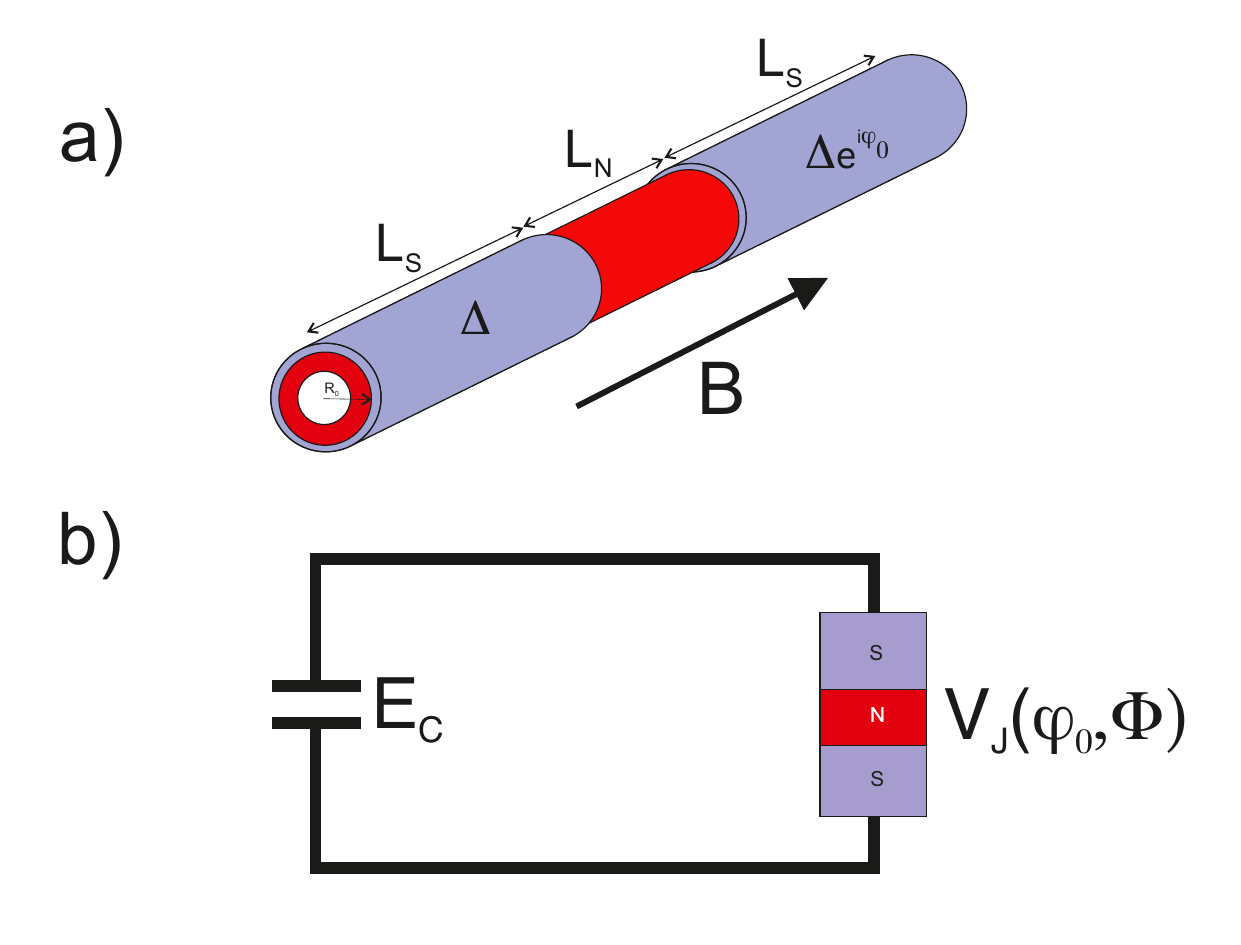}
\caption{(a) Schematic illustration of a full-shell NW Josephson
junction. A semiconducting core (red) of radius $R_0$ is wrapped
by a superconducting shell (mauve) of negligible thickness. A
middle unwrapped region defines a
superconducting-normal-superconducting (SNS) Josephson junction:
two S regions of length $L_S$ are separated by a N region of
length $L_N$, where $\Delta$ is the pairing potential and
$\varphi_0$ is the superconducting phase difference. In the
hollow-core approximation the electrons are confined near the
surface of the NW, and under an applied axial magnetic field $B$
the NW is threaded by a magnetic flux $\Phi=\pi R^2_0B$. (b)
Superconducting qubit circuit where the Josephson element is the
SNS junction in (a). The circuit consists of a capacitor with a
charging energy $E_c$, and a flux dependent Josephson potential
$V_J(\varphi_0,\Phi)$ whose Fourier decomposition contains higher
harmonics, Eq.~(\ref{fourier2}).}\label{cartoon}
\end{figure}

An interferometer formed by two Josephson junctions, gate-tuned
into balance and frustrated by a half-quantum of magnetic flux,
yields a $\cos 2\varphi_0$ Josephson
element~\cite{Ciaccia24,Valentini:24,PhysRevLett.133.186303,PhysRevResearch.6.033281,PRXQuantum.3.030303,PhysRevLett.129.267702,10.21468/SciPostPhys.16.1.030,LeblancNatComm2025}.
This scheme has been successfully incorporated into a
superconducting circuit to demonstrate parity-protected regimes
characterized by an enhanced $T_1$ time~\cite{protected_exper}.
More complex geometries involving four Josephson junctions have
also been explored, and the resulting Josephson potential has been
precisely tuned with a combination of gates and magnetic
frustration~\cite{PhysRevX.15.011021}.

In the present work we theoretically explore the possibility of
tuning the higher harmonics content in a \emph{single} Josephson
junction and creating a DW Josephson potential. Our proposal is
based on the flux tunability of a Josephson junction formed in a
full-shell nanowire (NW)~\cite{model}. Although some experiments
have explored qubits in full-shell
NWs~\cite{PhysRevLett.125.156804,PhysRevLett.126.047701,PhysRevB.108.L121406},
there is no theoretical work that investigates the influence of
the full-shell geometry on the qubit properties and, in
particular, the possibility of obtaining parity-protected regimes.
Importantly, we show that the Josephson potential can be tuned
with an applied magnetic flux, $\Phi$, threading the NW cross
section, and specifically demonstrate realistic regimes where the
Josephson potential changes from a single well (SW), with a
minimum located at $\varphi_0=0$, to a DW with minima at
$\varphi_0=0$ and $\varphi_0=\pi$.

Our study shows that the flux tunability can be accomplished
inside the so-called first Little-Parks (LP) lobe around $\Phi
\approx \Phi_0=h/2e$~\cite{LP1,LP2,LP3,LP4}. The strongest SW
potential is formed at the centre of the LP lobe ($\Phi=\Phi_0$),
and the DW is formed at two flux values located symmetrically
around $\Phi=\Phi_0$. The SW potential can define a usual
gate-tunable transmon qubit (gatemon), whereas the DW can define a
$\cos 2\varphi_0$ parity-protected qubit. We also present
estimates of both $T_1$ and $T_2$ times due to various noise
channels and find coherence times similar to those measured in
experiments~\cite{protected_exper}. Our overall analysis suggests
that control and readout of a full-shell NW parity-protected qubit
should be feasible.

The paper is organized as follows. In Sec.~\ref{model} we present
the model of the Josephson junction formed in a full-shell NW. In
Sec.~\ref{wells} we study how the applied magnetic flux can be
used to tune the Josephson potential and give rise to different
regimes. In the same section we employ a simplified junction model
based on approximate expressions for the Andreev bound states
(ABSs). This model provides valuable insight into the flux
dependence of the exact Josephson potential. The coherence
properties of the resulting qubits are examined in
Sec.~\ref{qubits}, and finally the conclusions of our work are
summarized in Sec.~\ref{conclusions}. Some more technical parts
are presented in appendices. The Bogoliubov-de-Gennes (BdG)
Hamiltonian of the full-shell NW is detailed in
Appendix~\ref{app:full-shell}, while the flux modulation of the
pairing potential due to the LP effect is described in
Appendix~\ref{app:flux-modulation}. Additional examples of the
Josephson potential, including deviations from the perfect
cylindrical symmetry, are presented in
Appendix~\ref{app:examples}. In Appendix~\ref{app:SOC} we
investigate the Josephson potential with spin-orbit coupling and
show that beyond the DW a triple-well potential can also arise in
the topological regime due to the appearance of Majorana states.
In Appendix~\ref{app:analytical-minimal} we present various
approximate results derived from a simplified junction model.
Finally, in Appendix~\ref{app:further-analysis} the coherence
times of the parity protected qubit are further explored and an
approximate model is presented to extract the qubit levels.

\section{Full-shell NW Josephson junction}\label{model}

\subsection{Full-shell NW Hamiltonian}

Full-shell hybrid NWs are semiconducting NWs fully wrapped by a
thin superconducting shell as shown in Fig.~\ref{cartoon}(a). We
consider the hollow-core approximation which assumes that the
electrons are confined near the surface of the NW. As detailed in
Appendix~\ref{app:full-shell}, without spin-orbit coupling the BdG
Hamiltonian describing the proximitized NW consists of the two
matrices
\begin{equation}\label{Hammain}
H^{\pm} = \left(\begin{array}{cc}
  \frac{p_z^2}{2m^*} + \mu_e^{\pm} & \Delta\\
\Delta^{*}& -\frac{p_z^2}{2m^*} + \mu_h^{\pm}    \\
\end{array}\right),
\end{equation}
where $p_z=-i\hbar\partial_z$ is the momentum operator along the
NW, $\Delta$ is the superconducting pairing potential, and $m^{*}$
is the effective mass of the NW.

An applied magnetic flux $\Phi$ through the cross-section of the
NW causes the superconducting phase in the shell to acquire a
quantized winding number $n$ (also called fluxoid) around the NW
axis. Defining the normalized flux $n_\Phi=\Phi/\Phi_0$, with
$\Phi_0=h/2e$, the winding number reads $n = \left
\lfloor{n_\Phi}\right \rceil$, and deviations from integer fluxes
are measured through the variable
\begin{equation}
\phi = n - n_\Phi = n - \frac{\Phi}{\Phi_0}.
\end{equation}
Tuning $\Phi$ results in winding jumps accompanied by a repeated
suppression and recovery of the superconducting pairing potential,
$\Delta = \Delta(\Phi)$ (Appendix~\ref{app:flux-modulation}).
These jumps form the so-called LP lobes~\footnote{Essentially the
same mechanism underlies the LP effect, a modulation of the
transition temperature, $T_c$, of a superconducting cylinder with
magnetic flux leading to a reentrant destruction of
superconductivity near odd half-integer multiples of $\Phi_0$
\cite{LP1,LP2,LP3,LP4}.} associated with each $n$, with $\phi=0$
corresponding to the middle of the lobe~\cite{lobes}. While in
realistic configurations the LP effect is always present, when the
coherence length $\xi$ of the superconducting shell is much
smaller than the radius $R_0$ of the NW the pairing potential is
approximately flux independent,
$\Delta(\Phi)\approx\Delta(0)=\Delta_0$. In the present work, we
study the physics of the Josephson junction in the first LP lobe,
namely $n=1$ and $0.5\le\Phi/\Phi_0\le 1.5$.

For a NW of radius $R_0$ the BdG energy levels are characterized
by the angular momentum number $m_j$, and the flux dependent
chemical potentials which, without spin-orbit coupling, can be
written as~\cite{model,giavaras24}
\begin{equation}\label{potentials}
\mu^\pm_{e} (\phi) = -\mu+\frac{\hbar^2}{8m^{*}R^{2}_0}(1\mp 2m_j
\pm \phi)^2,
\end{equation}
where $\mu$ is the flux independent chemical potential of the NW.
Similar chemical potentials, $\mu^\pm_{h} (\phi)$, can be defined
for the holes (Appendix~\ref{app:full-shell}). Due to these
effective chemical potentials the BdG energy levels shift with
flux, and near the center of a lobe this shift is approximately
linear
\begin{equation}\label{linear-phi}
\mu^\pm_{e} (\phi)\approx \pm
\frac{\hbar^2}{4m^{*}R^{2}_0}(1\mp2m_j)\phi.
\end{equation}
The first term ($m_j$ independent) can be interpreted as a
Zeeman-like contribution and the second one has an orbital origin
describing the coupling between the magnetic field and the angular
momentum~\cite{model, giavaras24, PhysRevB.100.155431}. Away from
the lobe's center the quadratic $\phi^2$ corrections start to play
a role.

\subsection{Defining the Josephson potential}

The Josephson junction shown in Fig.~\ref{cartoon}(a) consists of
a normal (N) region of length $L_N$ between two full-shell hybrid
NWs defining two superconducting (S) regions of length $L_S$. In
the N region the pairing potential is zero while in the left and
right S regions the pairing potentials are equal to $\Delta$ and
$\Delta e^{i\varphi_0} $ respectively with $\varphi_0$ being the
superconducting phase difference. The Josephson potential,
$V_{J}(\varphi_0)$, is calculated from the free-energy of the
Josephson junction, thus in the limit of zero temperature
$V_{J}(\varphi_0)$ can be written as the sum of different angular
momenta $m_j$ contributions
\begin{equation}\label{definition}
V_{J}(\varphi_0) = \sum_{m_j} V^{m_j}_{J}(\varphi_0) = \sum_{m_j}
\sum_{k<0} E^{m_j}_{k}(\varphi_0).
\end{equation}
Here, $E^{m_j}_{k}$ denotes the eigenenergies of the Josephson
junction that correspond to $m_j$ and for each $\varphi_0$ we sum
the eigenenergies below zero ($k<0$) until
$V^{m_j}_{J}(\varphi_0)$ converges. This means we only need to
consider $E^{m_j}_{k}$ with a significant $\varphi_0$-dispersion.
The eigenenergies in Eq.~(\ref{definition}) are obtained from the
numerical solution of the BdG Hamiltonian, Eq.~(\ref{Hammain}).

\begin{figure}
\includegraphics[width=3.cm, angle=270]{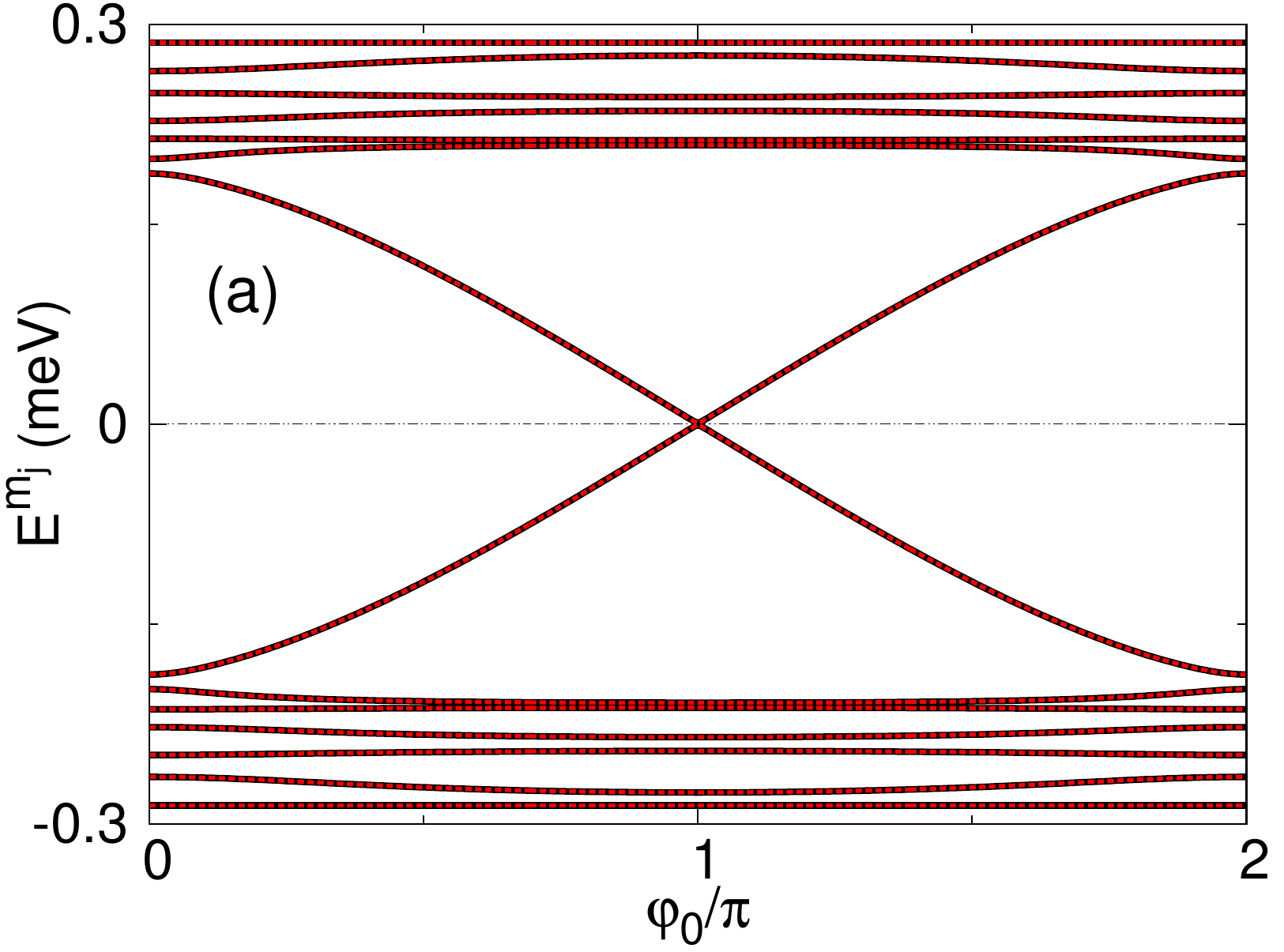}
\includegraphics[width=3.cm, angle=270]{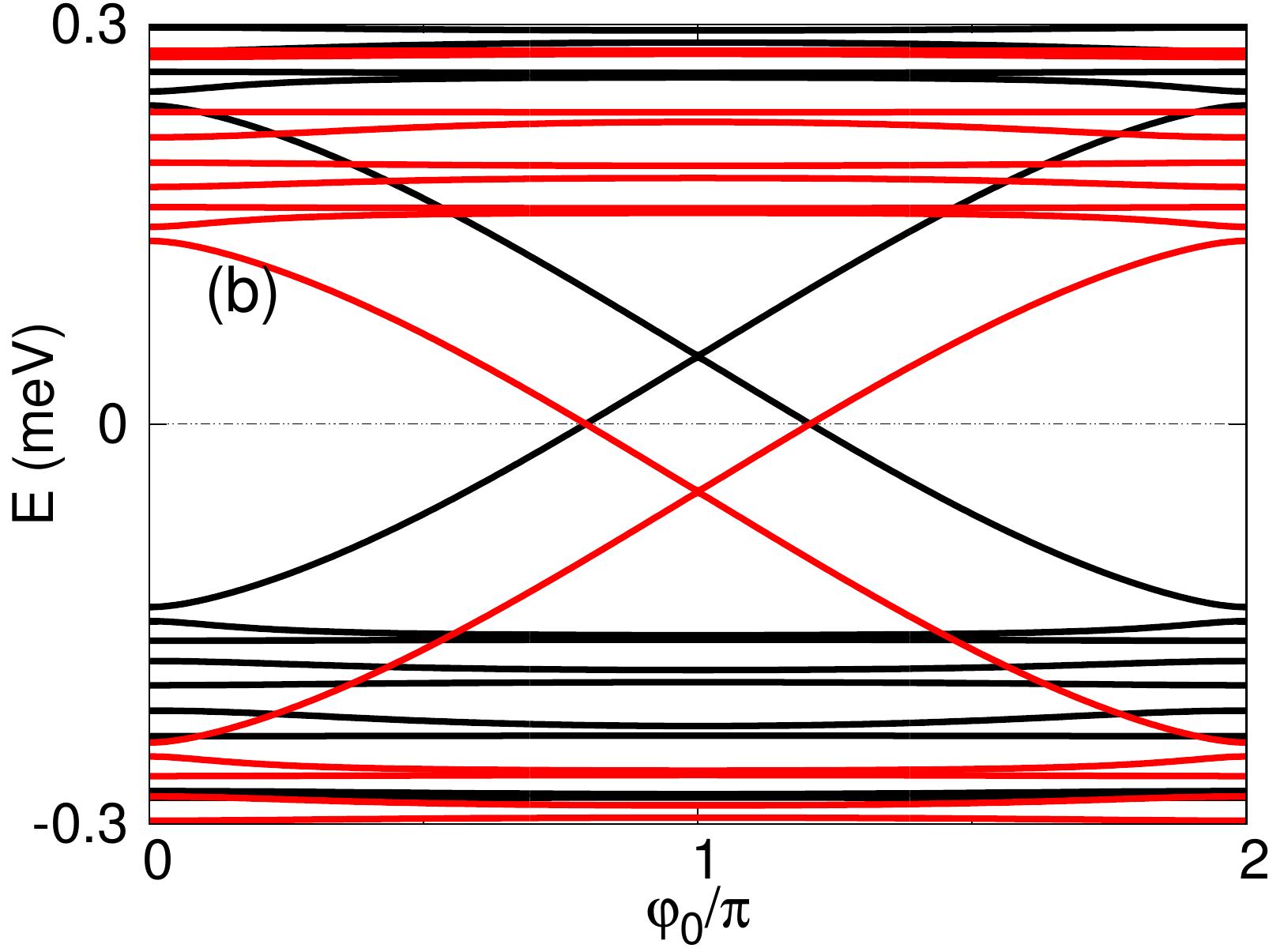}\\
\includegraphics[width=3.cm, angle=270]{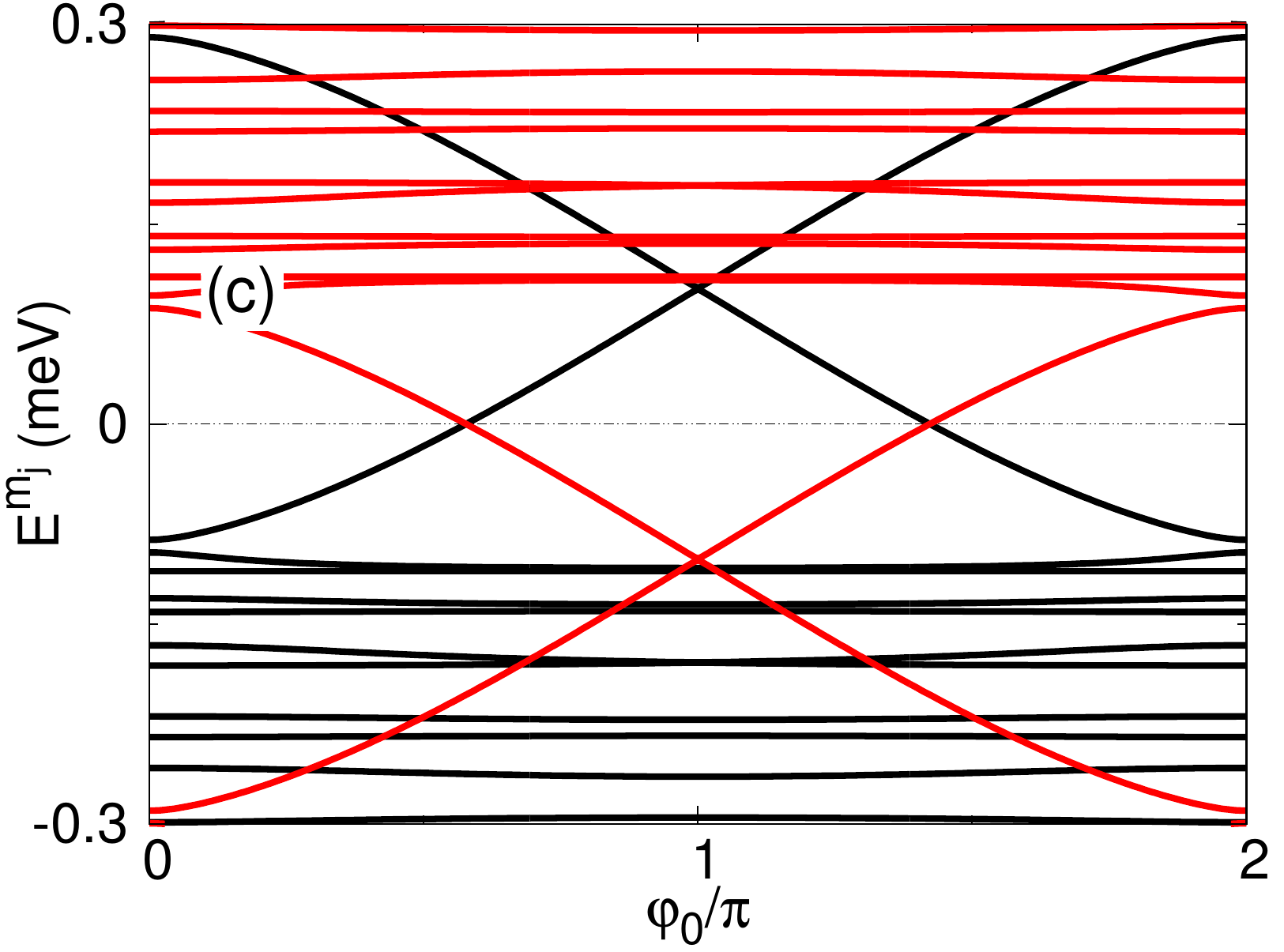}
\includegraphics[width=3.cm, angle=270]{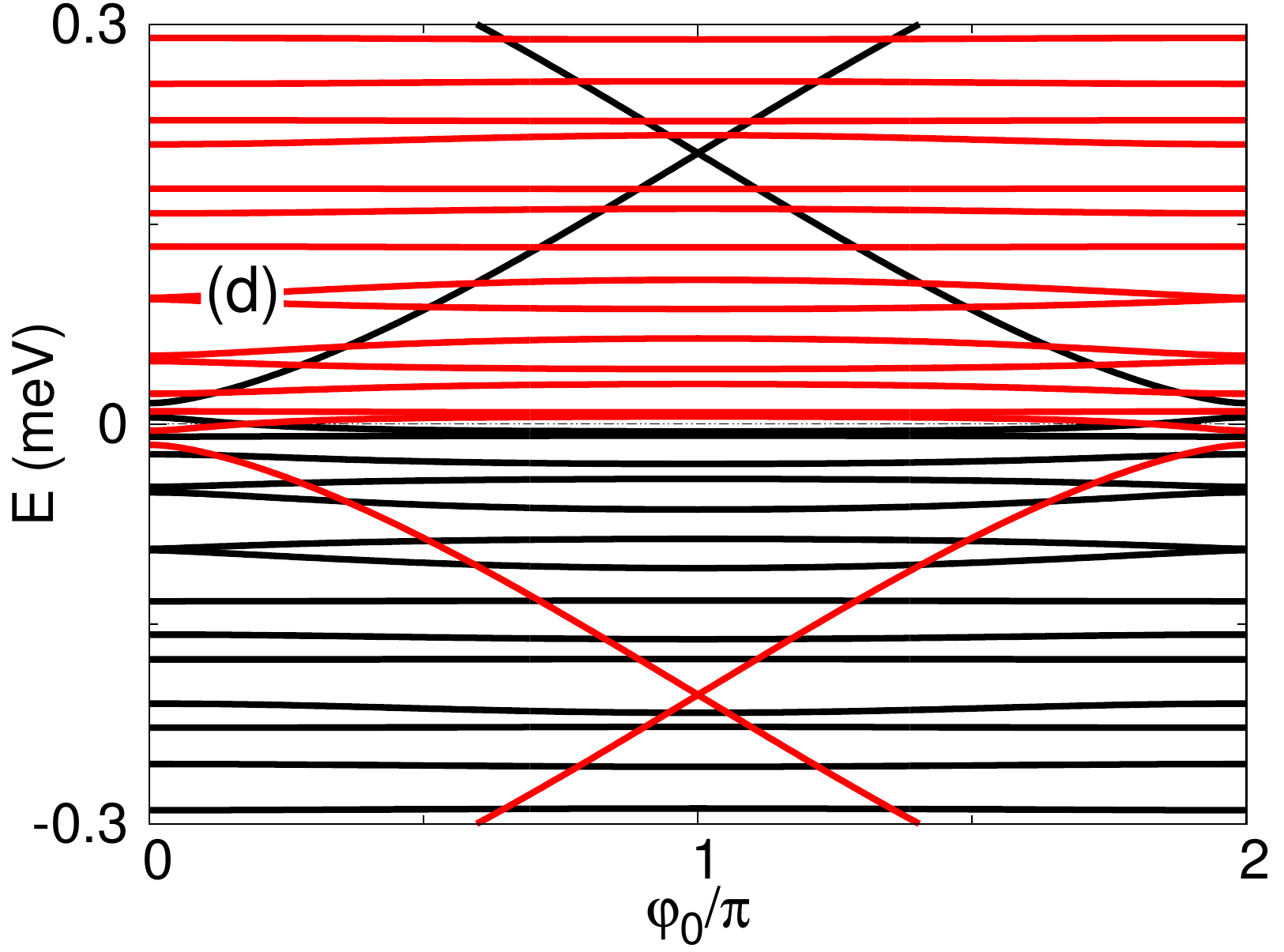}
\caption{BdG energy levels $E^{m_j}$ for $m_j=1$ and $m_j=-1$ as a
function of superconducting phase at different fluxes:
$\Phi/\Phi_0=1$ (a), 1.1 (b), 1.2 (c), and 1.4 (d).
$E^{-1}=-E^{1}$ due to electron-hole symmetry. Parameters:
$L_S=2000$ nm, $L_N=100$ nm, $R_0=50$ nm, and
$\Delta=\Delta_0=0.2$ meV.}\label{andreev}
\end{figure}

Figure~\ref{andreev} illustrates a representative example of the
BdG energy levels $E^{m_j}_k$ at different fluxes. Increasing the
magnetic flux shifts the $m_j=1$ levels upwards and eventually the
subgap mode crosses zero energy, Fig.~\ref{andreev}(d). The
corresponding $m_j=-1$ levels shift downwards and satisfy
$E^{-1}_k=-E^{1}_k$, thus the zero energy crossing leads to a so
called $0-\pi$ transition where positive and negative BdG levels
are exchanged. The flux-induced shift of the BdG levels can be
understood from the linear $\phi$ term in Eq.~(\ref{linear-phi}).
Near the centre of a lobe this term has the dominant role, whereas
away from the centre the quadratic $\phi^2$ term is also
significant and the flux dependence of the BdG levels is more
complicated~\footnote{See Ref.~\onlinecite{giavaras24} for a
detailed description of the flux dependence of the BdG spectrum
and the resulting flux-tunable critical current within the
hollow-core model. More advanced numerical simulations are
presented in Ref.~\onlinecite{paya_JJ}.}. Away from the lobe's
centre the Josephson potential can drastically change compared
with that at $\Phi/\Phi_0=1$ provided the flux-induced shift of
the Andreev levels is large. This remark indicates that a nanowire
with a smaller radius is advantageous to induce a larger shift and
produce a DW potential.

In the next sections, it is helpful to look at the flux dependence
of the Josephson potential. For this reason we combine
Eqs.~(\ref{fourier}) and (\ref{definition}), and we now explicitly
denote the flux dependence in the BdG energy levels and Fourier
decomposition
\begin{equation}\label{fourier2}
\begin{split}
V_{J}(\varphi_0,\Phi) = & \sum_{m_j} \sum_{k<0}
E^{m_j}_{k}(\varphi_0,\Phi) = \\
&\sum_{M \geqslant 1} E_{J,M}(\Phi)\cos(M\varphi_0).
\end{split}
\end{equation}
In Sec.~\ref{wells} we determine the flux dependence of
$E_{J,M}(\Phi)$ for different junctions, and demonstrate how the
applied magnetic flux can tune the Josephson potential from a SW
to a DW.

\section{From single to double well regime}\label{wells}

\subsection{Examples of Josephson potentials}

In this subsection we present some characteristic examples of the
Josephson potential while additional results as well as a brief
investigation about the role of deviations from the perfect
cylindrical symmetry can be found in Appendix~\ref{app:examples}.
To simplify the presentation we focus on low chemical potentials
$\mu$ so that only the $m_j=0$, $\pm1$ modes contribute to the
Josephson potential. Including additional $m_j$ modes only affects
the quantitative properties of the DW. We consider a junction
without spin-orbit coupling, $\alpha_{\rm so} = 0$, which leads to
degenerate BdG levels~\cite{giavaras24}, and explore the
$\alpha_{\rm so} \ne 0$ case in Appendix~\ref{app:SOC}. We also
assume the thickness of the superconducting shell to be much
smaller than the radius of the NW. Under this assumption the flux
modulation of the pairing potential due to the LP effect is
symmetric with respect to the centre of the lobe, $\Phi/\Phi_0=1$.
Finally, we consider the chemical potential $\mu$ to be the same
in the S and N regions. This choice reduces the number of
parameters, but does not affect the main conclusions. In a more
realistic configuration the number of modes in the junction can be
adjusted by tuning the chemical potential within the N
region~\cite{giavaras24}.

\begin{figure}
\includegraphics[width=4.0cm, angle=270]{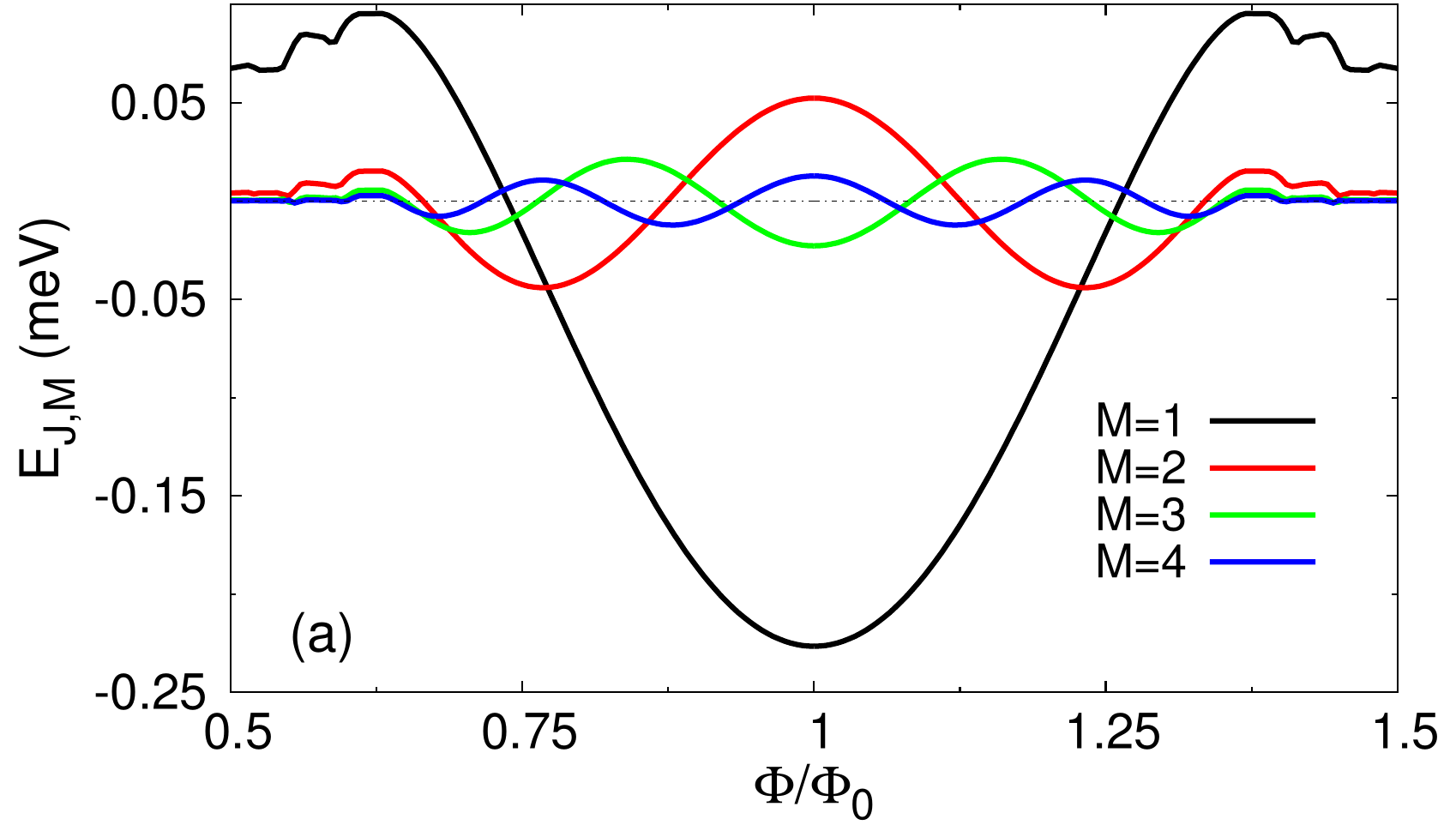}\\
\includegraphics[width=3.0cm, angle=270]{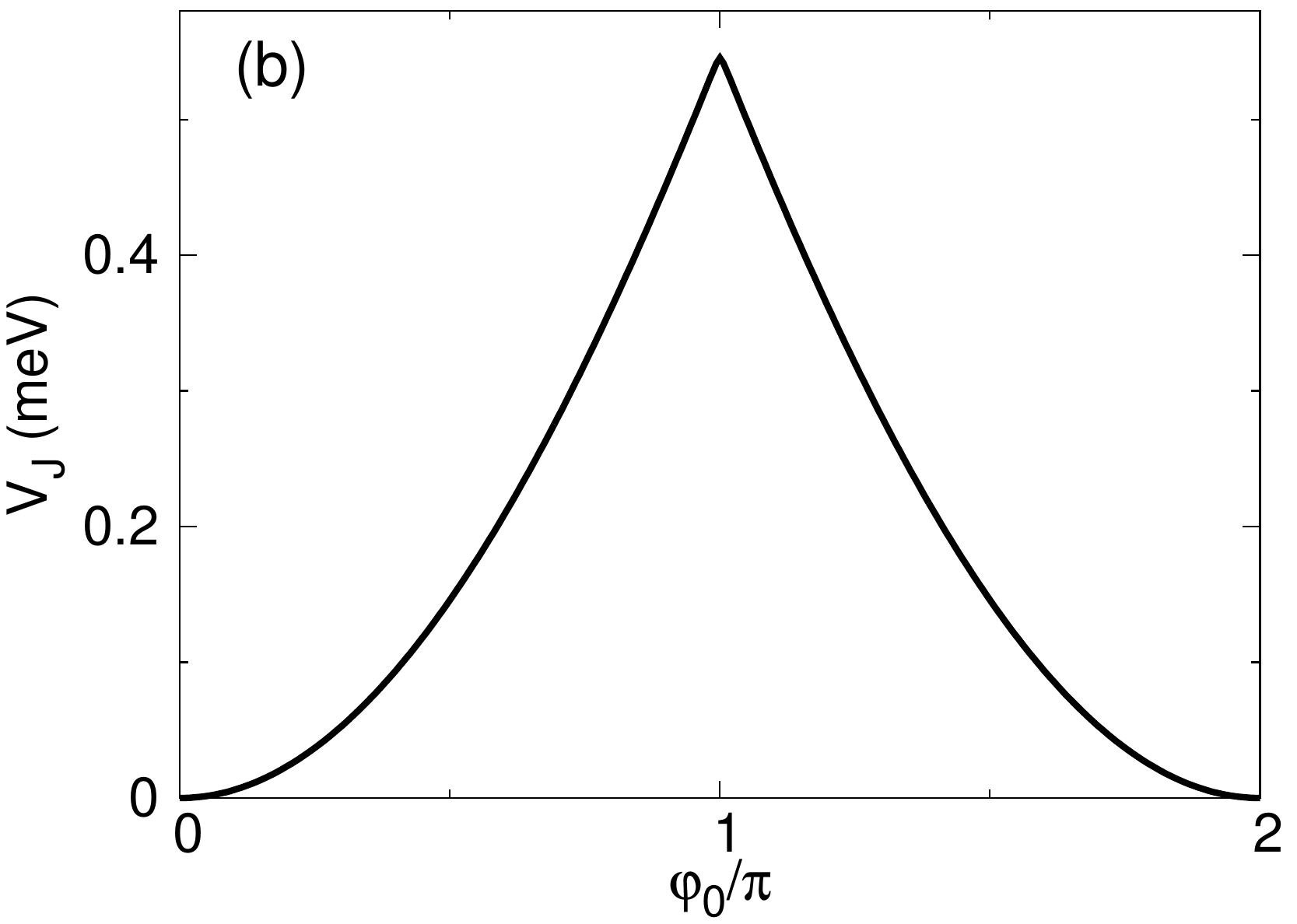}
\includegraphics[width=3.0cm, angle=270]{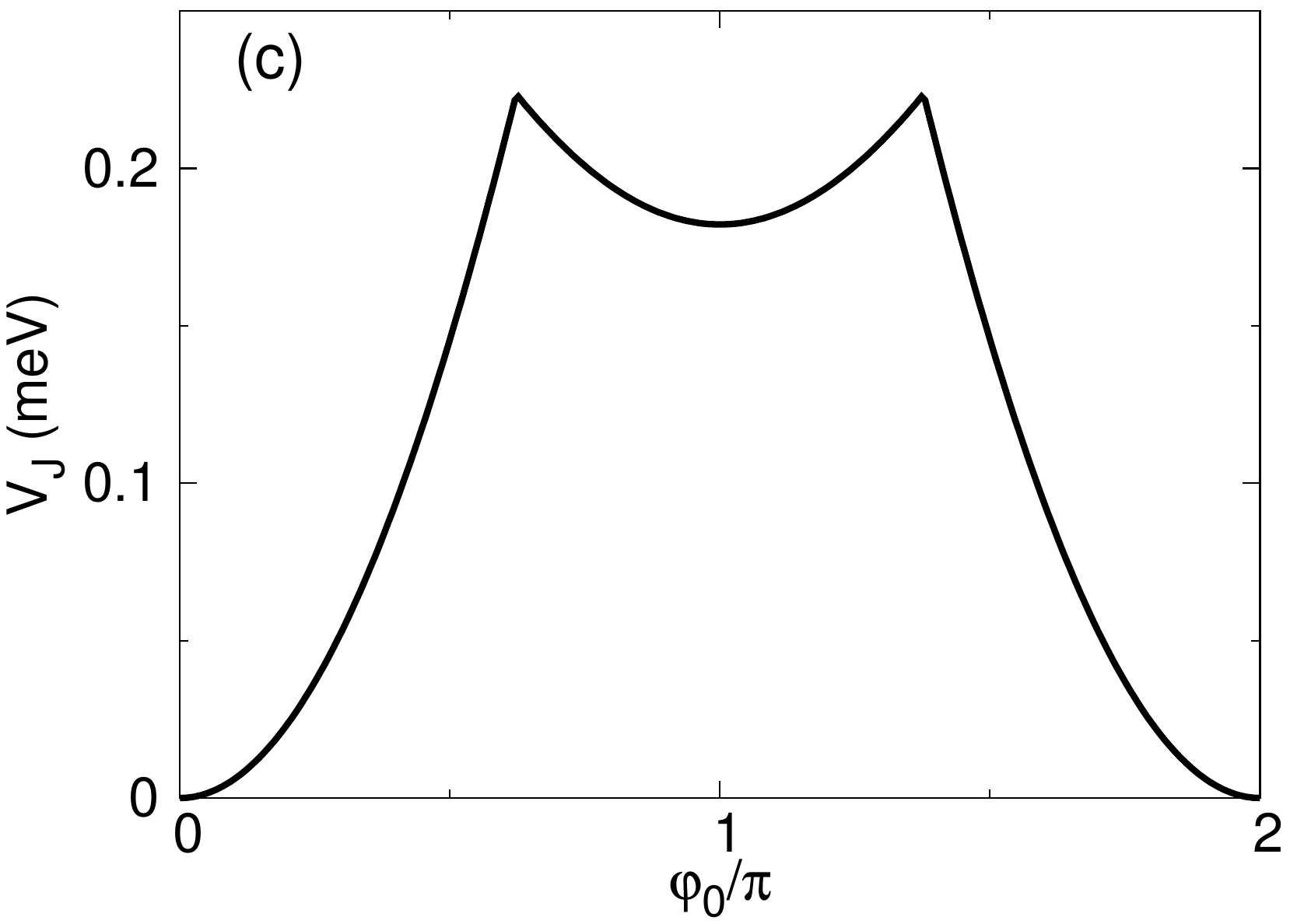}\\
\includegraphics[width=3.0cm, angle=270]{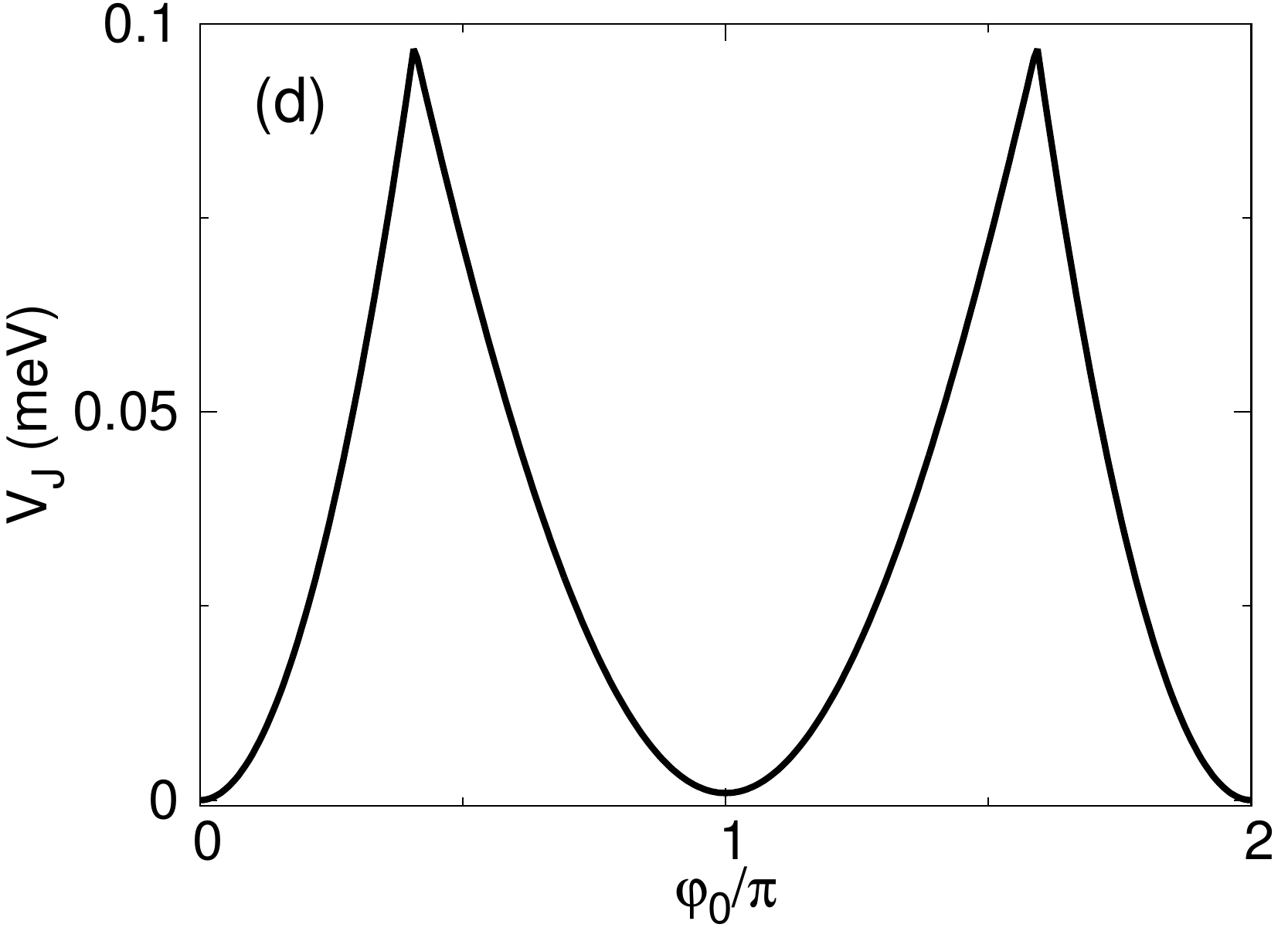}
\includegraphics[width=3.0cm, angle=270]{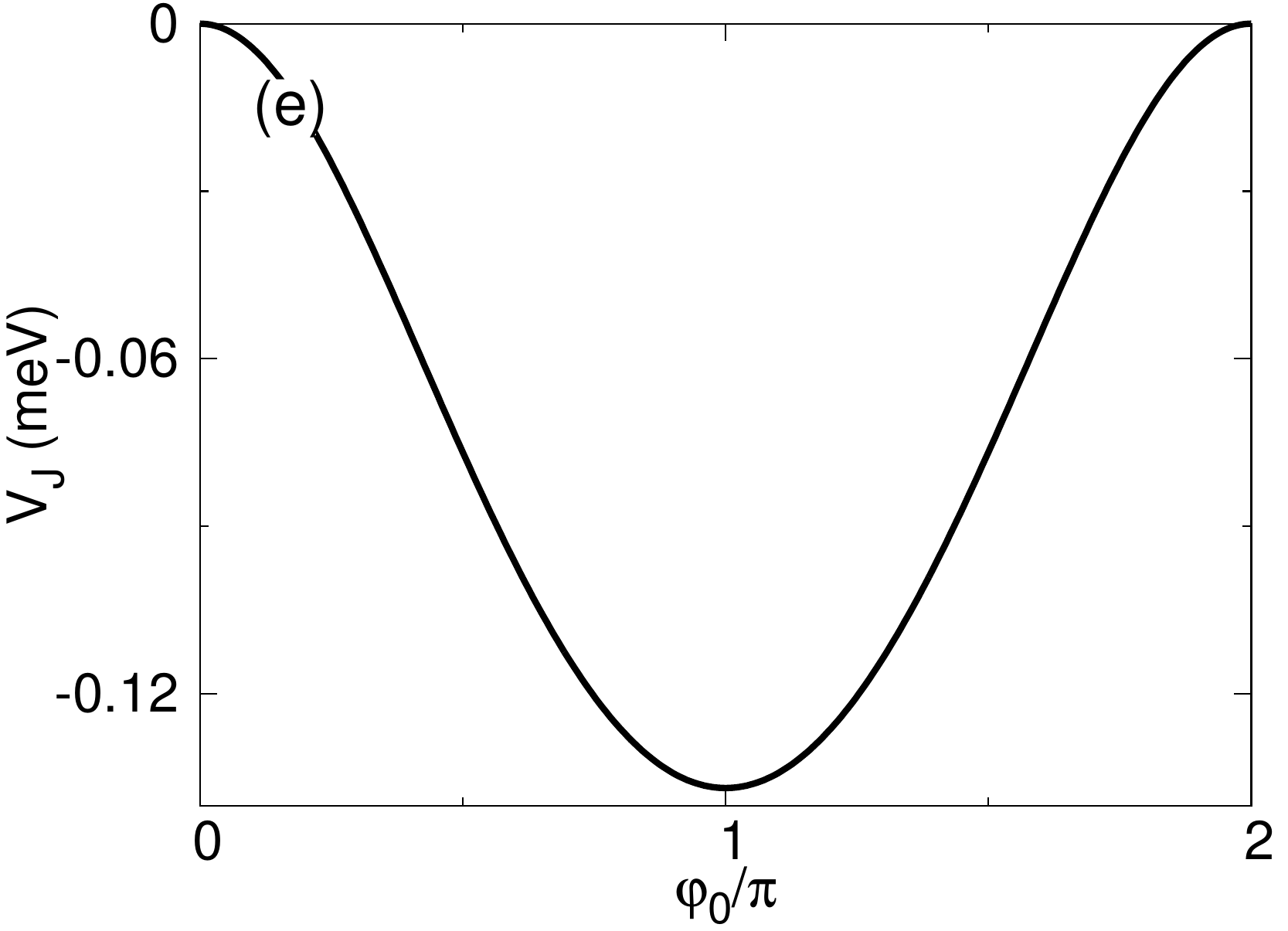}
\caption{Fourier components of the Josephson potential for $M=1-4$ as
a function of magnetic flux (a). Josephson potential at different
fluxes: $\Phi/\Phi_0=1$ (b), 1.18 (c), 1.26 (d), and 1.5 (e).
Parameters: $L_S=2000$ nm, $L_N=100$ nm, $R_0=50$ nm, and
$\Delta=\Delta_0=0.2$ meV.}\label{dwR50}
\end{figure}

The first few Fourier components of the Josephson potential as a
function of the magnetic flux are shown in Fig.~\ref{dwR50}(a).
Figures~\ref{dwR50}(b-e) show the resulting Josephson potentials
at different fluxes. At the centre of the LP lobe,
$\Phi/\Phi_0=1$, the component $E_{J,1}$ dominates giving rise to
a SW Josephson potential. By tuning the flux away from
$\Phi/\Phi_0=1$ the $E_{J,1}$ component decreases and $E_{J,2}$
becomes the dominant one, with $E_{J,1}\approx 0$ at $\Phi/\Phi_0
\approx 1.26$ and $0.74$. At these fluxes a DW potential with
equal minima $V_J(\varphi_0=0) \approx V_J(\varphi_0=\pi)$ is
formed, Fig.~\ref{dwR50}(d). This condition entails the sum of the
odd Fourier components to be nearly zero. We refer to this DW
potential as \textit{symmetric}, although the two well widths
differ because in general $E_{J,M}\ne0$ for $M\ne2$. In
Sec.~\ref{qubits} we show that for a symmetric DW a quasi
degenerate parity-protected qubit can be defined. Notice that
Figs.~\ref{dwR50}(b) and (e) reveal that the Josephson potential
undergoes a $0-\pi$ transition, characterized by the shift of the
SW potential from $\varphi_0=0$ to $\pi$. We emphasize that this
transition depends sensitively on the exact junction parameters,
but is not required to define a symmetric DW potential.

In full-shell NWs the LP effect leads to a flux dependent pairing
potential, $\Delta=\Delta(\Phi)$
(Appendix~\ref{app:flux-modulation}). The coherence length $\xi$
of the superconducting shell, as well as the radius $R_0$ of the
NW, determines the strength of the flux dependence. DWs formed
near the centre of a lobe are weakly affected by the flux
dependence of $\Delta$ and our numerical investigation in
Appendix~\ref{app:examples} demonstrates that the DW in
Fig.~\ref{dwR50}(d) remains to a very good approximation
unaffected by the LP effect when $\xi\approx 120$ nm (destructive
regime) or smaller.

\begin{figure}
\includegraphics[width=4.0cm, angle=270]{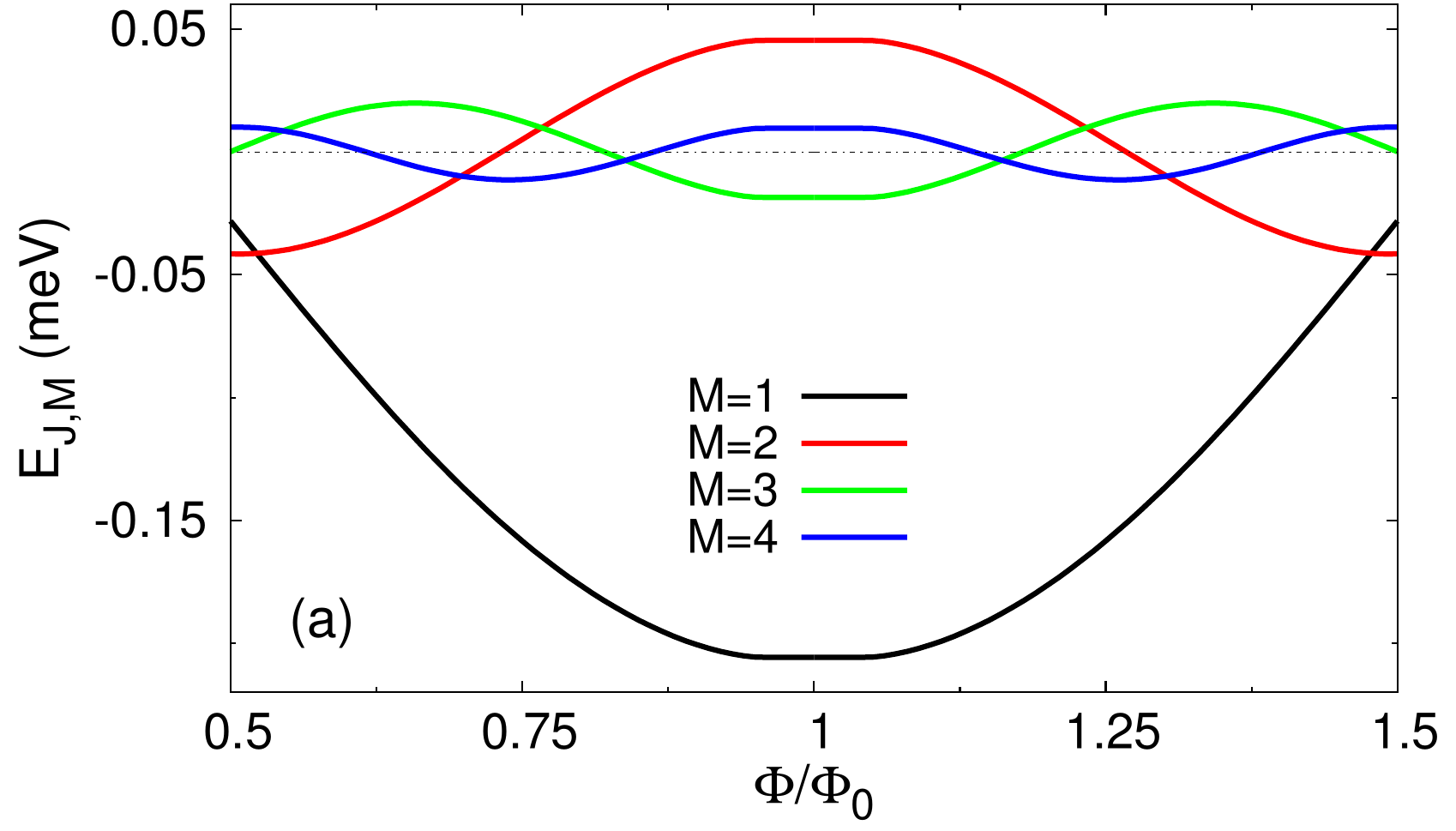}\\
\includegraphics[width=4.0cm, angle=270]{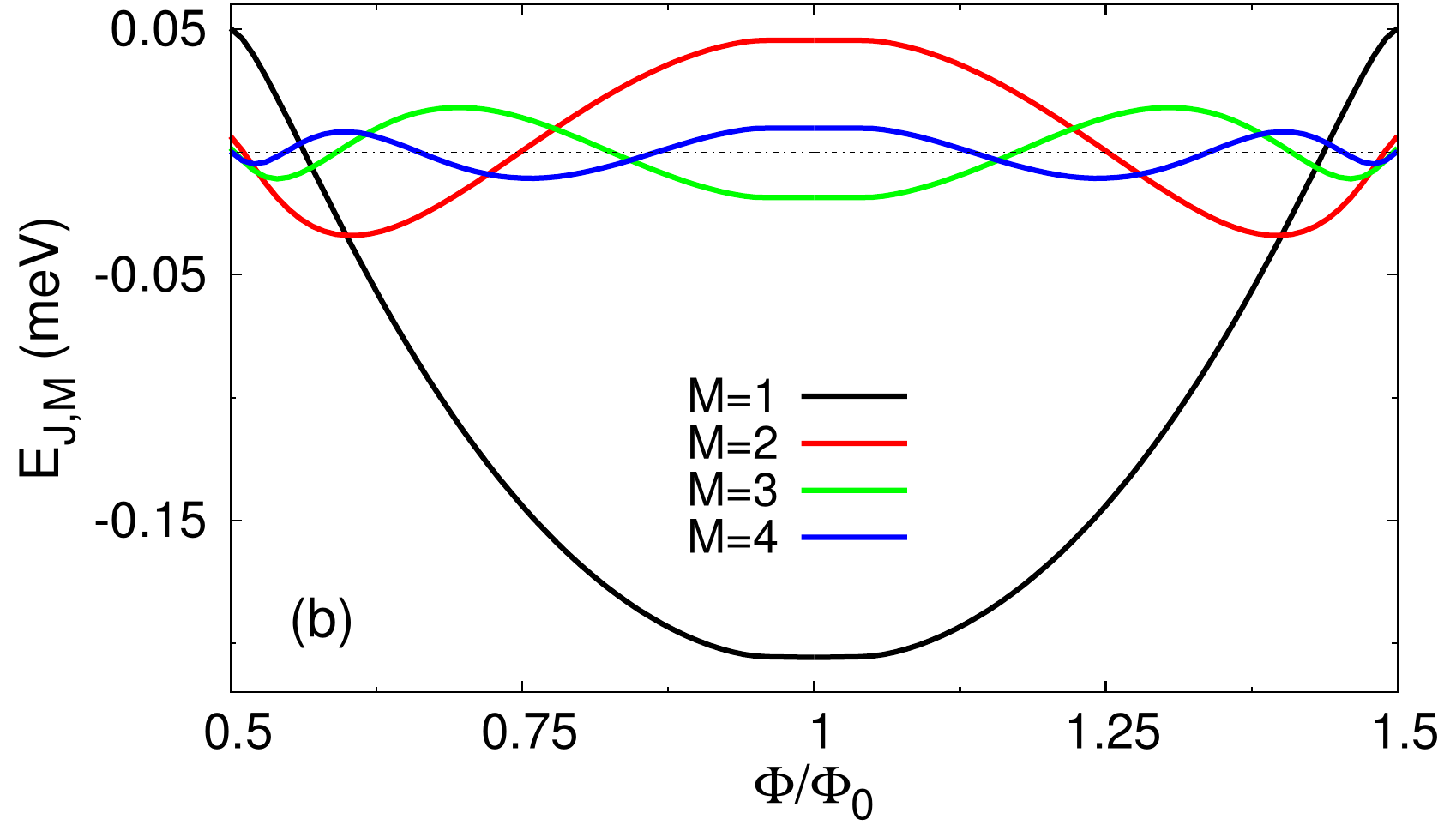}\\
\includegraphics[width=3.0cm, angle=270]{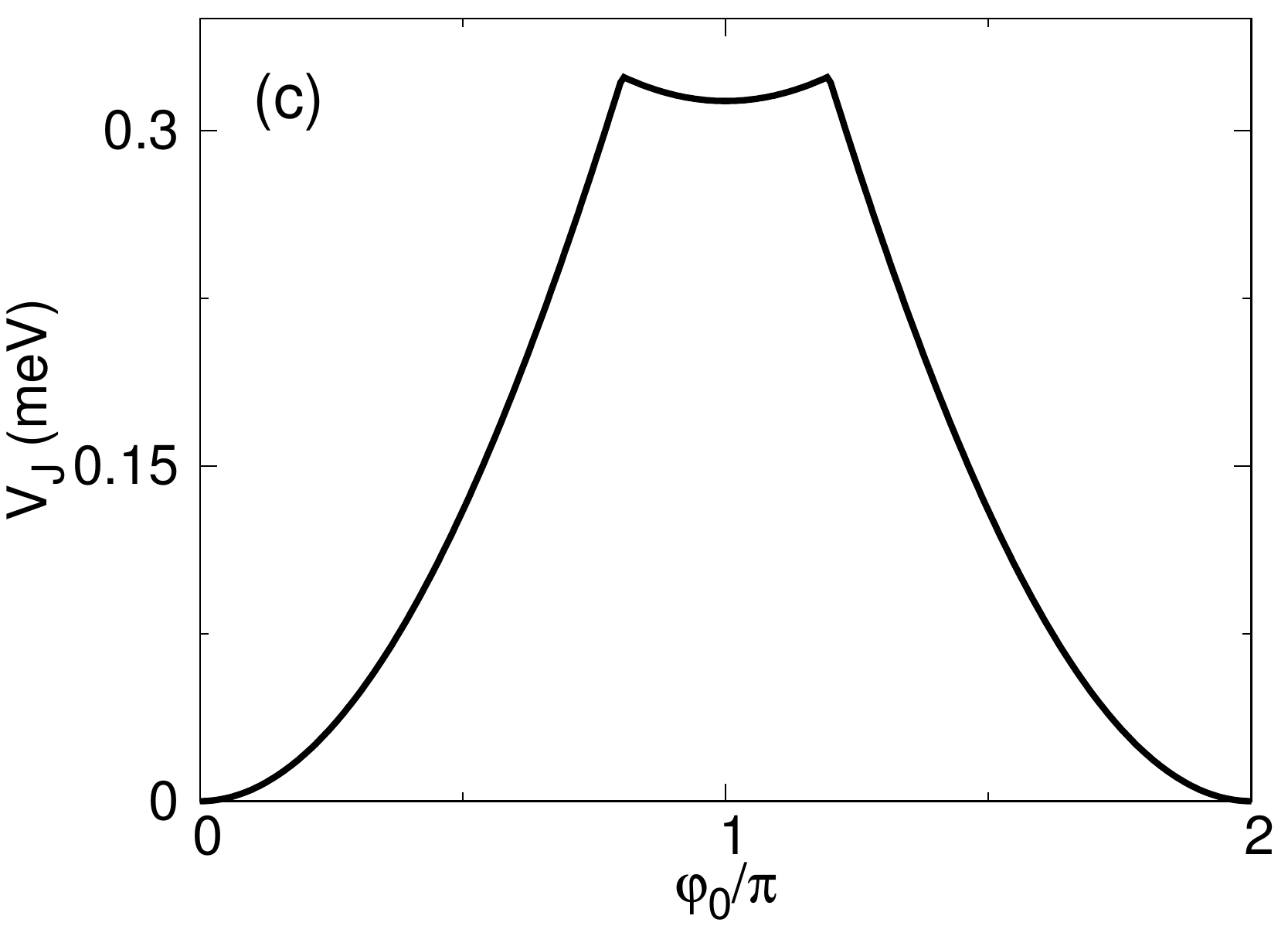}
\includegraphics[width=3.0cm, angle=270]{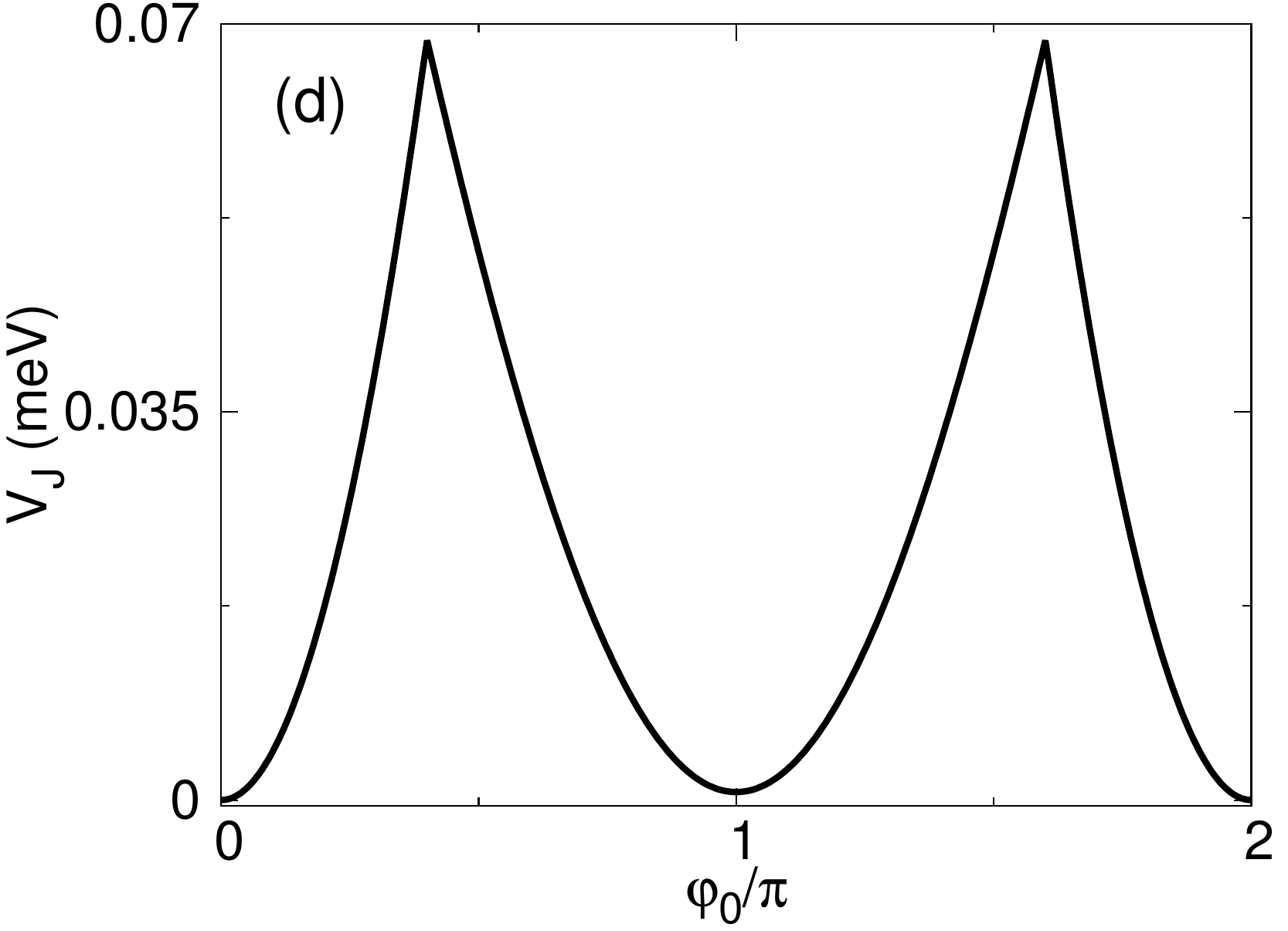}
\caption{Fourier components of the Josephson potential for $M=1-4$ as
a function of magnetic flux for a constant $\Delta=\Delta_0=0.2$ meV (a). As
in (a) but for a flux dependent $\Delta=\Delta(\Phi)$, and $\xi=110$ nm (b).
Josephson potential for $\Delta=\Delta(\Phi)$, and $\xi=110$ nm at
different fluxes: $\Phi/\Phi_0=1.2$ (c), and 1.44 (d). Parameters:
$L_S=2000$ nm, $L_N=100$ nm, and $R_0=75$ nm.}\label{dwR75}
\end{figure}

Although the SW potential at a lobe's centre can be found in any
junction, the formation of a symmetric DW is not guaranteed. One
such case is illustrated in Fig.~\ref{dwR75}(a) for a NW with
$R_0=75$ nm. Now $E_{J,1}$ is large within the entire LP lobe,
consequently, only highly asymmetric DWs are formed. However, when
we take into account the flux dependence of the pairing potential
[Fig.~\ref{dwR75}(b)] a symmetric DW is formed at $\Phi/\Phi_0
\approx 1.44$ and 0.56 [Fig.~\ref{dwR75}(d)]. This behaviour is
general for NWs with $R_0 \gtrsim 70 - 80$ nm and suitable values
of the coherence length $\xi$. A numerical example in
Appendix~\ref{app:analytical-minimal} demonstrates that a larger
$\xi$ is needed to form a DW potential as $R_0$ increases.

Finally, we comment on the junction length $L_N$. We have studied
junctions for different paraments and found that, in general,
longer junctions are more appropriate to define a DW potential.
Increasing $L_N$ shifts the DW regime closer to the centre of the
LP lobe, because the $\varphi_0$-dispersion of the energy levels
lying above the superconducting gap becomes stronger. We present a
characteristic example in Appendix~\ref{app:examples} comparing
two junctions with $L_N=100$ nm and $200$ nm. Tuning the magnetic
flux forms a DW only for the longer junction.

\subsection{Simplified Josephson junction}\label{simplifiedmodel}

The results in Figs.~\ref{dwR50} and~\ref{dwR75} can be understood
from the flux dependence of the ABSs levels lying in the
superconducting gap as well as the energy levels lying above the
gap. The latter are important to form the DW potential, provided
their overall $\varphi_0$-dispersion is opposite to that of the
ABSs subgap levels. To overcome the difficulties due to the
complexity of the BdG Hamiltonian, we here develop a simplified
model. Specifically, we assume for $m_j=1$ the ABS subgap level
\begin{equation}\label{aprox1}
\begin{split}
E_{\pm, s}(\varphi_0,\Phi) = \pm \Delta(\Phi) \sqrt{ 1 - \tau_s
\sin^2(\varphi_0/2) } + \gamma(\Phi),
\end{split}
\end{equation}
where $0<\tau_s \le 1$ models the transparency of the Josephson
junction, and the term
\begin{equation}\label{gamma}
\gamma(\Phi) = - \left( 1 - \frac{\Phi}{\Phi_0} \right)
\frac{\hbar^2}{4 m^{*}R^{2}_0},
\end{equation}
captures the linear shift due to the applied magnetic flux. For
$\gamma(\Phi)=0$ we recover the standard short-junction
formula~\cite{PhysRevLett.67.3836}. We also consider a level which
lies above the gap
\begin{equation}\label{aprox2}
\begin{split}
E_{\pm, a}(\varphi_0,\Phi) =  \pm \Delta(\Phi) \sqrt{ 1 - \tau_a
\sin^2(\varphi_0/2) } + \gamma(\Phi) \mp \beta.
\end{split}
\end{equation}
The exact value of the flux independent constant $\beta$ is not
important and we take $\beta \gtrsim 2 \Delta_0$ to satisfy the
usual regime $E_{+,a} < E_{-,s}$. The levels $E_{-, s}$ and $E_{+,
a}$ have opposite $\varphi_0$-dispersions and we take the limit
$\tau_a < \tau_s$ which models the most usual case in a Josephson
junction. The role of $E_{\pm, a}$ is to capture all the levels
lying above the gap in the corresponding BdG calculation and have
opposite $\varphi_0$-dispersion with respect to $E_{\mp, s}$.

Within the simplified model analytical conditions can be derived
and some simple limits can be identified. Using the definition,
Eq.~(\ref{fourier2}), and with the BdG levels
$E^{m_j}_{k}(\varphi_0,\Phi)$ being $E_{\pm, a}(\varphi_0,\Phi)$
and $E_{\pm, s}(\varphi_0,\Phi)$ we can show that at
$\Phi/\Phi_0=1$ the Josephson potential has a single minimum at
$\varphi_0=0$. By increasing $\Phi$ the Fourier coefficients vary
with flux (Appendix~\ref{app:analytical-minimal}) in the same way
as in Fig.~\ref{dwR75}(b), thus the shape of the potential can
drastically change. Equation~(\ref{fourier2}) suggests that only
when $E_{-s}(\pi,\Phi)>0$ the applied magnetic flux starts to
modify the SW potential. Therefore, away from the lobe's centre
the required $\Phi$ to observe a change in the Josephson potential
satisfies $\gamma(\Phi) / \Delta(\Phi)
> \sqrt{1-\tau_{s}}$. By further increasing $\Phi$, the necessary
condition leading to a symmetric DW potential is
\begin{equation}\label{dwcondition}
E_{+, a}(\pi,\Phi) = E_{+, a}(0,\Phi) + E_{-, s}(0,\Phi),
\end{equation}
which guarantees that the Josephson potential has equal minima at
$\varphi_0=0$ and $\pi$. Equation~(\ref{dwcondition}) reveals the
significance of the level $E_{\pm, a}$; if this is not taken into
account Eq.~(\ref{dwcondition}) cannot be satisfied. This can be
be understood by taking the tunneling limit $\tau_a\rightarrow0$
that leads to a non dispersive level $E_{+, a}$. The flux $\Phi =
\Phi_{\rm DW}$ satisfying Eq.~(\ref{dwcondition}) is given by the
expression
\begin{equation}\label{crit}
\frac{ \gamma( \Phi_{\rm DW} ) } { \Delta( \Phi_{\rm DW}) } =
\sqrt{ 1 - \tau_a}.
\end{equation}
Because $\tau_a<1$ Eq.~(\ref{crit}) imposes the constraint
$\gamma(\Phi_{\rm DW})<\Delta(\Phi_{\rm DW})$, and simultaneously
the junction parameters need to be controlled so that $\Phi_{\rm
DW}$ lies within the first LP lobe. To obtain a physically
acceptable $\Phi_{\rm DW}$ a NW with a smaller radius $R_0$ is
advantageous and a longer junction is also preferable since a
larger $\tau_a$ is expected. Interestingly, the formation of the
DW potential does not necessarily require a highly transparent
junction. This can be understood from the fact that $\tau_s$ is
not explicitly involved in Eq.~(\ref{crit}). However, a higher
$\tau_s$ can be advantageous to induce stronger DW potential
barriers and thus reduce the qubit frequency.

A more detailed analysis of the simplified model, together with
numerical examples, is given in
Appendix~\ref{app:analytical-minimal}. The key point in our
analysis is to derive the condition which the junction needs to
satisfy in order to form a DW potential. This condition is given
by Eqs.~(\ref{dwcondition}) and ~(\ref{crit}). Provided these two
equations are satisfied the SW at $\Phi=\Phi_0$ is transformed to
a DW at $\Phi=\Phi_{\rm DW}$.

\section{Flux tunable parity-protected qubit}\label{qubits}

\subsection{General formalism}

We now explore our proposed qubit system based on the
superconducting circuit schematically shown in
Fig.~\ref{cartoon}(b). The qubit Hamiltonian consists of the
charging energy as well as the Josephson potential energy,
\begin{equation}
\begin{split}
H = 4 E_{c} (\hat{N} - N_g )^2 + V_J(\hat{\varphi}_0,\Phi),
\end{split}
\end{equation}
where $\hat{N}$ is the Cooper-pair number operator that is
conjugate to $\hat{\varphi}_0$, $E_c$ is the charging energy, and
$N_g$ is an offset charge. In the Cooper-pair number basis $| N
\rangle$ the Hamiltonian has the form
\begin{equation}\label{qubitham}
\begin{split}
H = & \sum_N 4 E_{c} (N - N_g )^2 | N \rangle \langle N| \\
 & + \sum_{N, M} \frac{1}{2} E_{J,M} (\Phi)| N \rangle \langle N+M| +
\text{H.c.} ,
\end{split}
\end{equation}
where the flux dependent coefficients $E_{J,M}$ are defined in
Eq.~(\ref{fourier2}) and $N = 0, \pm1, \ldots$ is the number of
(excess) Cooper pairs. The two lowest eigenstates of this
Hamiltonian define the qubit states $|\psi_0 \rangle$ and $|\psi_1
\rangle$ which are flux-tunable due to the flux dependence of
$E_{J,M}$.

We calculate the relaxation times $T^{x}_1$ due to different noise
channels $x$, from the corresponding rates $\Gamma^{x}_1$ derived
from the golden rule
\begin{equation}\label{T1}
\frac{1}{T^{x}_{1}} =\Gamma^{x}_1  = \frac{1}{\hbar^2} \left|
\left\langle \psi_0 \left| \frac{\partial H }{\partial x} \right|
\psi_1 \right\rangle \right|^2 S_{x}(\omega_{10}),
\end{equation}
where $S_{x}(\omega_{10})$ is the spectral density at the qubit
frequency $\omega_{10} = (E_{1}-E_{0})/\hbar$. In our qubit system
$x$ represents either the magnetic flux, $x=\Phi$, or the charge
offset, $x=N_g$, or the chemical potential, $x=\mu$, and at low
frequencies a $1/f$-noise is expected
\begin{equation}
S_{x}(\omega_{10}) = A^{2}_{x} \left( \frac{ 2\pi }{ \omega_{10}}
\right) \times 1 \text{Hz}.
\end{equation}
In superconducting qubits the noise amplitudes $A_{x}$ can vary by
orders of magnitude. The precise values of these amplitudes for
qubits based on full-shell NWs are unknown, therefore we take some
typical values from the superconducting
literature~\cite{review_qubits, transmon} $A^{2}_{\Phi} = (10^{-6}
\Phi_0)^2$ Hz$^{-1}$ and $A^{2}_{N_g} = (10^{-4} e)^{2}$
Hz$^{-1}$.

Quasiparticle poisoning can also lead to relaxation and the
relevant time is derived from the expression~\cite{poisoning}
\begin{equation}\label{T1-qp}
\frac{1}{T^{p}_{1}} =\Gamma^{{p}}_{1} = \left| \left\langle \psi_0
\left| \sin(\hat{\varphi}_0/2)
 \right| \psi_1 \right\rangle \right|^2
S_{{p}}(\omega_{10}),
\end{equation}
with the spectral density given by
\begin{equation}
S_{{p}}(\omega_{10}) = \frac{ \omega_{10} g_{T} }{ 2 \pi g_K}
\left( \frac{2 \Delta}{ \hbar \omega_{10}}  \right)^{3/2}
\chi_{{p}}.
\end{equation}
Here, $\chi_{{p}}$ is the quasiparticle density normalised to the
Cooper-pair density, and we assume the value $\chi_{{p}}=
10^{-7}$. The parameters $g_K$ and $g_T$ correspond to the
conductance quantum and the junction conductance
respectively~\cite{poisoning}.

Within a density matrix approach we can calculate the dephasing
time $T^{x}_{2}$ of the qubit from the expression~\cite{grosz}
\begin{equation}\label{T2}
T^{x}_2 = \left[  2 D^{2}_{x} A^{2}_{x} |\ln ( \omega_{ir} t ) | +
2 D^{2}_{xx} A^4_{x} \kappa \right]^{-1/2},
\end{equation}
with $D_{x} = \partial \omega_{10} / \partial x $, $D_{xx} =
\partial^2 \omega_{10} / \partial x^2 $ and
\begin{equation}
\kappa = \ln^{2} \left( \frac{\omega_{uv}}{\omega_{ir}} \right) +
2 \ln^{2} ( \omega_{ir} t ).
\end{equation}
We assume a measurement time-scale of $t = 10$ $\mu$s and for $T
\lesssim 15$ mK the low- and high-frequency cutoff values of the
noise are $\omega_{ir} / 2\pi \approx $ 1 Hz and $\omega_{uv} /
2\pi \approx $ 3 GHz respectively.

\begin{figure}
\includegraphics[width=3.2cm, angle=270]{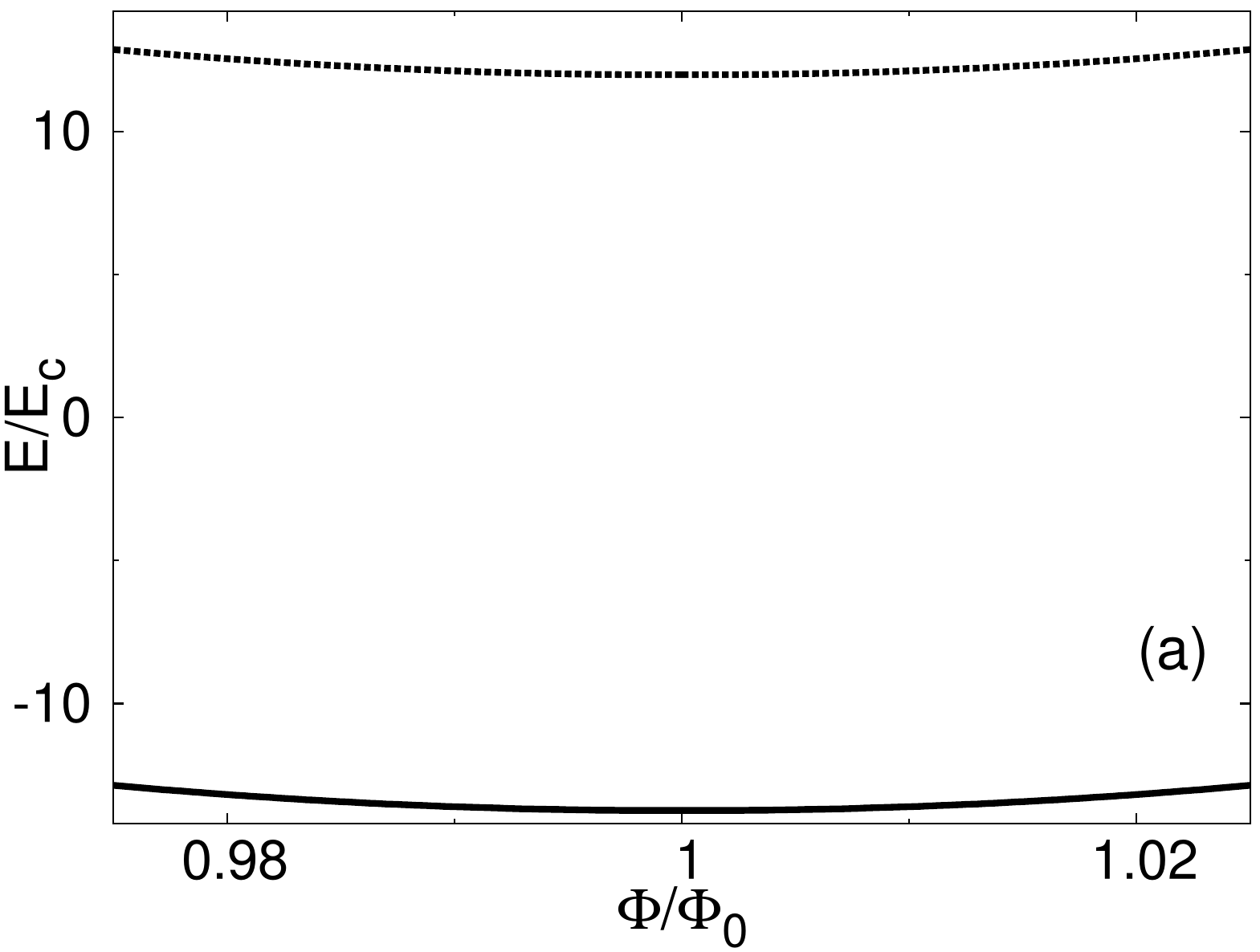}
\includegraphics[width=3.2cm, angle=270]{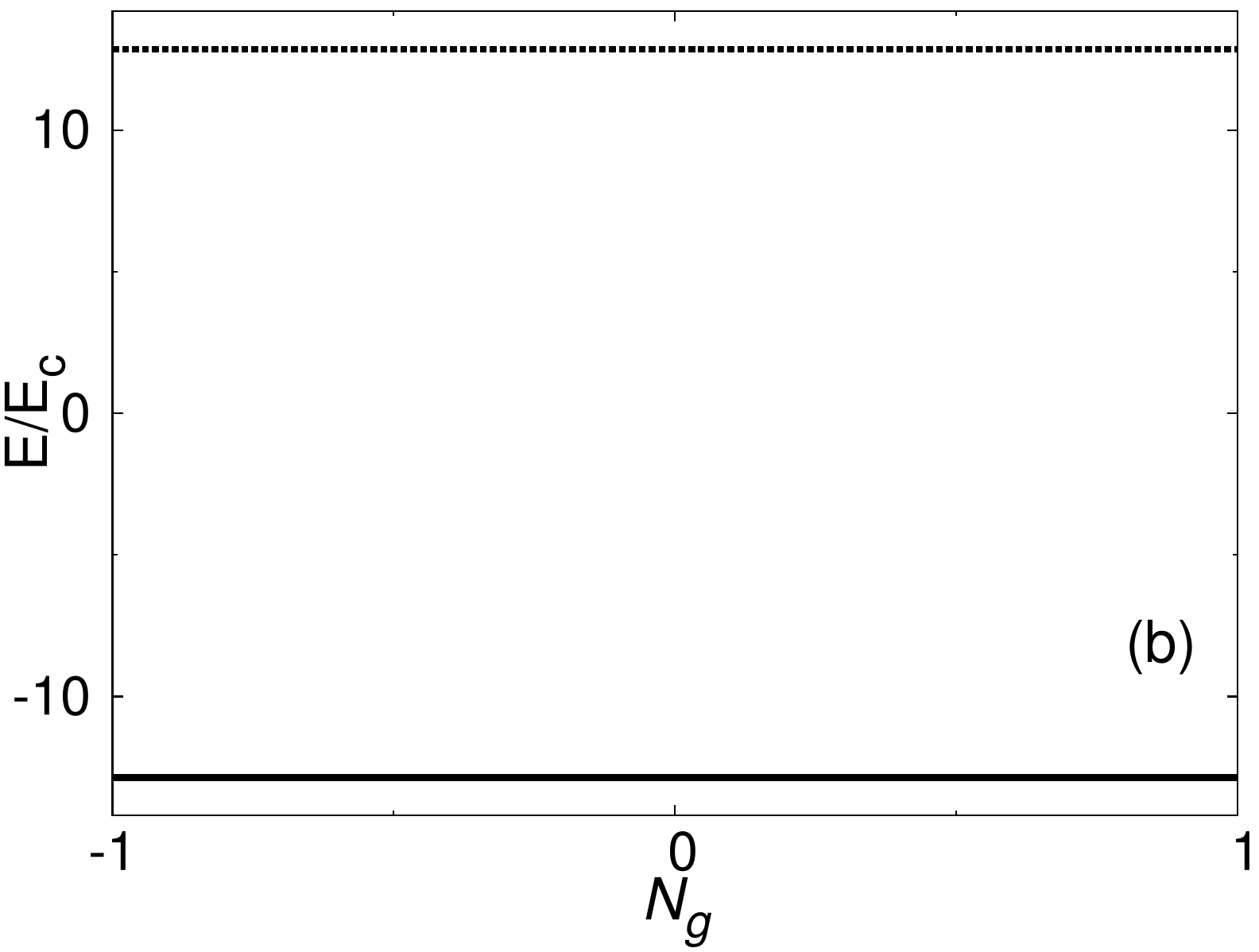}\\
\includegraphics[width=3.2cm, angle=270]{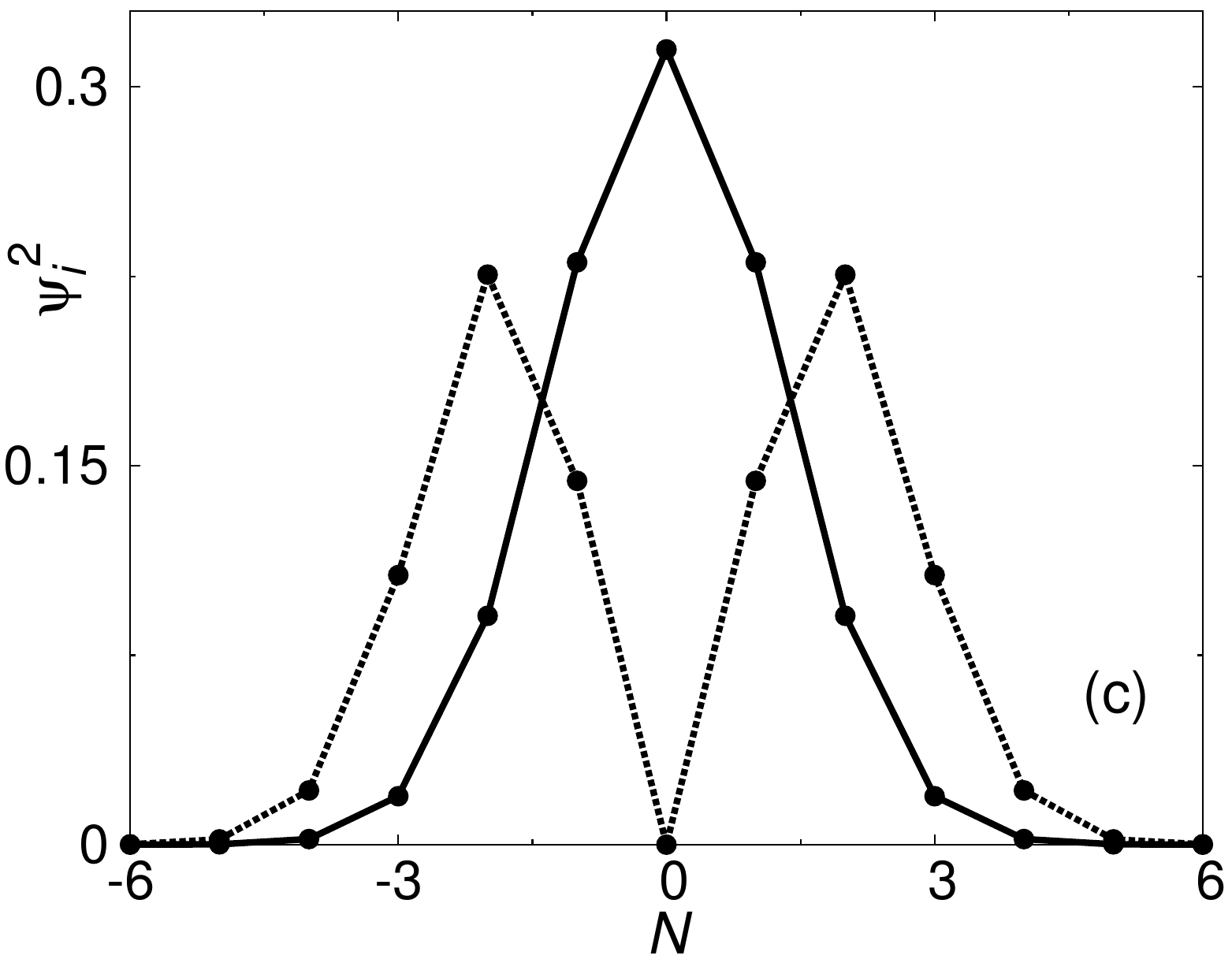}
\includegraphics[width=3.2cm, angle=270]{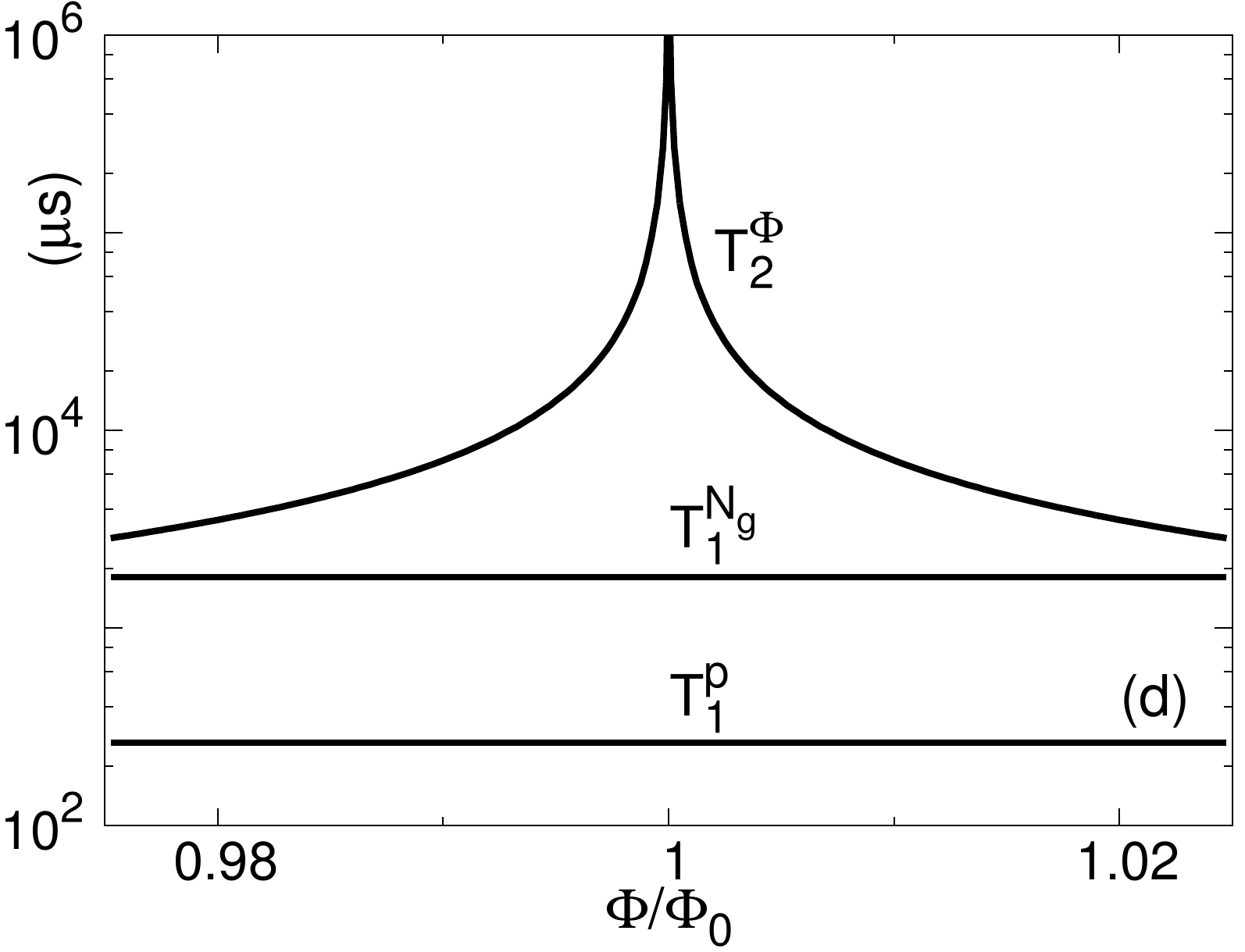}
\caption{(a) Energy levels of gatemon qubit (single well) as a
function of magnetic flux near the lobe's centre. (b) Qubit levels
as a function of $N_g$ at $\Phi/\Phi_0=1$. (c) Distributions of
qubit states at $\Phi/\Phi_0=1$. (d) Limiting timescales due to
$N_g$ ($T^{N_g}_1$) and $\Phi$ ($T^{\Phi}_2$) fluctuations, and
quasiparticle relaxation ($T^{p}_1$). In all frames $E_c=0.35$ GHz
and $N_g=0$ unless otherwise stated. For clarity in (a) and (b)
the energies are shifted near $E=0$.}\label{SWregime}
\end{figure}

\subsection{Qubit properties}

For all the calculations in this subsection we choose the
Josephson junction studied in Fig.~\ref{dwR50}, and explore the
qubit properties at different fluxes in the SW and DW regimes.
Figure~\ref{SWregime} summarizes the results in the SW regime.
Near the lobe's centre the Josephson potential changes slightly
with $\Phi$, hence, the flux dependence of the qubit levels
$E_{0}$ and $E_{1}$ is very weak, Fig.~\ref{SWregime}(a). In
particular, in the limit $\Phi/\Phi_0 \rightarrow 1$ the qubit
frequency remains to a good approximation constant and the
dephasing time $T^{\Phi}_{2}$ is of the order of seconds. The
exact value of $T^{\Phi}_{2}$ at $\Phi/\Phi_0 = 1$ cannot be
precisely resolved due to numerical round off errors producing
spurious derivatives $D_{\Phi}$ and $D_{\Phi\Phi}$; therefore we
have introduced a cutoff value of one second. Away from the lobe's
centre, $D_{\Phi}$ increases producing a shorter $T^{\Phi}_{2}$.
In the limit $E_{J,1} \gg E_c$, the qubit levels are only weakly
dependent on $N_g$~\cite{transmon}, Fig.~\ref{SWregime}(b), making
the qubit immune to charge dephasing. In Fig.~\ref{SWregime}(d) we
derive $T^{\Phi}_{2}$ based on fluctuations in the applied
magnetic flux at a fixed chemical potential. We can also consider
dephasing due to fluctuations in the chemical potential. These
fluctuations might be the result of disorder along the NW. We
assume that $A_{\mu} \sim A_{\Phi}$ and find $T^{\mu}_{2}\approx
35$ $\mu$s ($<T^{\Phi}_{2}$) at the lobe's center. Thus, this is
the limiting dephasing time for the gatemon qubit.

At the lobe's centre the gatemon qubit is expected to be linearly
insensitive to relaxation due to flux induced noise because
$\partial E_{J,M} / \partial \Phi = 0$, Fig.~\ref{dwR50}(a). For
$N_g=0$ this property is also valid away from the lobe's centre,
although in general $\partial E_{J,M} / \partial \Phi \ne 0$. This
can be understood from the symmetries of the qubit states $|\psi_0
(N) \rangle = |\psi_0 (-N)\rangle$ and $|\psi_1 (N) \rangle = -
|\psi_1 (-N)\rangle$. Consequently, the relevant limiting
relaxation times for the gatemon qubit are $T^{N_{g}}_{1}$ and
$T^{{p}}_{1}$, plotted in Fig.~\ref{SWregime}(d). We have
confirmed that the results in Fig.~\ref{SWregime} are general and
in the regime $E_{J,1}\gg E_c$ apply equally well to NWs with
different radii.

\begin{figure}
\includegraphics[width=3.2cm, angle=270]{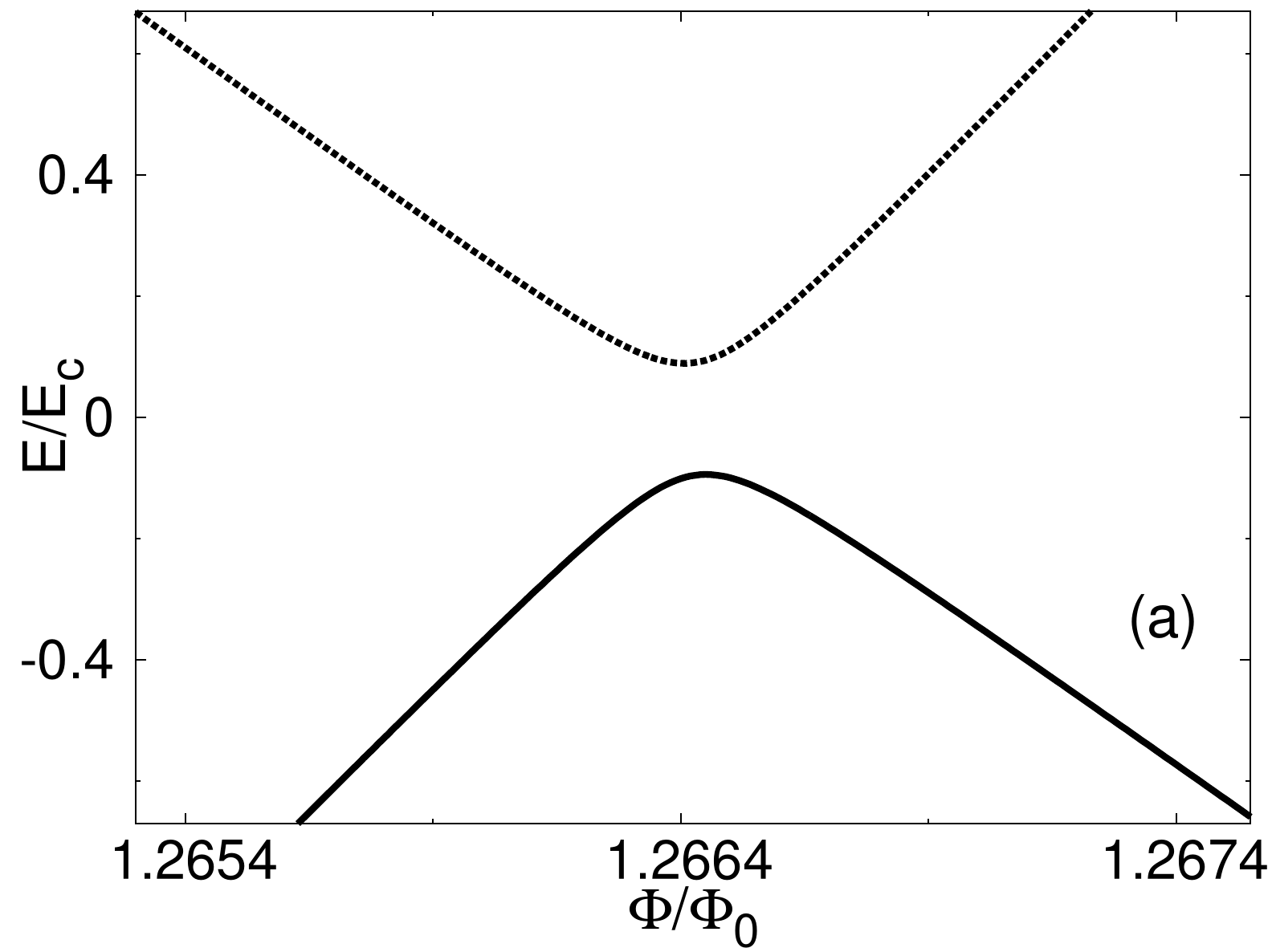}
\includegraphics[width=3.2cm, angle=270]{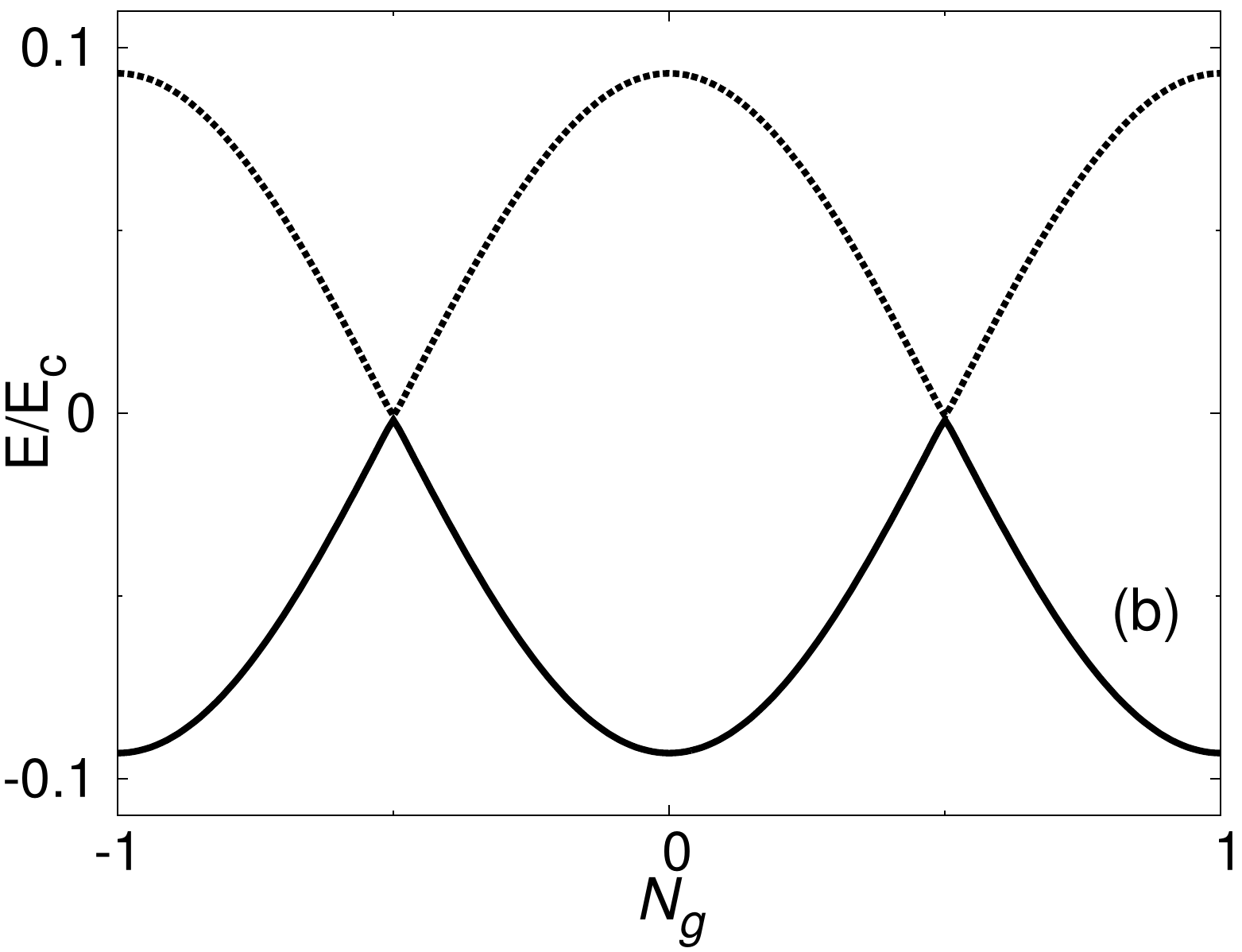}\\
\includegraphics[width=3.2cm, angle=270]{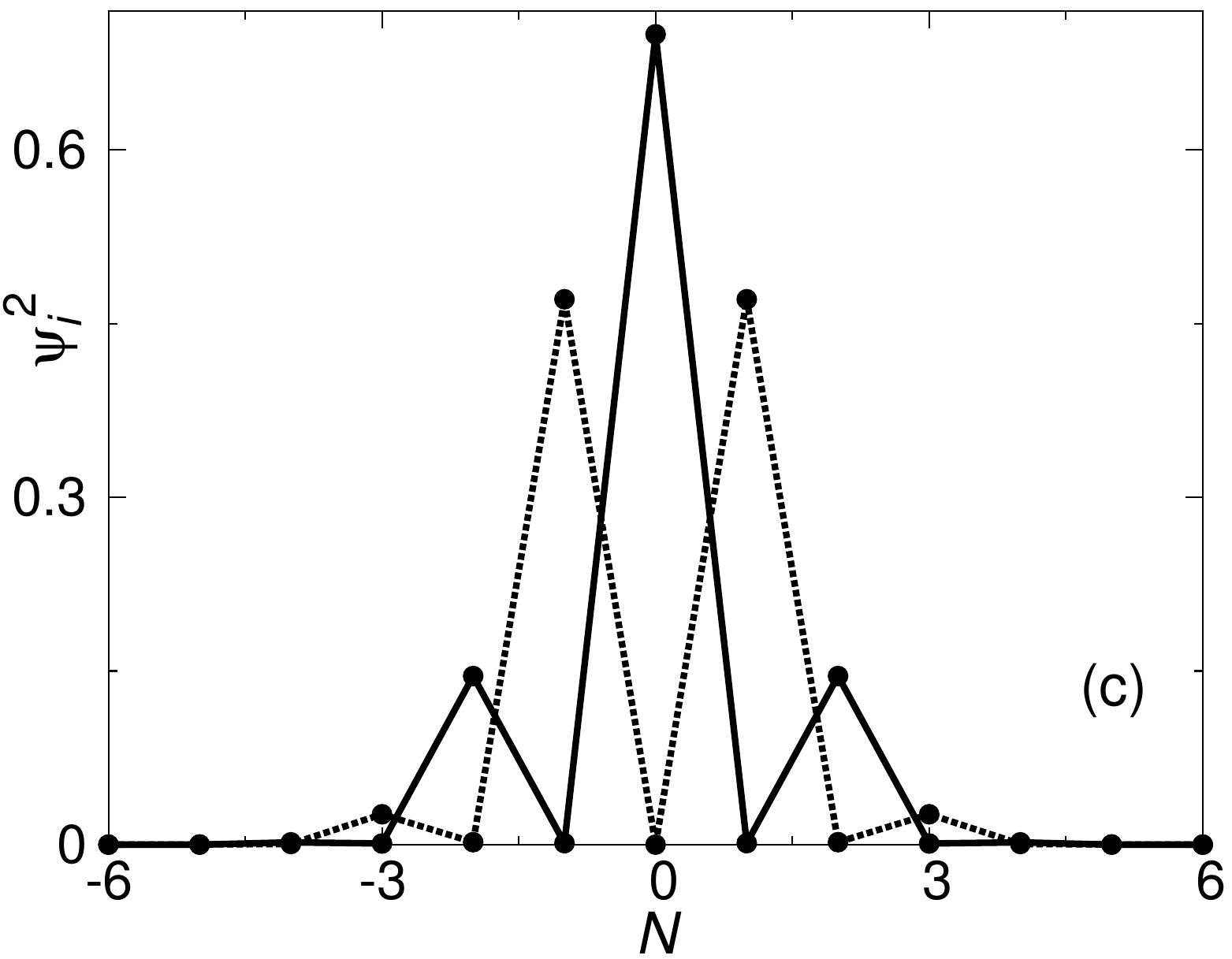}
\includegraphics[width=3.2cm, angle=270]{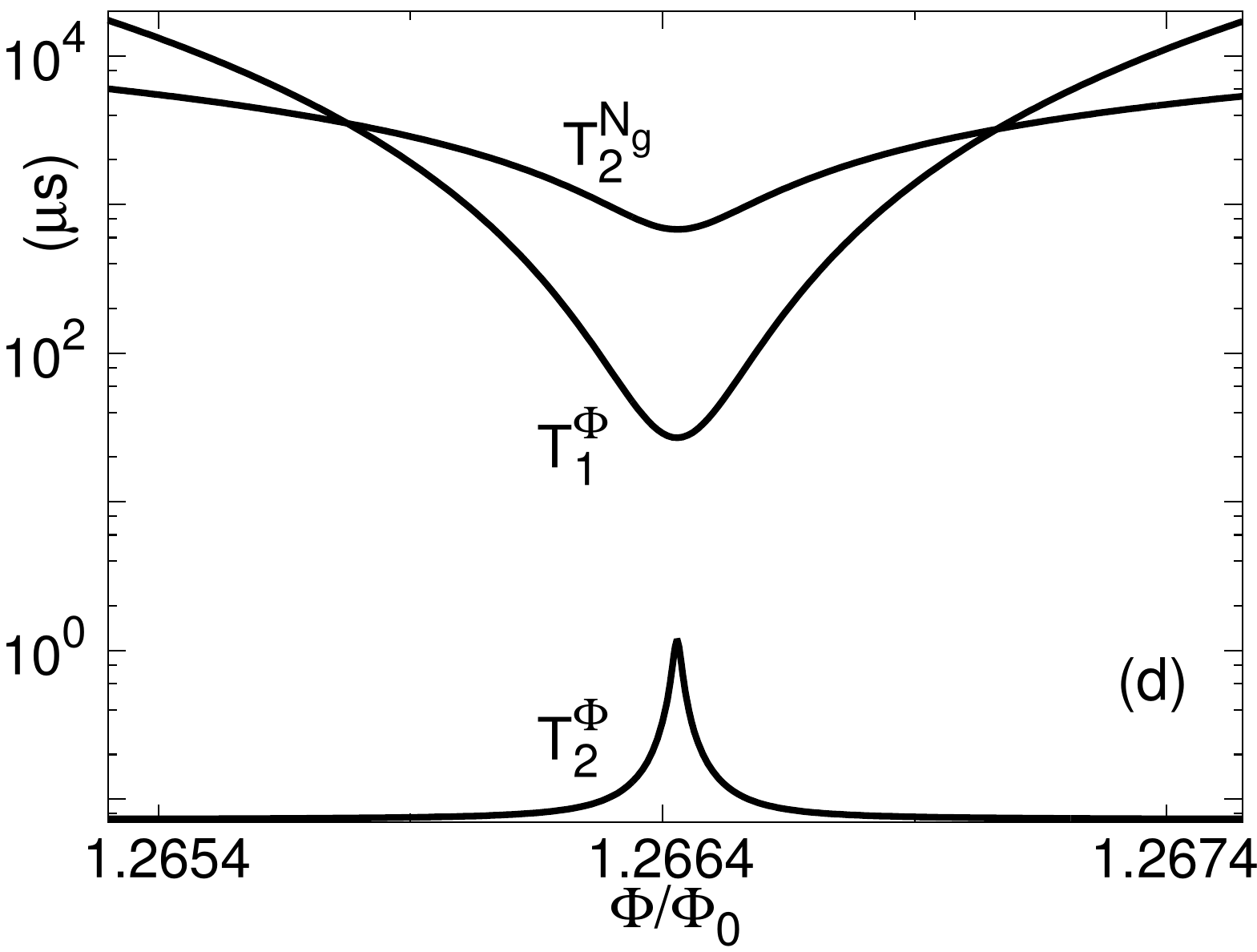}
\caption{(a) Energy levels of parity-protected qubit (double well)
as a function of magnetic flux forming an anticrossing at
$\Phi/\Phi_0 \approx 1.2664$. (b) Qubit levels as a function of
$N_g$ at anticrossing. (c) Distributions of qubit states at
anticrossing. (d) Limiting timescales. In all frames $E_c=0.35$
GHz and $N_g=0$ unless otherwise stated. For clarity in (a) and
(b) the energies are shifted near $E=0$.}\label{DWregime}
\end{figure}

Our above estimates using standard superconducting qubit
parameters exceed by more than two orders of magnitude the short
relaxation times measured in gatemon qubits based on full-shell NW
junctions, $T_1\approx 0.5-1\,\mu$s, in the first lobe
\cite{PhysRevLett.125.156804}. While the microscopic origin of
such short relaxation times is still unclear, subsequent
experiments~\cite{PhysRevB.108.L121406} have measured parity
lifetimes of the order of $100$~ms at $\Phi=0$. These are
consistent with low quasiparticle poisoning and far from limiting
the coherence time of the qubit. In the first lobe, however,
quasiparticle poisoning is rapidly enhanced, possibly due to the
increased density of subgap states (see also
Ref.~\onlinecite{Valentini:N22}), which may explain the short
$T_1^p$.

Figure~\ref{DWregime} summarizes the results in the DW regime. The
qubit levels $E_{0}$ and $E_{1}$ form an anticrossing as a
function of the magnetic flux with an energy gap proportional to
$E_c$ (Appendix~\ref{app:further-analysis}). Hence, small charging
energies lead to a quasi-degenerate qubit. From a slightly
different point of view, this two-level system essentially
operates as a flux qubit with the charging energy governing the
tunneling term between the minima. At temperatures $T \lesssim
15$~mK the qubit is isolated from the excited states since these
lie above the anticrossing point by $\sim 6$ GHz. As illustrated
in Fig.~\ref{DWregime}(b), the value of the avoided crossing
depends on $N_g$ but larger ratios $E_{J,2}/E_c$ weaken this
dependence. For the parameters in Fig.~\ref{DWregime}(b) a
charging energy of $\lesssim 0.1$ GHz leads to nearly flat qubit
levels.

At the anticrossing point defined in the DW regime the qubit
eigenstates are also parity eigenstates, which satisfy $\hat{P}|
\psi_0 \rangle = | \psi_0 \rangle$ and $\hat{P}| \psi_1 \rangle =
- | \psi_1 \rangle$, with $\hat{P}=\exp(i\hat{N}\pi)$ being the
Cooper-pair parity operator. Due to this symmetry, the qubit
eigenstates have the characteristic ``even-odd'' probability
distributions, plotted in Fig.~\ref{DWregime}(c). These
distributions lead to the matrix element $\langle \psi_0 | \hat{N}
| \psi_1 \rangle = 0$ and, because $\partial H /\partial N_g = - 8
E_c \hat{N}$ ($N_g=0$), the relaxation due to charge induced noise
is suppressed, $\Gamma^{N_{g}}_1=0$. This property is one of the
main advantages of the parity-protected qubit in the DW compared
to the gatemon qubit in the SW. The even-odd distributions at the
anticrossing point are exact when the DW is defined uniquely by
$E_{J,2}$, namely, $V_{J}(\varphi_0) = E_{J,2}\cos(2\varphi_0)$,
thus $[\hat{P}, H]=0$. The qubit Hamiltonian
[Eq.~(\ref{qubitham})] contains various $E_{J,M}$ terms making the
even-odd distributions only approximately valid. The approximation
is nevertheless particularly accurate as seen in
Fig.~\ref{DWregime}(c) and remains applicable to the different
junction parameters we have explored.

The results for the coherence times of the parity-protected qubit
in the vicinity of the anticrossing point are presented in
Fig.~\ref{DWregime}(d). These results demonstrate that dephasing
due to flux induced noise $T^{\Phi}_{2}$ is the limiting time
scale. In contrast to the gatemon qubit, now both
$\Gamma^{N_{g}}_{1}$ and $\Gamma^{{p}}_{1}$ are suppressed because
of the DW symmetries; the corresponding matrix elements are zero.
The anticrossing point can be considered as a flux sweet spot
where $T^{\Phi}_{2}$ is an order of magnitude longer compared to
$T^{\Phi}_{2}$ away from the anticrossing. In addition, larger
values of $E_c$ tend to increase $T^{\Phi}_{2}$ as they produce a
smoother anticrossing. The results in Fig.~\ref{DWregime}(d) are
for $N_g=0$ which defines the charge sweet spot. For $N_g\ne0$ we
have observed no improvement in the coherence times while
simultaneously $\Gamma^{N_{g}}_{1}\ne0$ and
$\Gamma^{{p}}_{1}\ne0$. According to our calculations in
Appendix~\ref{app:further-analysis} the latter range from
$T^{{p}}_{1}> 1$ s near $N_g=0$ to $T^{{p}}_{1}\approx 10$~ms as
we approach $N_g=0.5$. For the parity-protected qubit we can also
consider relaxation and dephasing due to fluctuations in the
chemical potential at a fixed magnetic flux. Assuming again that
$A_{\mu} \sim A_{\Phi}$ and focusing on the anticrossing, we find
that the times $T^{\mu}_{1,2}$ are at least two orders of
magnitude longer than the corresponding $T^{\Phi}_{1,2}$. This
result indicates that magnetic flux fluctuations is the limiting
noise source.

Although our estimates are for specific NW parameters and noise
amplitudes, we expect the change from a $T^{\mu}_2$-limited qubit
in the SW regime to a $T^{\Phi}_2$-limited qubit in the DW regime
to be general enough. We also expect that longer coherence times
in the DW regime can be obtained with some simple parameter
optimization. Interestingly, a qubit formed with two NW-based
junctions in a SQUID configuration~\cite{protected_exper} can
reach relaxation times of $T_1 \approx 7$ $\mu$s in the
parity-protected regime, in contrast to $T_1 \approx 0.6$ $\mu$s
outside the protected regime. However, we emphasize that in this
SQUID the flux enters very differently as compared to our qubit
system~\footnote{It is useful to compare Eq.~(2) in
Ref.~\onlinecite{protected_exper}, where the flux enters as a
phase modulation, $\varphi_0\rightarrow
\varphi_0-2\pi\Phi/\Phi_0$, with our Eqs.~(\ref{aprox1}),
~(\ref{gamma}) and ~(\ref{aprox2}) where the flux enters as a
linear shift.}. In particular, in
Ref.~\onlinecite{protected_exper} the flux is tuned to
$\Phi/\Phi_0=1/2$, which in practice requires a relatively large
SQUID loop enhancing the sensitivity to flux noise. In contrast,
the flux in our qubit system enters through the axial magnetic
field applied to the NW. These considerations are expected to lead
to different behaviors against flux fluctuations, and hence in the
resulting $T^{\Phi}_{1}$ and $T^{\Phi}_{2}$ times. Having
clarified the above differences, we expect the physical behavior
of these two qubits to be relatively similar near the anticrossing
(see also Ref.~\onlinecite{PhysRevLett.115.127002}).

\section{Discussion and conclusions}\label{conclusions}

We considered a Josephson junction formed in a full-shell NW and
demonstrated that we can change the Josephson potential from a SW
to a DW potential by tuning the applied magnetic flux threading
the NW section. These potential wells can serve as the basis of a
gatemon qubit and a parity-protected qubit respectively. The
protected qubit studied in our work is defined in a single
Josephson junction and is arguably advantageous compared to more
complex geometries proposed earlier and based on interferometers.

Some of the conclusions of our study that could be relevant
towards the experimental implementation of these ideas include: i)
NWs with small radii, e.g. < 60 nm, are generally better to find a
DW regime and define a parity-protected qubit. In these NWs the DW
is formed away from the lobe's boundary, thus the flux modulation
of the pairing potential due to the LP effect has a negligible
role. For NWs with large radii the formation of a DW potential
depends on the strength of the LP effect and the coherence length
of the superconducting shell; ii) we have studied junctions with
lengths of $100-200$ nm and found that the DW is more likely to
arise in long junctions. The reason is that the energy levels
lying above the superconducting gap are actively involved in the
formation of the DW and their role becomes more significant as the
junction length increases.

Qubit experiments based on full-shell
NWs~\cite{PhysRevLett.125.156804,PhysRevLett.126.047701,PhysRevB.108.L121406}
typically find short time scales $T_1\approx 0.5-1\,\mu$s. This
might be caused by the quasiparticle poisoning enhancement near
$\Phi/\Phi_0 \approx 1$ due to the increased density of subgap
states~\cite{Valentini:N22}. It is interesting though, that the
density of the subgap states is gate-tunable and that recent
experiments have demonstrated their coherent control with
transitions yielding relaxation times $T_1\approx
3-5\,\mu$s~\cite{PhysRevLett.126.047701}. These encouraging
results demonstrate that the combined gate and flux tunability of
full-shell NW Josephson junctions could offer a wide range of
possibilities when designing qubits. Furthermore,
simulations~\cite{paya_JJ} beyond the hollow-core model show a
rich phenomenology that includes $0$, $\pi$ and $\phi_0$ junctions
that could be exploited for qubit designs. Also, the flux
modulation of ABSs is not unique to full-shell NWs, since a
magnetic field parallel to the NW causes orbital
effects~\cite{PhysRevB.100.155431}, and thus a flux-dependent
qubit frequency~\cite{PhysRevB.108.L020505}.

\begin{acknowledgments}
We acknowledge the support of the Spanish Ministry of Science
through Grants PID2021-125343NB-I00 and TED2021-130292B-C43 and
PID2022-140552NA-I00 funded by MCIN/AEI/10.13039/501100011033,
``ERDF A way of making Europe'', the Spanish Comunidad de Madrid
``Talento Program'' (Project No. 2022-T1/IND-24070), and European
Union NextGenerationEU/PRTR. We acknowledge the support of the
CSIC's Quantum Technologies Platform (QTEP). Support from the CSIC
Interdisciplinary Thematic Platform (PTI+) on Quantum Technologies
(PTI-QTEP+) is also acknowledged.
\end{acknowledgments}

\setcounter{secnumdepth}{0} 

\setcounter{secnumdepth}{1}

\appendix

\section{Detailed description of full-shell nanowire}
\label{app:full-shell}

We consider a semiconducting NW with cylindrical symmetry and
assume within the hollow-core approximation~\cite{model} that the
electrons are confined near the surface of the NW. Working in
cylindrical coordinates $(r, \varphi, z)$ with the unit vectors
$({\hat e_r},{\hat e_\varphi},{\hat e_z})$ the hollow-core
approximation allows us to fix the radial coordinate, $r=R_0$,
where $R_0$ is the radius of the NW. As a result, the flux
threading the cross-section of the NW is $\Phi=\pi B R_0^2$ with
$B$ being the magnitude of the applied magnetic field along the
direction of the NW; ${\vec B}=B{\hat e_z}$. The magnetic vector
potential in the azimuthal direction is
\begin{equation}
\vec{A}=A_{\varphi}{\hat e_\varphi}=\frac{\Phi}{2\pi R_0} {\hat
e_\varphi},
\end{equation}
and the Hamiltonian of the NW is
\begin{equation}\label{NW}
H_0 = \frac{(\vec{p}+e \vec{A} )^2}{2m^*}-\mu+H_{\text{SO}},
\end{equation}
where $m^{*}$ is the effective mass, $\mu$ is the chemical
potential, and the momentum operator has the form
\begin{equation}
\vec{p}=(p_\varphi,p_z)=\left(-\frac{i\hbar}{R_0}\partial_\varphi,
-i\hbar\partial_z \right).
\end{equation}
The Rashba spin-orbit (SO) Hamiltonian is
\begin{equation}
H_{\text{SO}}= \frac{\alpha_{\rm so}}{\hbar}
\left[p_z\sigma_\varphi-(p_\varphi+eA_{\varphi})\sigma_z\right],
\end{equation}
and comes from radial inversion symmetry breaking, i.e.,
$\vec{\alpha}=\alpha_{\rm so}{\hat e_r}$. Finally
$\sigma_\varphi=(\sigma_y \cos\varphi-\sigma_x \sin\varphi)$ and
$\sigma_z$ are spin-1/2 Pauli matrices.

The semiconducting NW is wrapped by a thin superconducting shell
and due to the proximity effect it acquires superconducting
pairing terms. An applied magnetic flux modulates the induced
superconductivity via the Little-Parks effect which in turn
induces a winding of the superconducting phase in the shell around
the NW axis, $\Delta(\vec{r})=\Delta(r) e^{i n\varphi}$. The
amplitude $\Delta$, with $r=R_0$, and the winding number $n$
depend on the applied magnetic flux
(Appendix~\ref{app:flux-modulation}). Denoting by $\Phi_0=h/2e$
the flux quantum, the winding number is equal to the nearest
integer to $\Phi/\Phi_0$, i.e., $n=0$, $\pm 1$, $\pm 2$, $\ldots$
We can then use the variable $\phi = n -\Phi/\Phi_0$ to measure
deviations from integer fluxes, and also label with $n$ the
so-called LP lobes~\cite{lobes}.

The Bogoliubov-de-Gennes Hamiltonian describing the proximitized
NW can be written as decoupled Hamiltonians characterized by the
eigenvalues $m_j$ of a generalized angular momentum
operator~\cite{model}
$J_z=-i\partial_\varphi+\frac{1}{2}\sigma_z+\frac{1}{2}n
\tilde{\tau}_z$ (with $\hbar=1$ and $\tilde{\tau}_{z}$ acts in
Nambu space). For odd $n$, $m_j=0$, $\pm1$,~$\ldots$ and for even
$n$, $m_j=\pm 1/2$, $\pm 3/2$,~$\ldots$ The final BdG Hamiltonian
for a given $m_j$ can be written in the compact matrix form
\begin{equation}\label{Hmatrix}
H_{\rm BdG} = \left(\begin{array}{cc}
   H^+  & H^z_{\text{SO}} \\
  -H^z_{\text{SO}} & H^-  \\
\end{array}\right),
\end{equation}
with the $2\times2$ matrices
\begin{equation}
H^{z}_{\rm SO} = \left(\begin{array}{cc}
 -\alpha_{\rm so}\partial_z    & 0 \\
    0 &  \alpha_{\rm so}\partial_z \\
\end{array}\right),
\end{equation}
and
\begin{equation}\label{HA}
H^+ = \left(\begin{array}{cc}
  \frac{p_z^2}{2m^*} + \mu_e^+ & \Delta\\
\Delta^{*}& -\frac{p_z^2}{2m^*} + \mu_h^+    \\
\end{array}\right).
\end{equation}
The Hamiltonian $H^-$ has the same form as $H^+$ but with
$\mu_{e}^+\rightarrow \mu_{e}^-$ and $\mu_{h}^+\rightarrow
\mu_{h}^-$. The flux dependent effective potentials are
\begin{eqnarray}\label{effe}
\mu_{e(h)}^+ (\phi) &=& \mu^{0+}_{e(h)}+ \delta^+(\phi) \pm \frac{\hbar^2}{8m^{*}R^{2}_{0}} \phi^2  ,\nonumber\\
\mu_{e(h)}^- (\phi) &=& \mu^{0-}_{e(h)}+ \delta^-(\phi) \pm
\frac{\hbar^2}{8m^{*}R^{2}_{0}} \phi^2,
\end{eqnarray}
and at the centre of a lobe ($\phi=0$) we have
\begin{equation}
\mu^{0\pm}_{e} = - \mu + \frac{(1\mp 2m_j)^2
\hbar^2}{8m^{*}R_0^{2}} + \frac{(1\mp 2m_j)\alpha_{\rm so}}{2R_0},
\end{equation}
with $\mu^{0+}_{h}=-\mu^{0+}_{e}$ and
$\mu^{0-}_{h}=-\mu^{0-}_{e}$. Finally, the linear terms are
\begin{equation}
\delta^\pm (\phi) = \frac{(-2m_j \pm 1)\hbar^2}{4m^{*}R^{2}_0}\phi
\pm \frac{\alpha_{\rm so}}{2 R_0}\phi,
\end{equation}
which dominate for small fluxes near the centre of a lobe. For
each $m_j$, the BdG Hamiltonian produces two subgap modes
originating from the Hamiltonians $H^+$ and $H^-$. At low chemical
potentials only one mode lies in the superconducting gap, while an
increase of the order of $\mu^{0-}_{e}-\mu^{0+}_{e}$ puts the
second mode into the gap~\cite{giavaras24}. For $\alpha_{\rm
so}=0$ degenerate BdG energy levels occur because
$\mu^{0+}_e(m_j+1)=\mu^{0-}_{e}(m_j)$~\cite{giavaras24}.

To obtain the eigenvalues of $H_{\rm BdG}$, we expand the BdG
solution in electron-hole basis states $| u_i \rangle$
($\Delta=0$), calculate the matrix elements $\langle u_j| H_{\rm
BdG} | u_i \rangle$, and finally diagonalize the resulting BdG
matrix. The average energy spacing of the basis states with
respect to $\Delta$ determine the required number of basis states
(a few hundreds). This number decreases in the presence of the
Little-Parks effect making the method more powerful. The exact BdG
model is needed in order to account for the energy levels which
lie outside the superconducting gap. We emphasize that without
these levels the Josephson potential cannot be accurately
computed.

\section{Flux modulation of pairing potential}
\label{app:flux-modulation}

\begin{figure}
\includegraphics[width=4.5cm, angle=270]{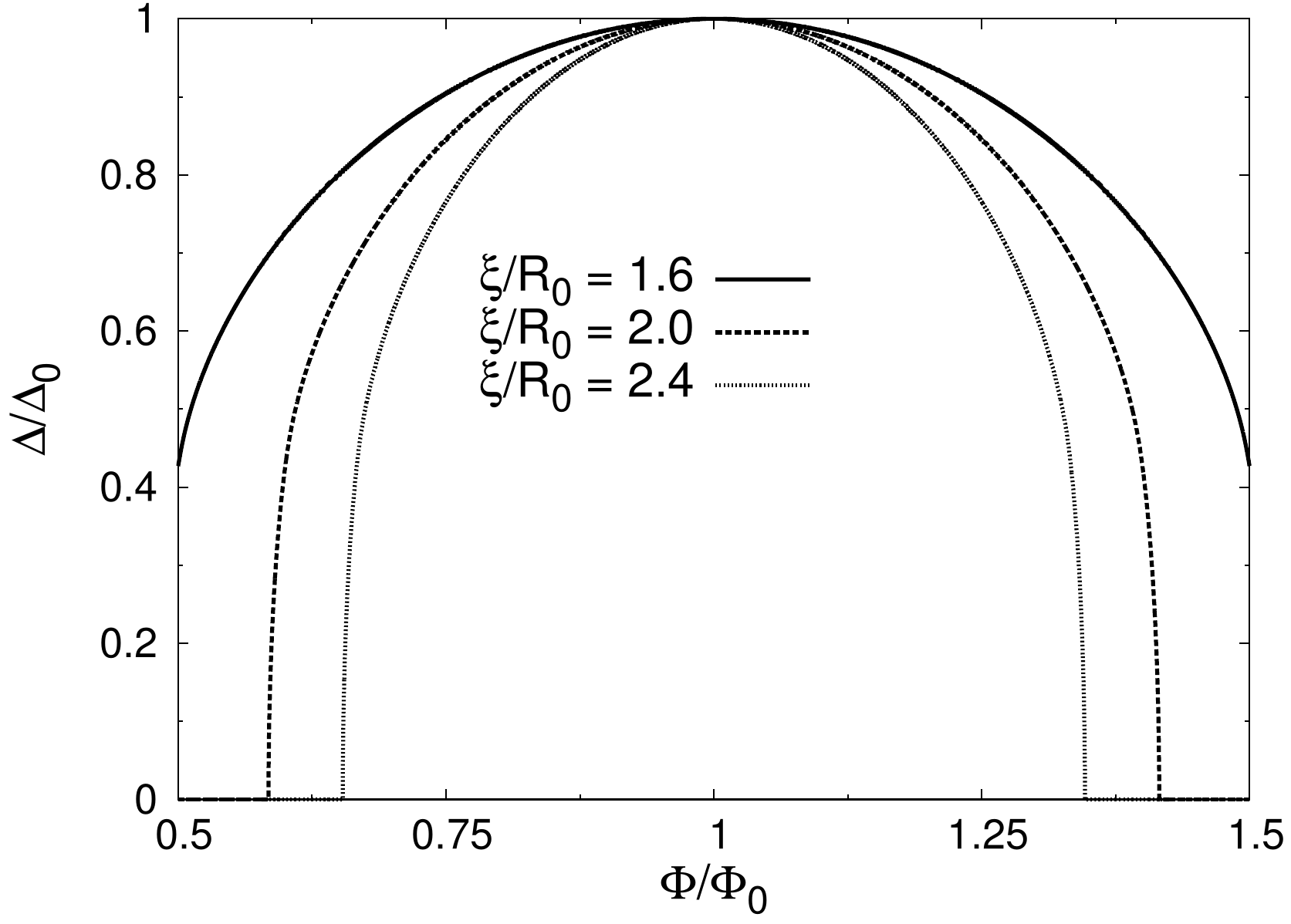}
\caption{Pairing potential as a function of magnetic flux for
different ratios $\xi/R_0$, where $\xi$ is the coherence length of
the superconducting shell and $R_0$ is the radius of the
semiconducting NW.}\label{Dpair}
\end{figure}

In full-shell NWs the LP effect results in a flux dependent
pairing potential term, $\Delta = \Delta(\Phi)$. The flux
dependence can be derived from the expression~\cite{skalski}
\begin{eqnarray}\label{delta}
\ln\frac{\Delta}{\Delta_0} &=&
-\frac{\pi}{4}\frac{\Lambda}{\Delta}, \quad \Lambda \le \Delta, \notag\\
\ln\frac{\Delta}{\Delta_0} &=& - \ln\left( \frac{\Lambda}{\Delta}
+ \sqrt{ (\Lambda/\Delta)^2 - 1}\right) +
\frac{\sqrt{(\Lambda/\Delta)^2-1}}{2(\Lambda/\Delta)} \notag\\
&&-\frac{\Lambda}{2\Delta}\arctan\frac{1}{\sqrt{(\Lambda/\Delta)^2-1}},
\quad \Lambda \ge \Delta.
\end{eqnarray}
$\Delta_0$ denotes the value of $\Delta$ at zero flux and
$\Lambda$ is the flux-dependent pair-breaking term $\Lambda =
\Lambda(\Phi)$. Assuming that the thickness of the superconducting
shell is much smaller than the radius $R_0$ then, to a good
approximation
\begin{equation}
\Lambda(\Phi) \approx \frac{ 4 \xi^2 k_{B} T_{c}}{\pi R^{2}_{0}}
\left( n - \frac{\Phi}{\Phi_0} \right)^{2},
\end{equation}
where $n$ is the lobe index, $T_{c}$ is the zero-flux critical
temperature, $\xi$ is the coherence length of the superconducting
shell and $k_{B} T_{c} = \Delta_0/1.76$. Equation~(\ref{delta}) is
solved numerically and some typical examples of the pairing
potential as function of the magnetic flux are shown in
Fig.~\ref{Dpair}. Larger values of $\xi$ lead the junction to the
destructive regime where $\Delta=0$ for $\Lambda \gtrsim
\Delta_0/2$.

\section{Additional examples of Josephson potentials}
\label{app:examples}

\begin{figure}
\includegraphics[width=3.0cm, angle=270]{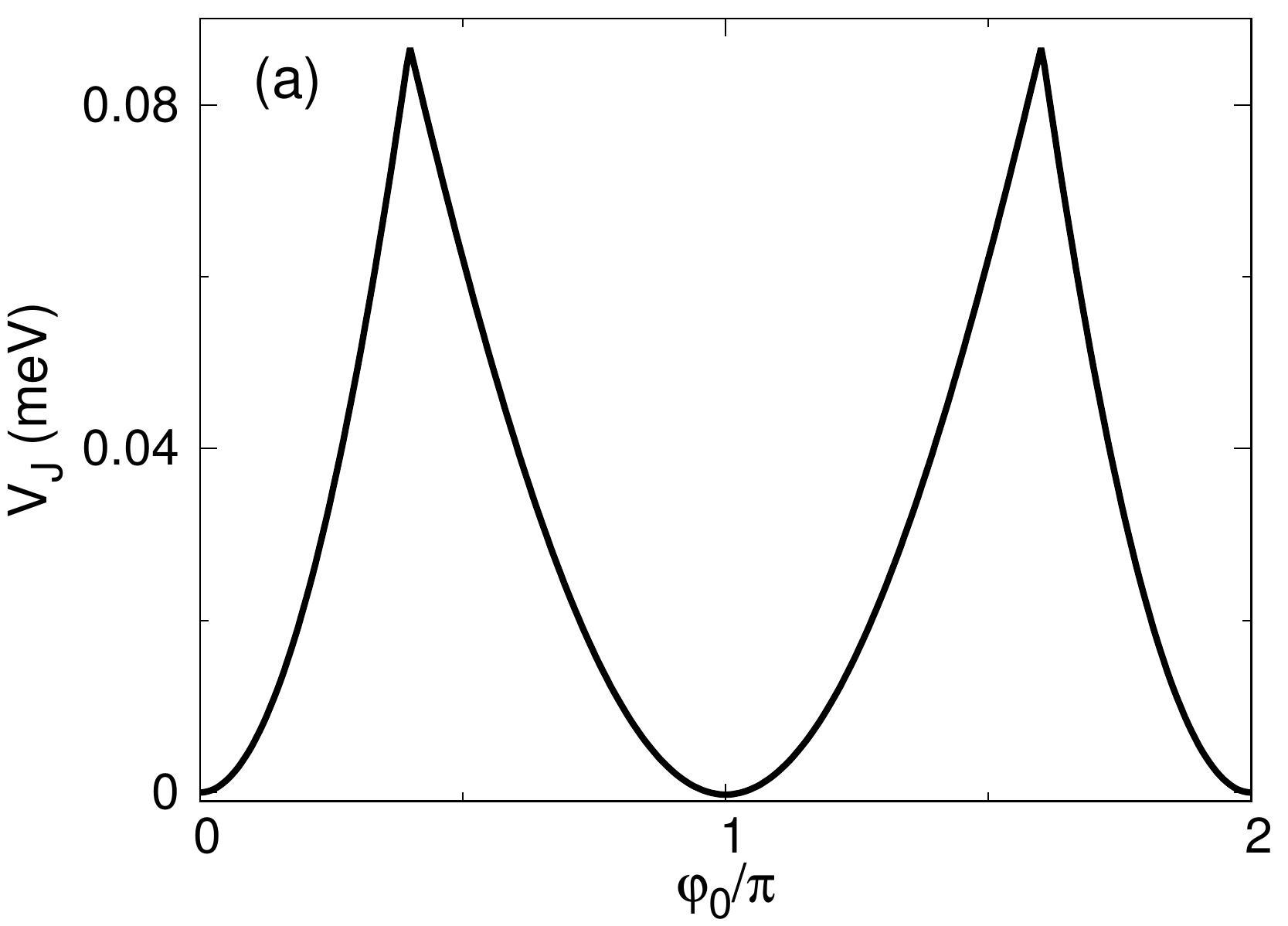}
\includegraphics[width=3.0cm, angle=270]{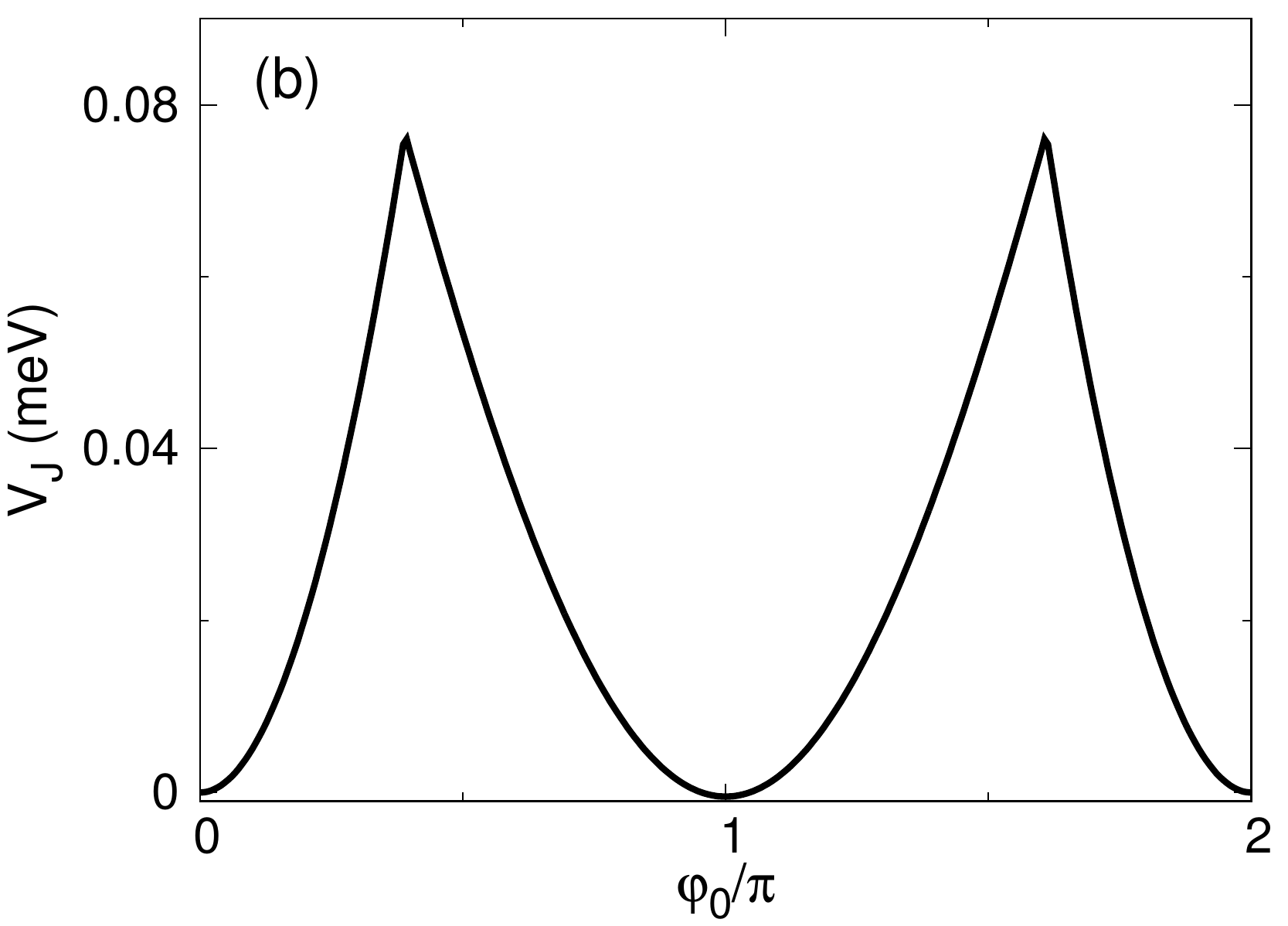}\\
\includegraphics[width=3.0cm, angle=270]{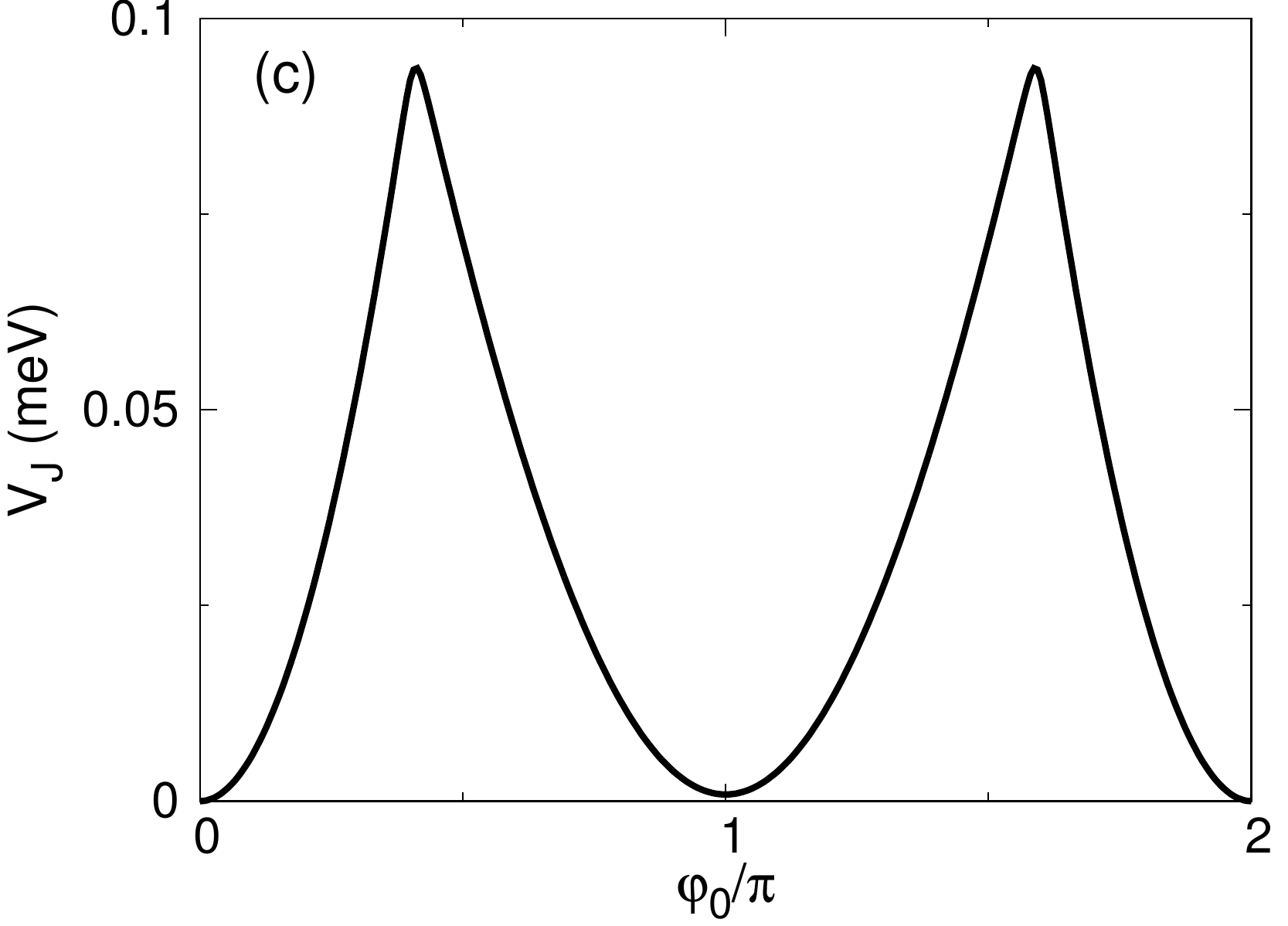}
\includegraphics[width=3.0cm, angle=270]{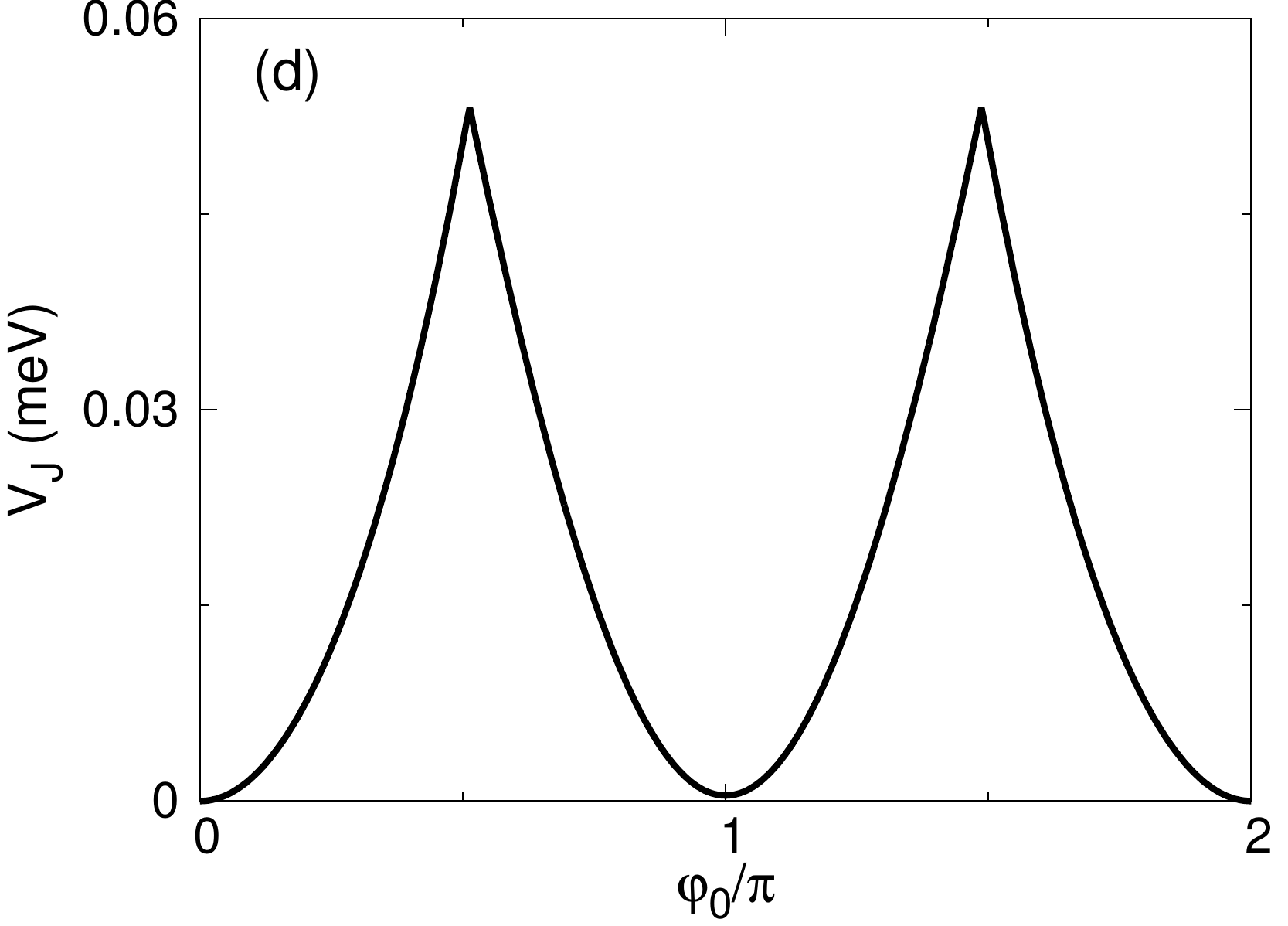}
\caption{Examples of double well Josephson potentials. (a)
$\Delta=\Delta(\Phi)$ with $\xi=80$ nm, (b) $\Delta=\Delta(\Phi)$
with $\xi=120$ nm, (c) $\Delta=\Delta_0=0.2$ meV at finite
temperature $T=15$ mK, and (d) $\Delta=\Delta_0=0.2$ meV including
a Zeeman term with $g$-factor $g=10$. Parameters: $L_S=2000$ nm,
$L_N=100$ nm, and $R_0=50$ nm.}\label{ksi}
\end{figure}

\begin{figure}
\includegraphics[width=4.0cm, angle=270]{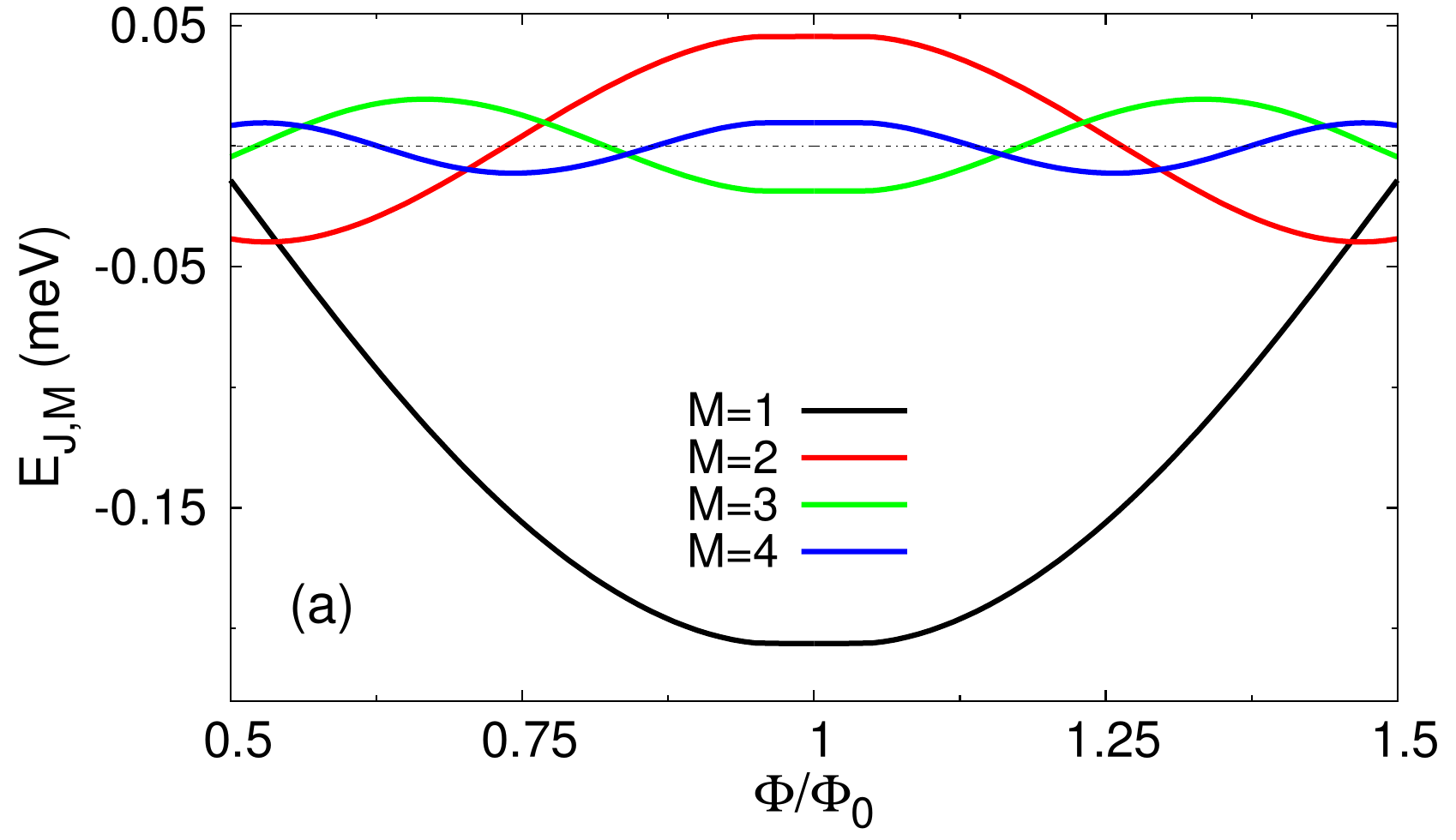}\\
\includegraphics[width=4.0cm, angle=270]{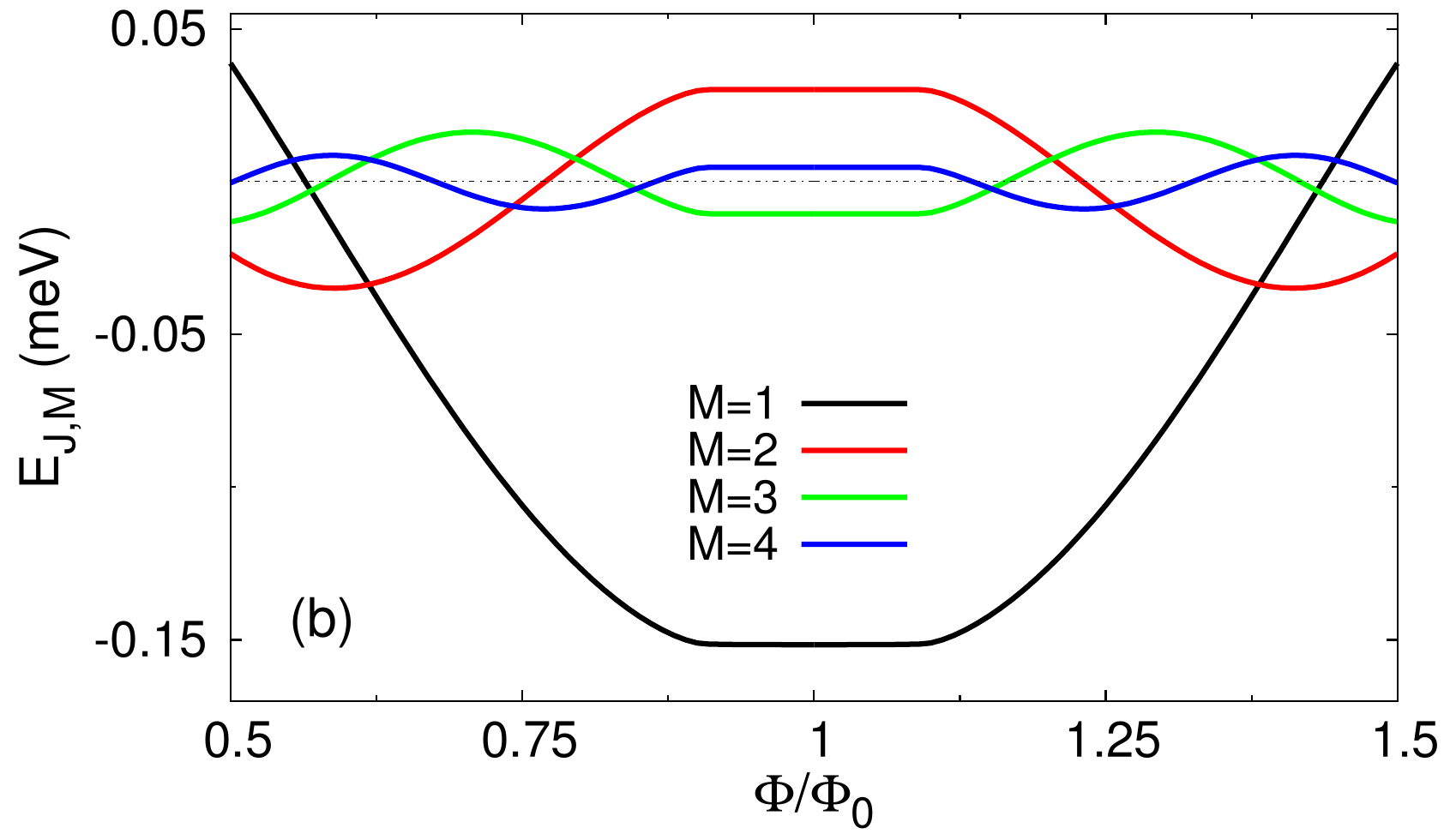}
\includegraphics[width=3.0cm, angle=270]{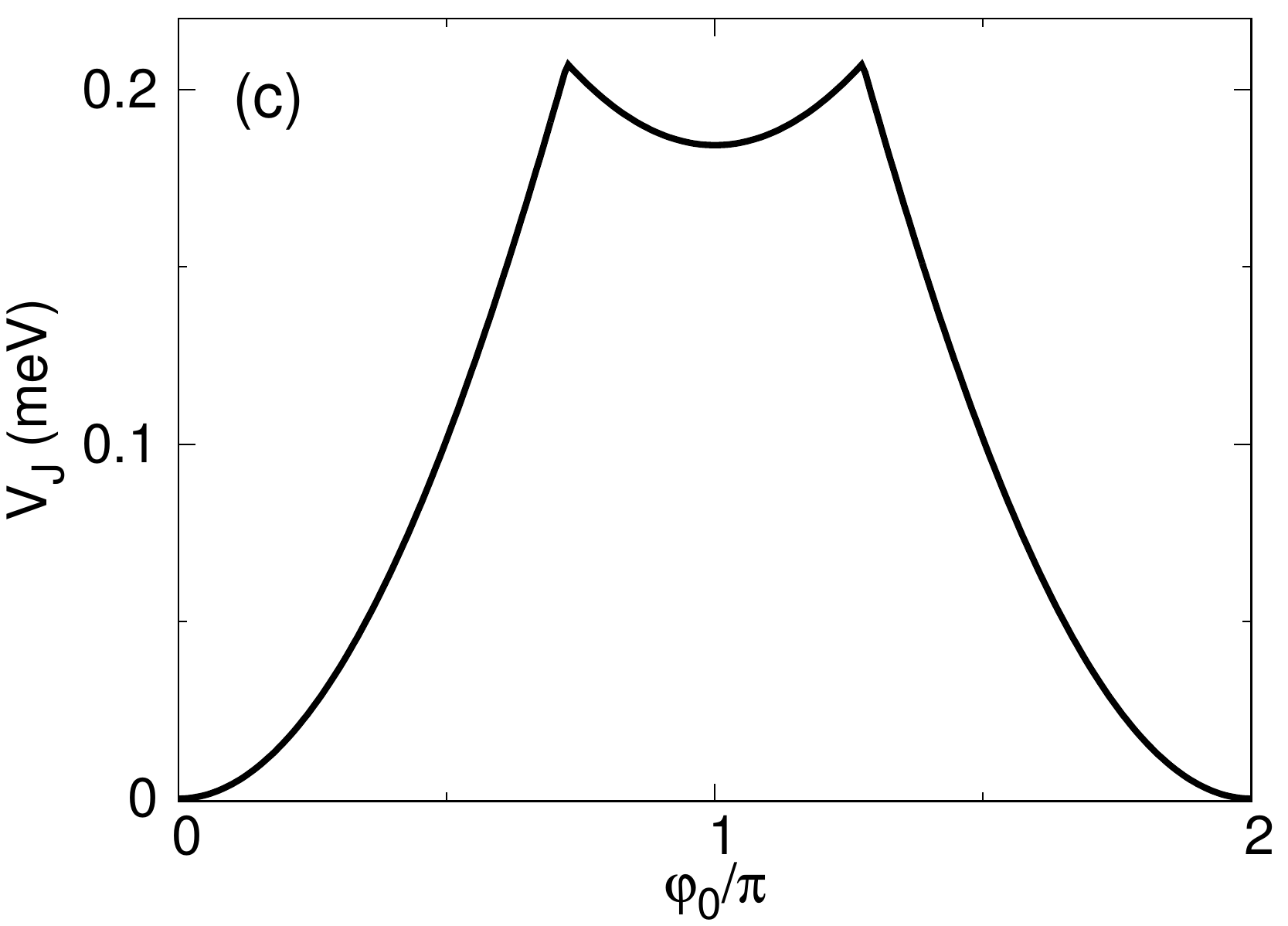}
\includegraphics[width=3.0cm, angle=270]{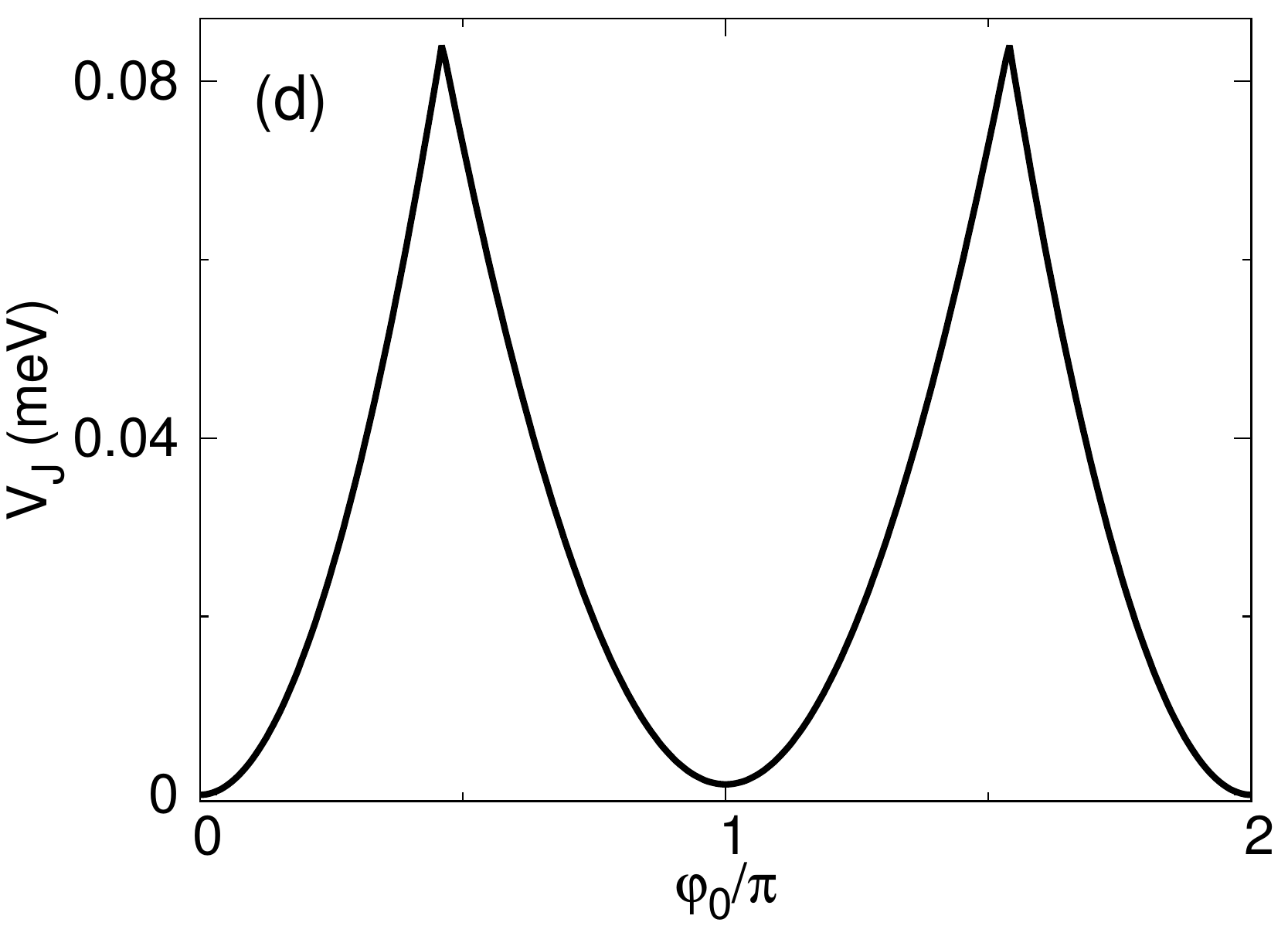}
\caption{Fourier components of Josephson potential for $M=1-4$ as
a function of magnetic flux for $\Delta=\Delta(\Phi)$, $\xi=50$
nm, and $L_N=100$ nm (a). As in (a) but for $L_N=200$ nm (b).
Josephson potential for $\Delta=\Delta(\Phi)$, $\xi=50$ nm, and
$L_N=200$ nm at different fluxes: $\Phi/\Phi_0=1.25$ (c), and 1.44
(d). Parameters: $L_S=2000$ nm, and $R_0=75$ nm.}\label{dwR75LN}
\end{figure}

In this appendix we present additional examples of the Josephon
potential. We consider a NW with $R_0=50$ nm and study the most
general case where $\Delta=\Delta(\Phi)$. We gradually tune the
junction to the destructive regime by increasing $\xi$ and compute
the Josephson potential from the BdG Hamiltonian. We focus on the
DW regime and show the results in Fig.~\ref{ksi}(a) and (b). It
can be seen that the LP effect induces a small decrease in the
potential barriers but the overall shape of the DW remains
unaffected even for relatively large values of $\xi$ ($120$ nm)
that lead the junction to the destructive regime. Our analysis
indicates that this is a general result, namely, for a NW with a
small radius the LP effect has no significant role.

The DW Josephson potentials studied in Sec.~\ref{wells} are for
zero temperature, however, a DW Josephson potential can also be
formed at finite temperatures. To demonstrate this argument we
focus on realistic temperatures for superconducting qubits, $T
\lesssim 30$ mK, and present in Fig.~\ref{ksi}(c) an example of a
DW potential at $T=15$ mK. Compared to $T=0$ K (see
Fig.~\ref{dwR50}) the finite temperature has a negligible effect.
This result is in agreement with an earlier study of the critical
current at $T\ne0$ K in full-shell NWs~\cite{giavaras24}. Finally,
we take into account the Zeeman effect due to the applied magnetic
field $B$. Here, we include the Zeeman term $g \mu_B B \sigma_z/2$
in the BdG Hamiltonian, Eq.~(\ref{Hmatrix}), where $g$ is the
Lande $g$-factor of the NW and $\mu_B$ is the Bohr magneton. We
have explored different junctions giving rise to DW Josephson
potentials in the first LP lobe; one example is presented in
Fig.~\ref{ksi}(d). Based on our numerical investigation we
conclude that including a small Zeeman term in the BdG Hamiltonian
tends to produce weaker Josephson potentials due to the
rearrangement of the flux dependent potentials $\mu_{e}^{\pm}$ and
$\mu_{h}^{\pm}$. However, following the general arguments given in
Sec.~\ref{wells} we can always form a flux tunable DW potential
for realistic $g$-factor values $g \lesssim 20$. Furthermore, with
a fine tuning of the parameters, such as $R_0$, $\xi$, $g$, and
$L_N$, the Zeeman term can lead to more symmetric DW potentials,
though this special case cannot be always guaranteed.

According to the description in Sec.~\ref{wells} a longer
Josephson junction can be advantageous to form a DW potential. In
Fig.~\ref{dwR75LN} we show one typical example comparing two
junctions with different $L_N$ lengths. Tuning the magnetic flux
forms a DW with nearly equal potential minima
[Fig.~\ref{dwR75LN}(d)] only for the longer junction ($L_N=200$
nm). This example confirms the significance of the energy levels
lying above the supeconducting gap.

Deviations from the cylindrical symmetry of the NW, for example
due to disorder, can affect the flux dependence of the BdG levels
and the resulting Josephson potential. To account for this aspect
we introduce a coupling between the BdG energy levels that come
from different angular numbers $m_j$. Specifically, in the BdG
Hamiltonian [Eq.~(\ref{Hmatrix})] we couple the electron states
with the same spin component but with different $m_j$. We consider
this coupling between nearest $m_j$ numbers and follow the same
procedure for the hole states. We explore the simplest regime
where the coupling strength, $q$, is assumed to be constant along
the entire length of the junction. Although this regime is
probably not the most realistic, it gives valuable insight into
how robust the Josephson potential is to weak asymmetries due to
possible disorder and/or geometrical imperfections in the NW.

For a NW with $R_0=50$ nm, $\Delta=\Delta_0=0.2$ meV, and $q=0$ a
DW potential is formed at $\Phi/\Phi_0 \approx 1.26$
(Sec.~\ref{wells}). Increasing $q$ gradually shifts the DW to
lower fluxes until the DW eventually reaches the lobe's centre. A
further increase in $q$ destroys the DW potential. This means that
an anticrossing point between the qubit levels as in
Fig.~\ref{DWregime}(a) can no longer be defined. For example,
$q=0.12$ meV results in a DW at $\Phi/\Phi_0 \approx 1.12$ but
$q=0.15$ meV gives no DW. According to Sec.~\ref{wells}, for a NW
with $R_0=75$ nm, $\Delta=\Delta_0=0.2$ meV, and $q=0$ there is no
DW. Our calculations show that $q=0.055$ meV gives rise to a DW at
the lobe's boundary, $\Phi/\Phi_0 \approx 1.5$. Increasing the
coupling strength to $q=0.12$ meV shifts the DW to a lower flux,
i.e., $\Phi/\Phi_0 \approx 1.19$ and eventually $q=0.15$ meV
destroys the DW potential. This numerical investigations
demonstrates that the Josephson potentials studied in
Sec.~\ref{wells} remain robust when a small coupling, $q \lesssim
\Delta_0/2$, is introduced (for any physical reason) between the
BdG energy levels of the original symmetric NW. We remark that
much stronger couplings can be assumed as long as $q$ is taken to
be nonzero only within a small region of the junction.

\begin{figure*}
\includegraphics[width=3.9cm, angle=270]{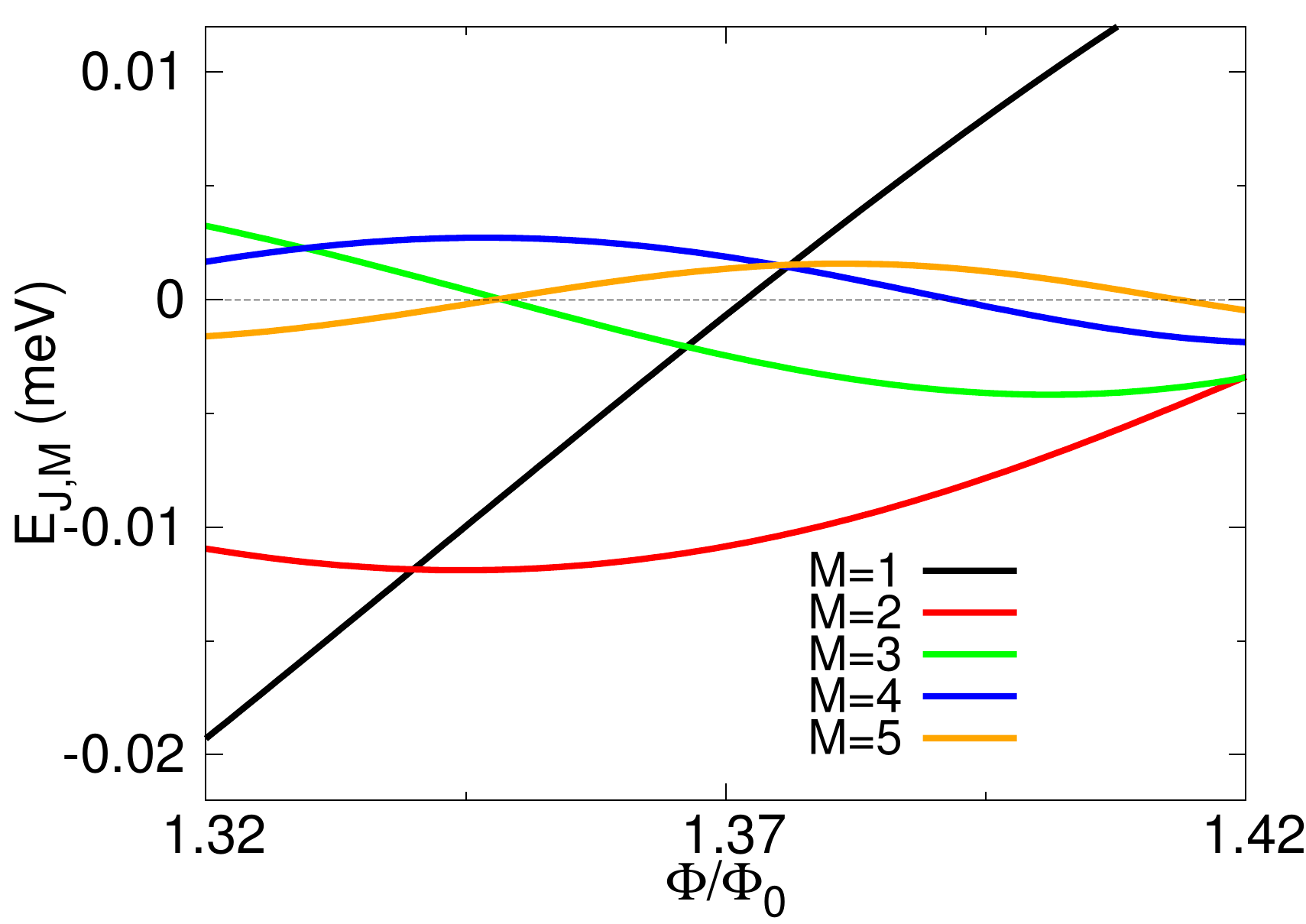}
\includegraphics[width=3.9cm, angle=270]{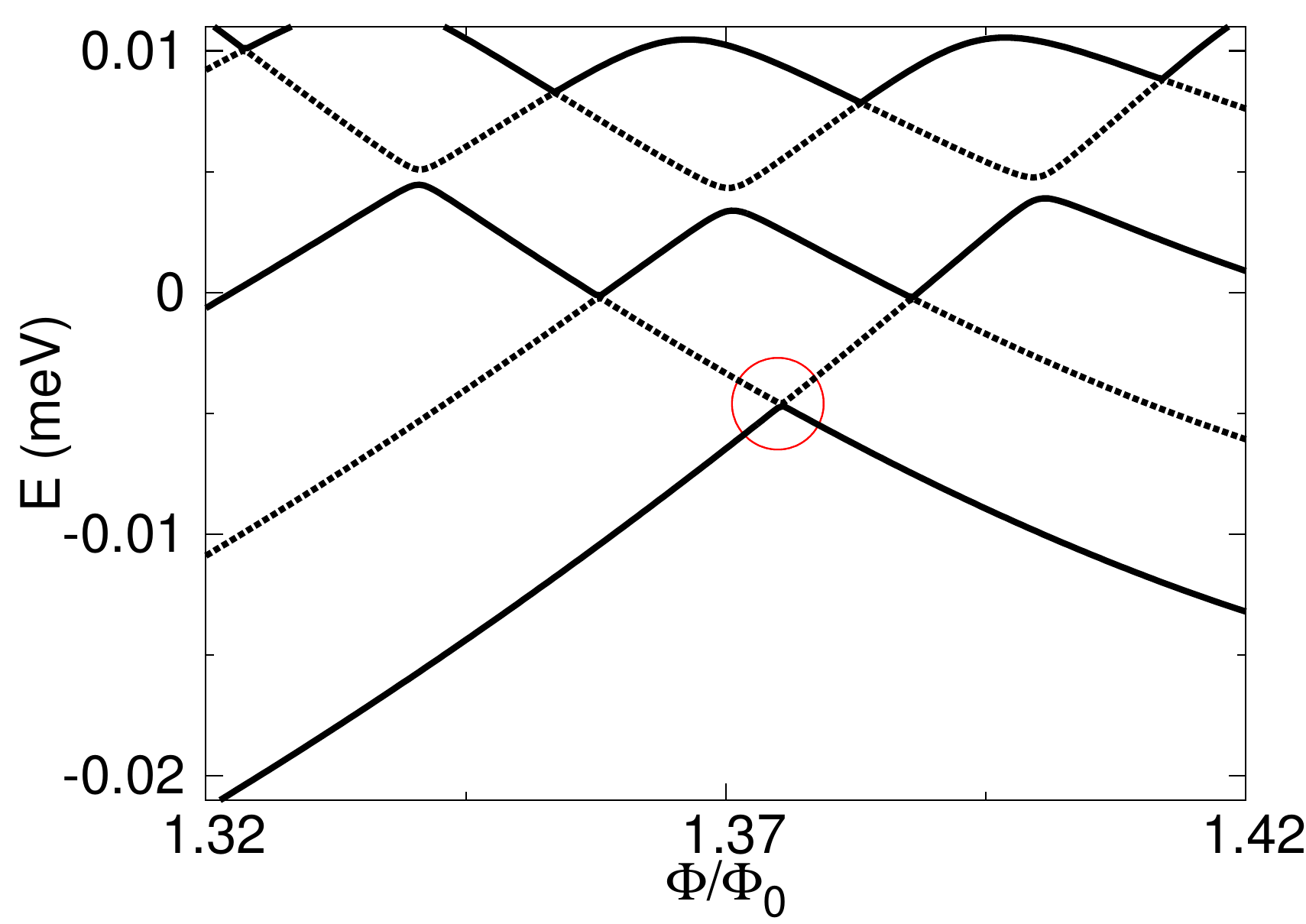}
\includegraphics[width=3.9cm, angle=270]{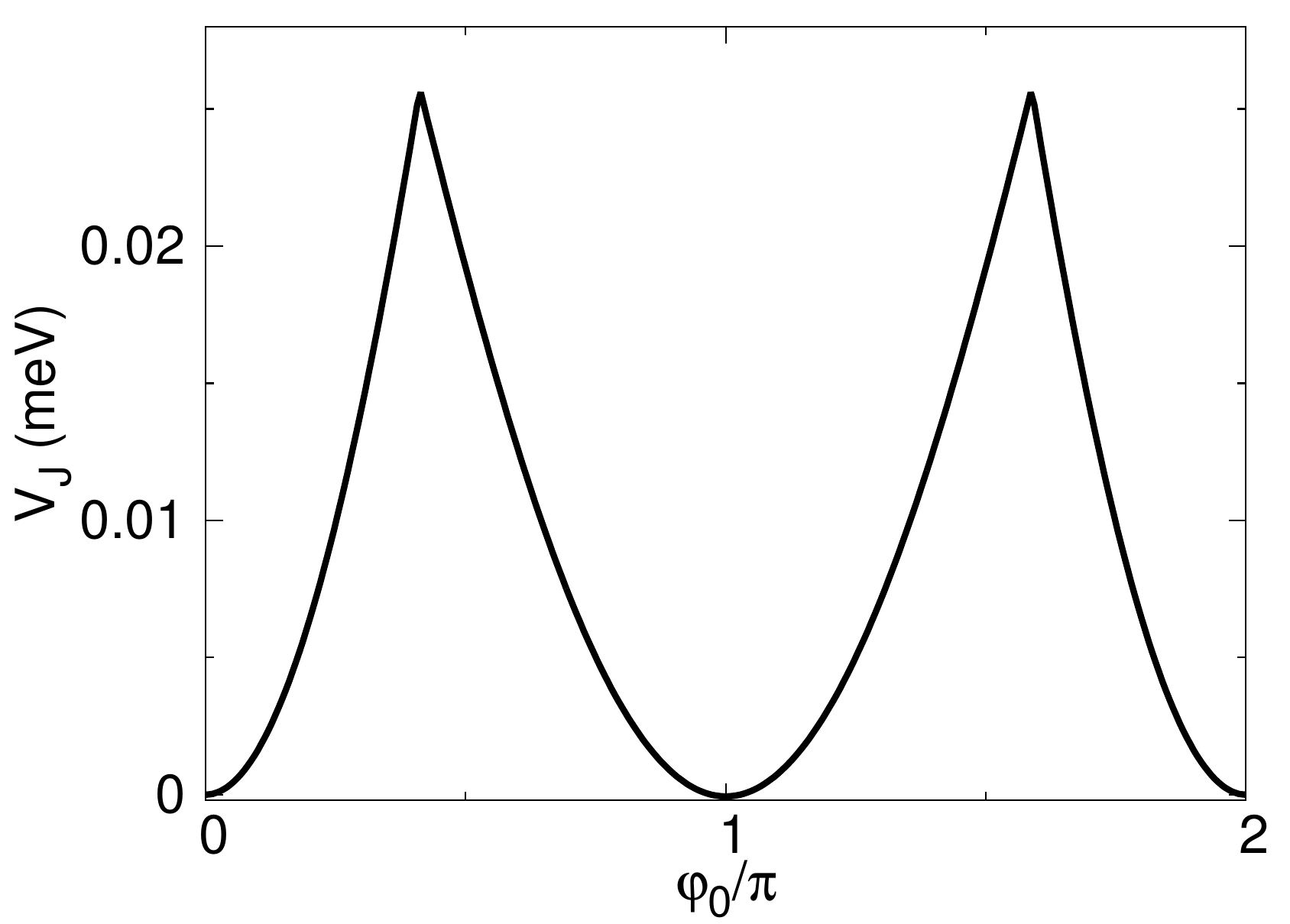}\\
\includegraphics[width=3.9cm, angle=270]{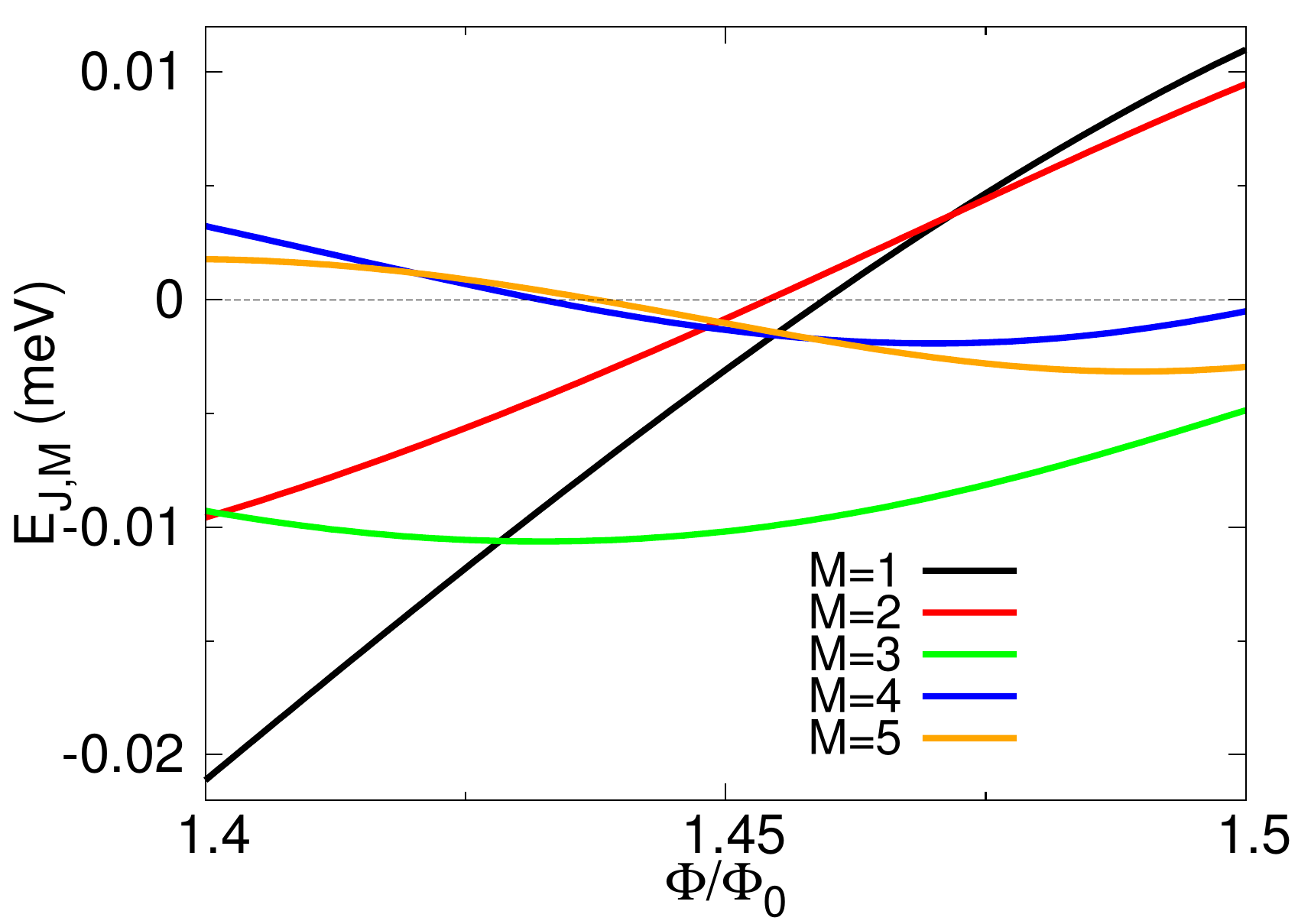}
\includegraphics[width=3.9cm, angle=270]{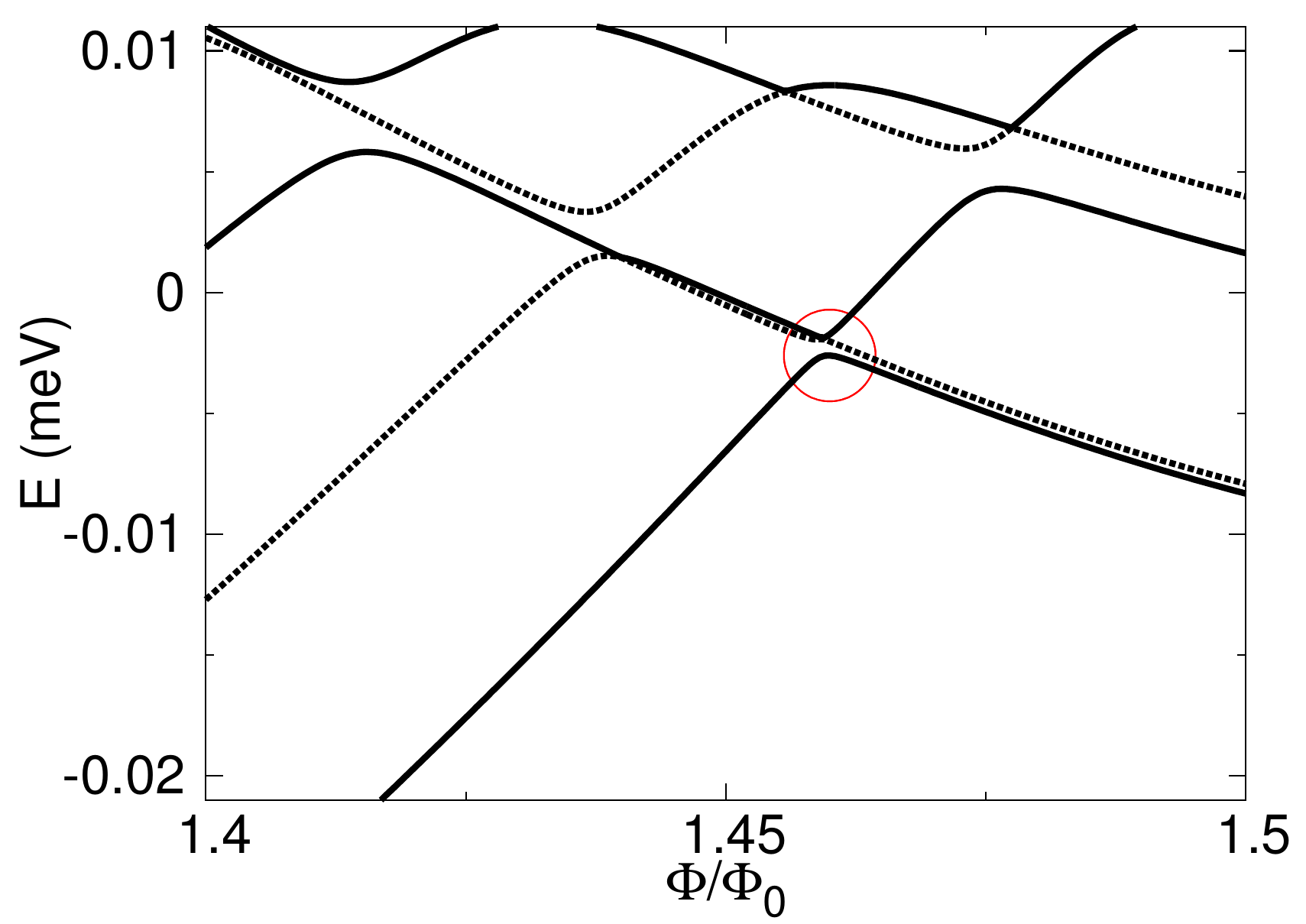}
\includegraphics[width=3.9cm, angle=270]{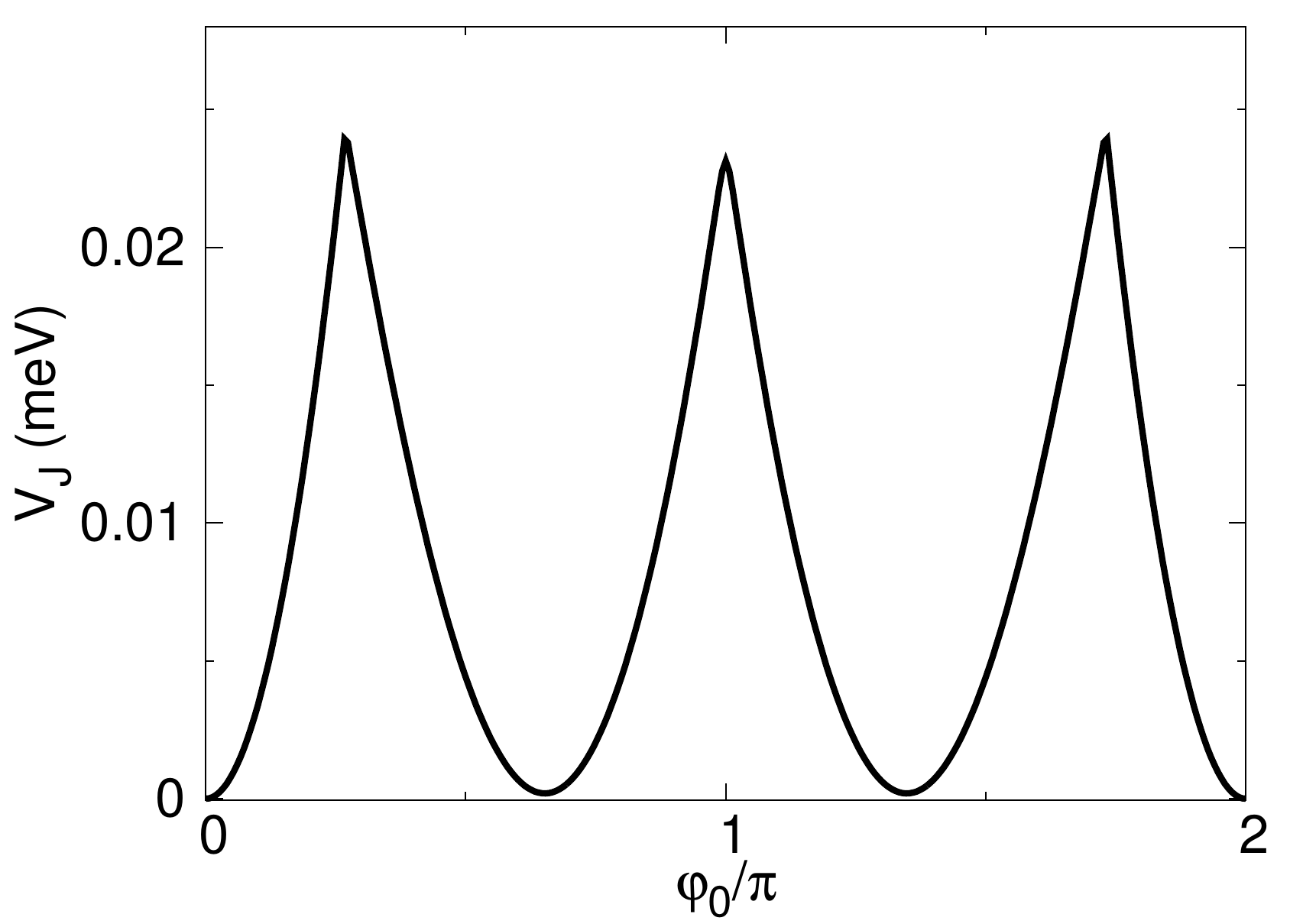}
\caption{Fourier components of Josephson potential, energy levels
of qubit Hamiltonian [Eq.~(\ref{qubitham})], and Josephson
potential at the flux defining the lowest anticrossing. Upper
frames: double well regime at $\mu=0.2$ meV. Lower frames: triple
well regime at $\mu=0.5$ meV. Parameters: $L_S=2000$ nm, $L_N=100$
nm, $R_0=50$ nm, $\alpha_{\rm so}=20$ meV nm, and
$\Delta=\Delta_0=0.2$ meV.}\label{so}
\end{figure*}

\section{Effect of spin-orbit coupling on Josephson potential}
\label{app:SOC}

A nonzero spin-orbit coupling in the radial direction,
$\alpha_{\rm so} \ne 0$, breaks the degeneracy of the $m_j=0$ and
$m_j=\pm1$ energy levels. In this appendix we demonstrate that a
symmetric DW potential can still be formed independent of this
degeneracy. We consider a relatively small coupling, $\alpha_{\rm
so}=20$ meV nm, and assume this coupling to be uniform along the
junction. Gate-induced electric fields in the N region may
introduce some degree of non-uniformity but this is not expected
to drastically modify the Josephson potential provided the length
of the N region is small.

In Fig.~\ref{so} we present a case where a DW Josephson potential
is formed. We see that the overall flux dependence of the Fourier
components $E_{J,M}$ is similar to the case with $\alpha_{\rm
so}=0$ [Fig.~\ref{dwR50}(a) and Fig.~\ref{dwR75}(a)] studied in
the main text. Furthermore, the DW potential leads to an
anticrossing between the two lowest qubit levels at
$\Phi/\Phi_0\approx 1.375$. We have confirmed that similar trends
can be found for different chemical potentials and spin-orbit
couplings.

Although it is not the main topic of our study a noteworthy
difference from the $\alpha_{\rm so}=0$ case can be observed when
the NW is in the topological regime and the $m_j=0$ mode gives
rise to a pair of Majorana states. The properties of these states
and their exact flux dependence in full-shell NWs have been
studied in detail in previous theoretical
works~\cite{model,PhysRevResearch.2.023171,paya_a, paya_b}.
Following the arguments in these works it can be shown that the
appearance of Majorana states can result in a triple well
Josephson potential as demonstrated in Fig.~\ref{so}. The
potential barrier centred at $\varphi_0=\pi$ is indicative of the
Majorana states. In the triple well regime the component $E_{J,3}$
is the dominant one and provided $E_c \ll E_{J,3}$ the system
Hamiltonian [Eq.~(\ref{qubitham})] has three quasidegenerate
ground levels. In Fig.~\ref{so} the charging energy is $E_c=0.1$
GHz and similar to the DW a smaller $E_c$ tends to close the
observed energy gap at $\Phi/\Phi_0\approx 1.46$. According to our
calculations the strength and symmetry of the triple well depend
sensitively on the exact junction parameters as well as the LP
effect. In addition, for an accurate treatment of the Majorana
states a three dimensional model seems to be
necessary~\cite{sanjose, paya_a, paya_b}. For these reasons we do
not consider the hollow-core model to be adequate to capture
details of the triple well potential, and therefore do not further
explore this regime in our work.

\section{Josephson potential derived from simplified model}
\label{app:analytical-minimal}

In this appendix we explore in some detail the DW Josephson
potential derived from the simplified junction model introduced in
Sec.~\ref{wells}. For clarity of presentation we focus on the
realistic limits $\tau_a \lesssim 0.5$ and $\tau_s \gtrsim 0.75$
which are in agreement with our BdG calculations. Focusing on the
range $(0, \pi)$ a simple investigation of the Josephson potential
shows that the angle $\theta$:
\begin{equation}\label{theta}
\theta = 2\sin^{-1} \sqrt{ \frac{1}{\tau_s} \left(1 - \frac{
\gamma^2(\Phi_{\rm DW}) }{\Delta^{2}(\Phi_{\rm DW}) } \right) },
\end{equation}
at which
\begin{equation}
E_{-, s}(\theta, \Phi_{\rm DW}) = 0,
\end{equation}
defines the position of the left potential barrier. The
corresponding barrier height is equal to
\begin{equation}\label{barrier}
V_{\rm B} = E_{+, a}(\theta,\Phi_{\rm DW}) - E_{+,
a}(\pi,\Phi_{\rm DW}),
\end{equation}
and this increases for smaller values of $\theta$. Substituting
Eq.~(\ref{crit}) into Eq.~(\ref{theta}) we derive that
\begin{equation}
\theta = 2\sin^{-1} \sqrt{ \frac{\tau_a}{\tau_s}  },
\end{equation}
and Eq.~(\ref{barrier}) gives the barrier height
\begin{equation}
V_{\rm B} = ( \sqrt{1- \tau^{2}_a/\tau_s } - \sqrt{1-\tau_a} )
\Delta(\Phi_{\rm DW}).
\end{equation}
Thus the limit $\tau_s\rightarrow 1$ leads to the maximum height,
and for the special case $\tau_s=2\tau_a$ the left potential
barrier is centred at $\theta=\pi/2$ leading to the most symmetric
DW profile.

The flux dependence of the Fourier components can be written as
follows
\begin{equation}\label{Fourflux}
\begin{split}
E_{J,M}(\Phi) = X_M(\Phi) +
E_{J,M}(\Phi_0)\frac{\Delta(\Phi)}{\Delta_0},
\end{split}
\end{equation}
with
\begin{equation}\label{X}
\begin{split}
X_M(\Phi) = - \frac{1}{\pi}\int^{\pi+\delta}_{\pi-\delta}
E_{-,s}(\varphi_0, \Phi) \cos(M\varphi_0) d \varphi_0,
\end{split}
\end{equation}
and $\delta=\delta(\Phi)$. If $E_{-,s} < 0$ $(>0)$ for any
$\varphi_0$ then $\delta = 0$ $(\pi)$; otherwise the value of
$\delta$ is found from the condition $E_{-, s}(\pi-\delta, \Phi) =
0$ with $\pi-\delta=\theta$ at $\Phi=\Phi_{\rm DW}$. The term
$X(\Phi)$ quantifies the effect of $\gamma(\Phi)$ on the Josephson
potential, thus, assuming $\gamma(\Phi)=0$ we have $\delta=0$. As
a result, the flux dependence of $E_{J,M}(\Phi)$ is determined
solely by the second term in Eq.~(\ref{Fourflux}) which is due to
the LP effect.

\begin{figure}
\includegraphics[width=8.00cm, angle=270]{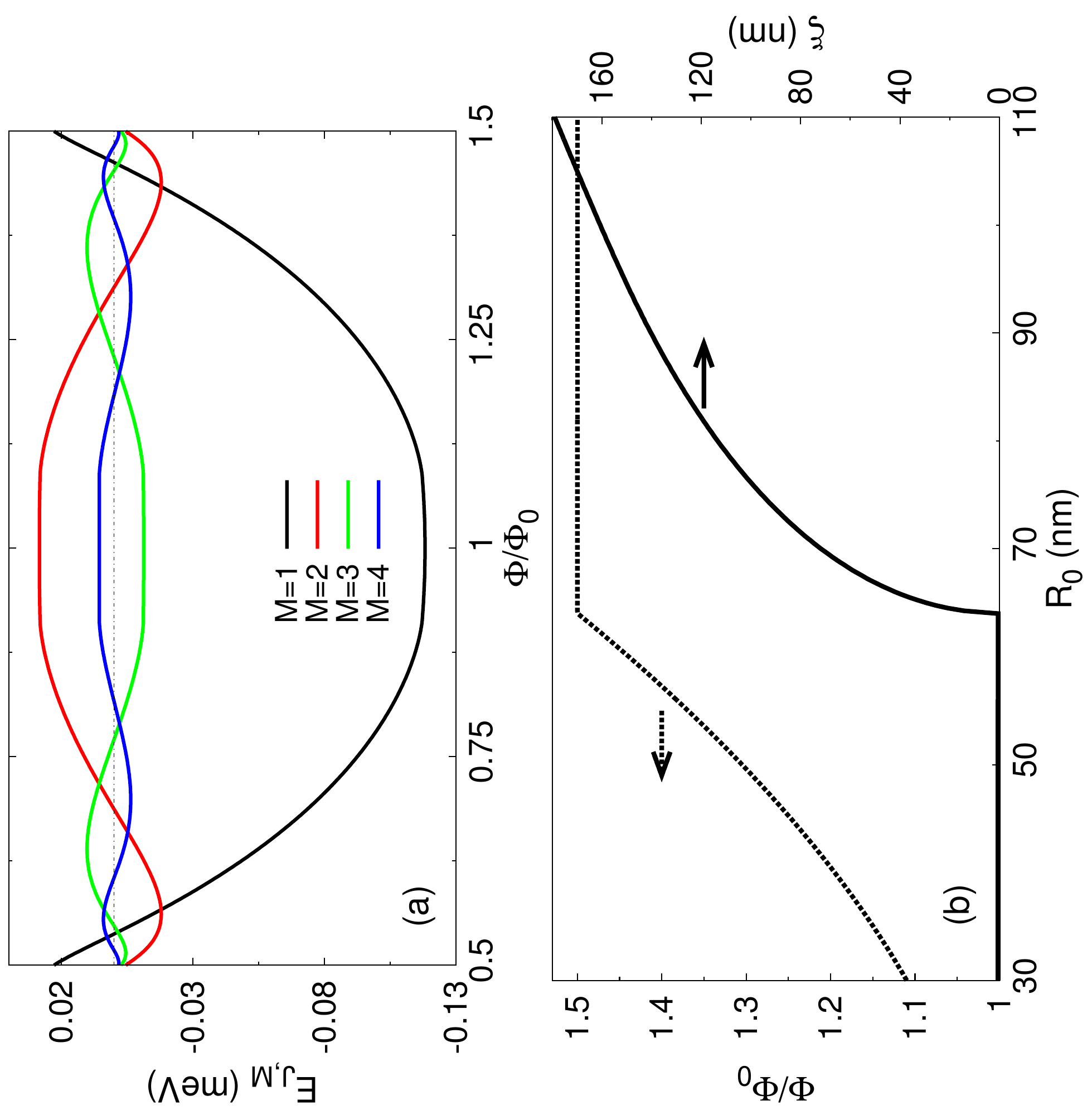}
\caption{(a) Fourier components of Josephson potential derived
from simplified model with $R_0=75$ nm, $\Delta=\Delta(\Phi)$, and
$\xi=110$ nm. (b) Left axis: Flux resulting in DW for
$\Delta=\Delta_0=0.2$ meV when $R_0 < 64$ nm. When $R_0 \gtrsim
64$ nm, $\Delta=\Delta(\Phi)$ is needed to define a DW. Right
axis: The required $\xi$ to form a DW at $\Phi/\Phi_0=1.5$ (a
smaller $\Phi$ requires a larger $\xi$).}\label{simpli}
\end{figure}

In the limit $\tau_s\rightarrow 1$ the modulation of the Josephson
potential due to $\gamma(\Phi)$ can be analytically computed and
the final result is
\begin{equation}
X_M(\Phi) = - \Delta(\Phi) \mathcal{I}_{M} + \gamma(\Phi)
\mathcal{J}_{M},
\end{equation}
with the flux dependent terms
\begin{equation}
\mathcal{I}_{M} = C_M  (  1  - \cos\frac{\delta}{2} \cos M\delta )
+ D_M \sin\frac{\delta}{2} \sin M\delta,
\end{equation}
\begin{equation}
\mathcal{J}_{M} = 2(-1)^{M+1} \frac{\sin M \delta}{\pi M},
\end{equation}
and the constants
\begin{equation}
C_M = \frac{(-1)^{M}}{\pi(  1/4 - M^2 )},  \qquad D_M = \frac{2
M(-1)^{M+1}}{\pi( 1/4 - M^2 )}.
\end{equation}
When $\gamma(\Phi) \approx 0$ we derive that
$\mathcal{I}_{M}=\mathcal{J}_{M}\approx 0$ thus the LP effect
dominates.

We now present some results obtained from the simplified model.
Assuming the parameters $\tau_s=0.99$, $\tau_a=0.40$, and
$\beta=0.46$ then for $\Delta(\Phi)=\Delta_0$ Eq.~(\ref{crit}) is
satisfied when $R_0=50$ nm. Thus, by tuning the flux a DW
potential is formed at $\Phi/\Phi_0 \approx 1.31$. In contrast,
when $R_0=75$ nm a DW cannot be formed because Eq.~(\ref{crit}) is
not satisfied at any flux value within the first lobe. However,
when the LP effect is taken into account with $\xi=110$ nm, then
Eq.~(\ref{crit}) is satisfied and a DW with equal potential minima
can be formed at $\Phi/\Phi_0 \approx 1.44$. This flux value is
sensitive to $\xi$ and typically decreases with $\xi$. This result
qualitatively agrees with the BdG model and demonstrates that the
LP effect can favor the formation of a DW potential.

A general observation, independent of the LP effect, is that the
SW potential at $\Phi/\Phi_0=1$ is determined mainly by $E_{-,s}$,
whereas the SW potential at $\Phi/\Phi_0\approx 1.5$ is the result
of $E_{+,a}$. For this reason the potential at $\Phi/\Phi_0
\approx 1.5$ is always swallower and shifted by $\pi$ compared
with that at $\Phi/\Phi_0 = 1$. This observation is in agreement
with the exact Josephson potential in Fig.~\ref{dwR50}, derived
from the BdG Hamiltonian.

In Fig.~\ref{simpli}(a) we plot the first few Fourier components
of the Josephson potential as a function of the magnetic flux for
one typical junction. The overall flux dependence has the same
characteristics as those obtained from the BdG model. When
considering the LP effect on the superconducting pairing a
symmetric DW potential is formed. As described in
Sec.~\ref{wells}, NWs with large radii can induce only asymmetric
DWs unless the coherence length $\xi$ of the shell has a suitable
value. In Fig.~\ref{simpli}(b), we quantify this value for
different radii by numerically solving Eq.~(\ref{crit}). For $R_0
< 64$ nm the value of $\xi$ is not relevant, namely, a constant
pairing potential suffices to form a symmetric DW. The NW with
$R_0=50$ nm studied in Sec.~\ref{wells} is one example of this
category. In contrast, for $R_0 \gtrsim 64$ nm the value of $\xi$
is relevant and rather large values are necessary to form a DW as
illustrated in Fig.~\ref{simpli}(b). The NW with $R_0=75$ nm
(Sec.~\ref{wells}) belongs to this category. In
Fig.~\ref{simpli}(b), we plot the value of $\xi$ required to form
the DW potential at the lobe's boundary, $\Phi/\Phi_0 = 1.5$. A
larger $\xi$ is needed to shift the DW potential closer to the
centre. Because the value of $\xi$ is not controllable, we
conclude that NWs with small radii should be advantageous to probe
a parity-protected qubit.

\begin{figure}
\includegraphics[width=3.1cm, angle=270]{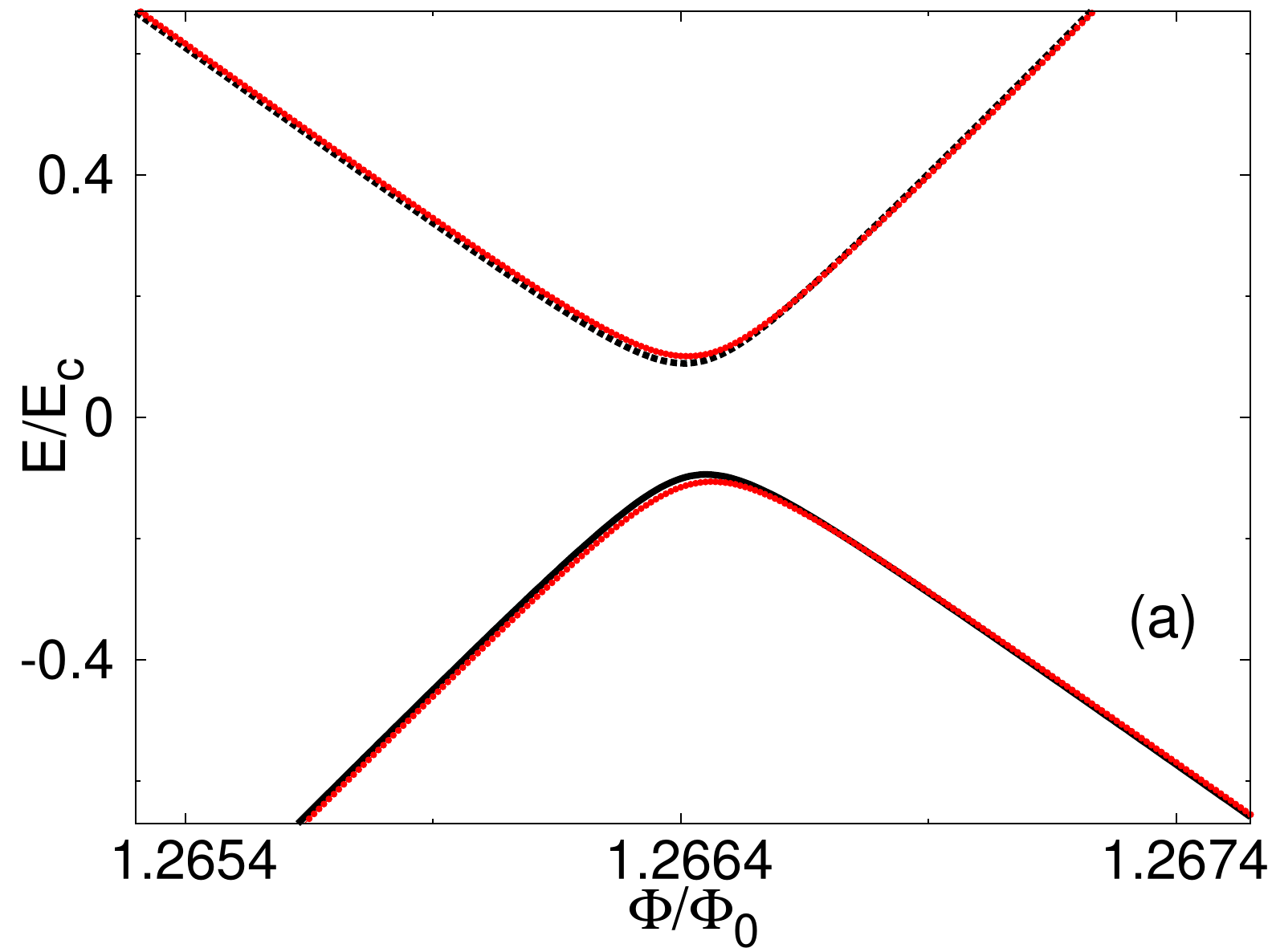}
\includegraphics[width=3.1cm, angle=270]{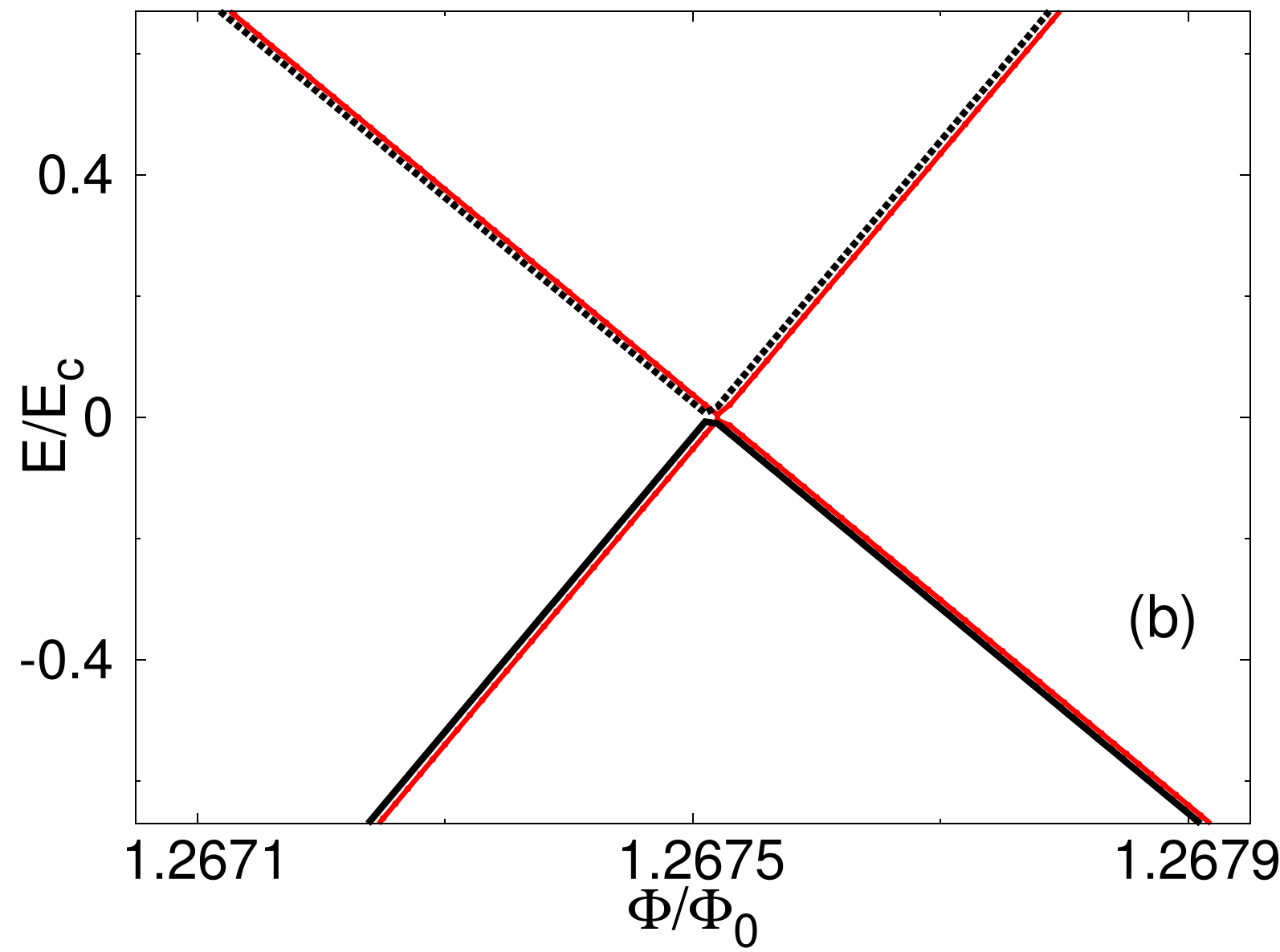}
\caption{Exact (black) and approximate (red) energy levels of
parity-protected qubit as a function of magnetic flux for charging energy:
$E_c=0.35$ GHz (a), and 0.12 GHz (b).}\label{gap}
\end{figure}

\begin{figure*}
\includegraphics[width=4.cm, angle=270]{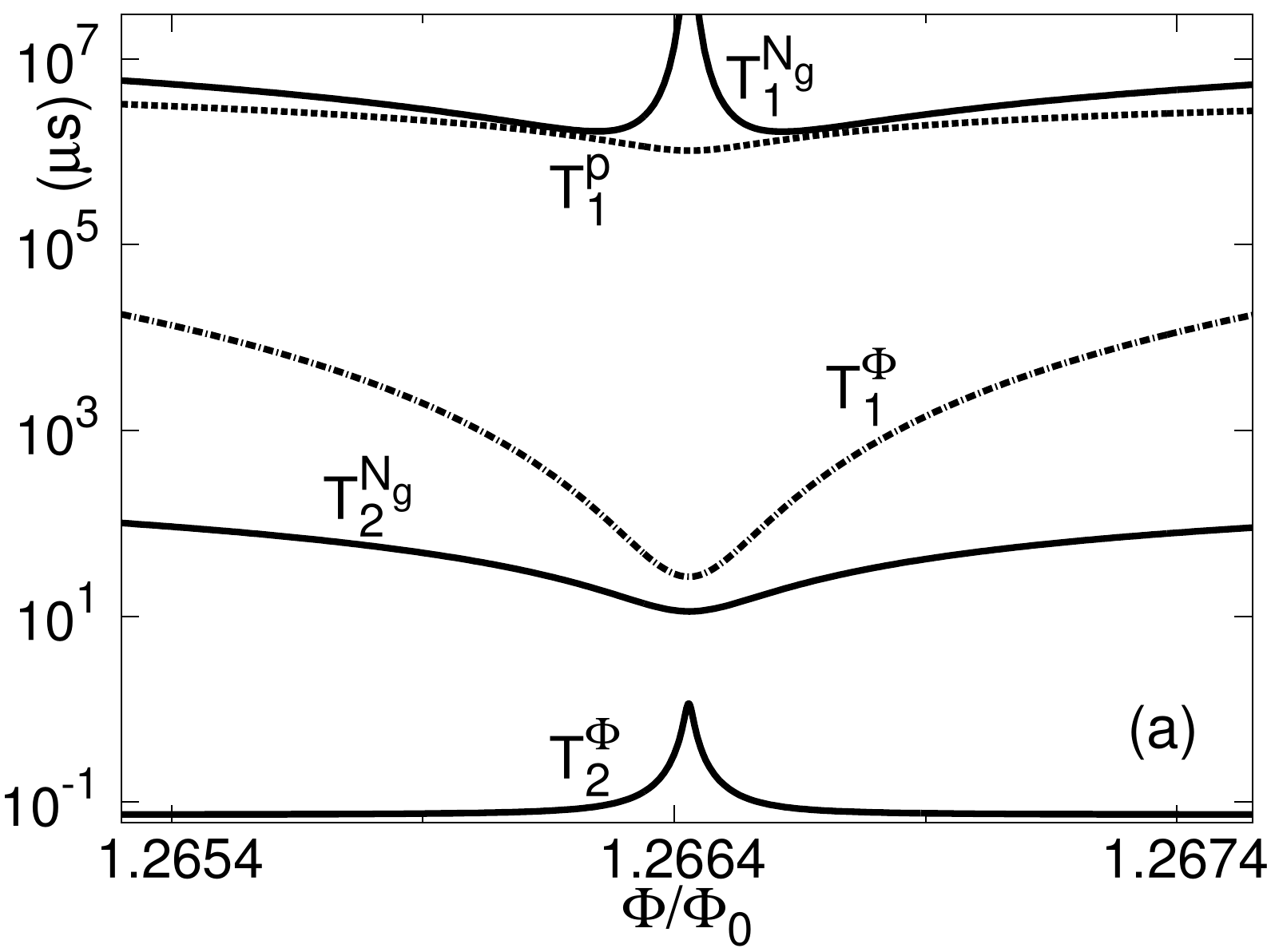}
\includegraphics[width=4.cm, angle=270]{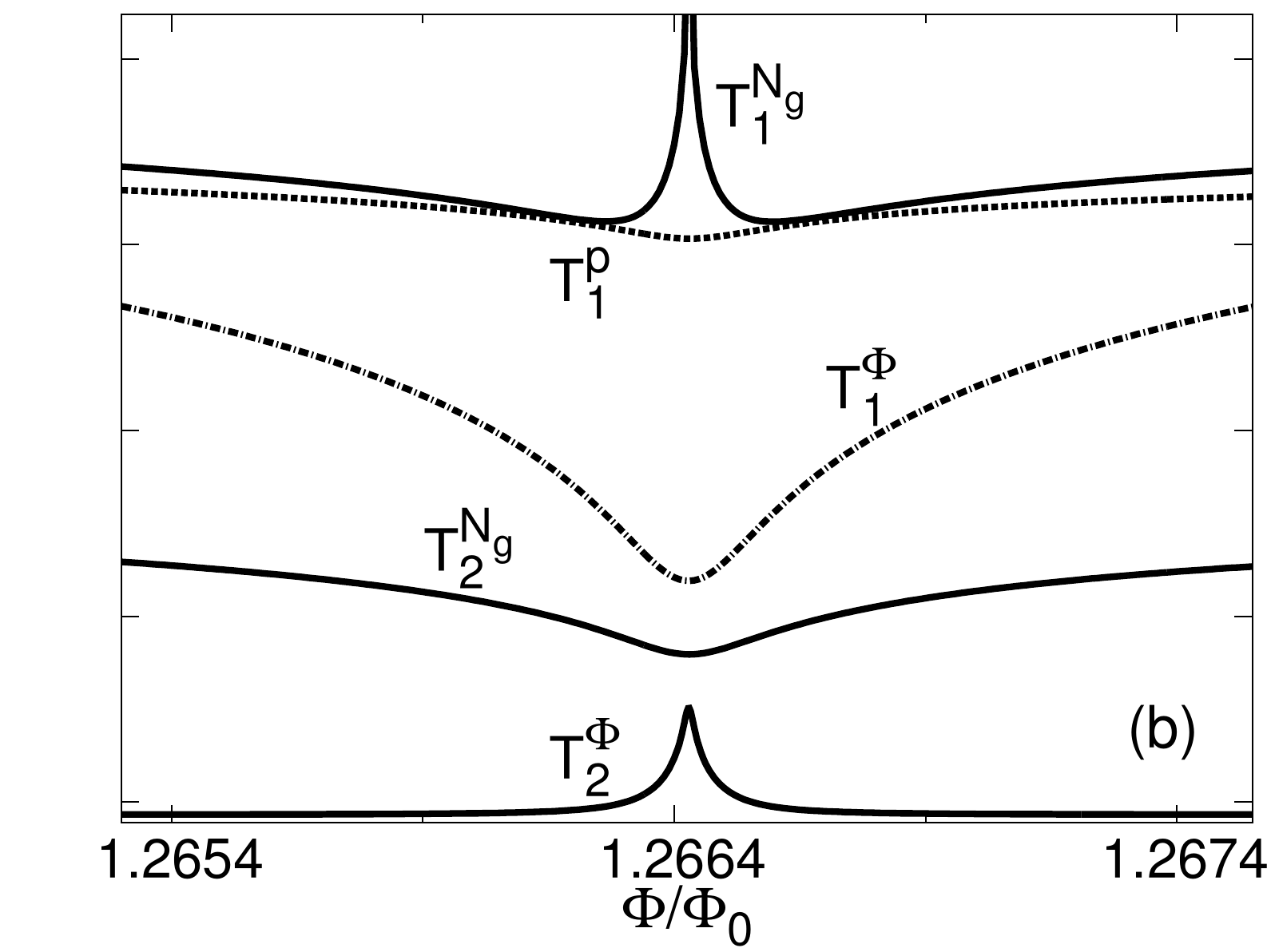}
\includegraphics[width=4.cm, angle=270]{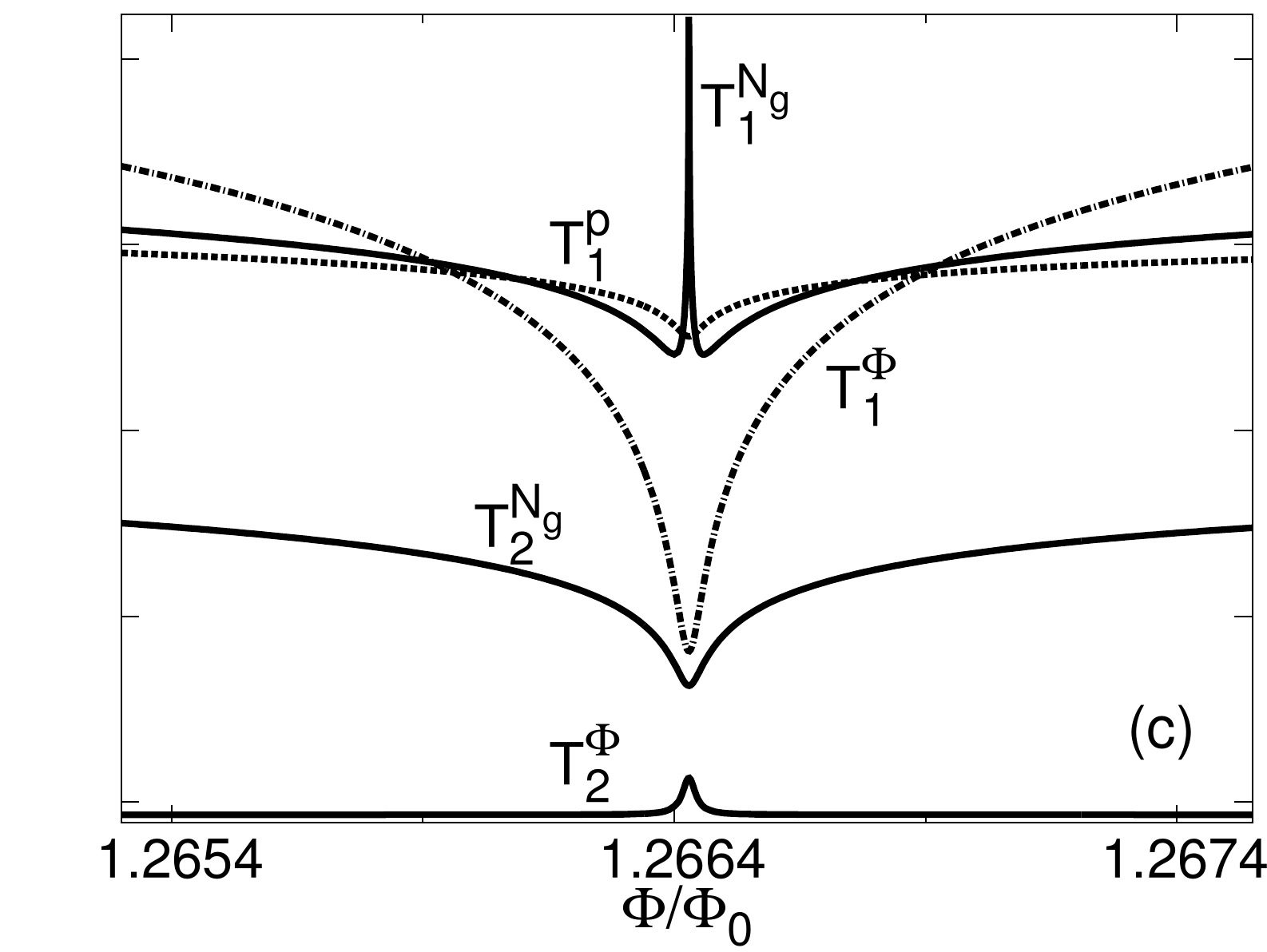}
\caption{As in Fig.~\ref{DWregime}(d) main text but for nonzero
offset charge: $N_g=0.05$ (a), 0.15 (b), and 0.45 (c).}\label{Ng}
\end{figure*}

\section{Further analysis of parity-protected qubit}
\label{app:further-analysis}

In the DW regime the ratio $E_{J,2}/E_c$ determines the
parity-protected qubit energy levels. These levels become quasi
degenerate as the charging energy decreases leading to the closing
of the anticrossing gap (Fig.~\ref{gap}). This result can be
understood by writing the qubit Hamiltonian, Eq.~(\ref{qubitham}),
in $\varphi_0$-space and noticing that the `kinetic term' is
proportional to $E_c$. Additional insight can be gained by using
an approximate model where the qubit states are written as a
superposition of two basis states. For this purpose we choose the
(lowest) SW states at $\Phi/\Phi_0=1$ and $\Phi/\Phi_0=1.5$ where
$E_{J,1}$ is the dominant term [Fig.~\ref{dwR50}(b) and
\ref{dwR50}(e)]. We denote the basis states at these two flux
values by $|y_1\rangle$ and $|y_2\rangle$ respectively. Despite
the simplicity of the model the approximate energies in
Fig.~\ref{gap} agree very well with the exact energies [calculated
from Eq.~(\ref{qubitham})] and correctly predict the closing of
the anticrossing gap. In the Cooper pair basis $| N \rangle$ we
write the states as $|y_k\rangle = \sum_{N} C^{k}_{N}| N \rangle$,
$k=1,\, 2$ so the gap derived from the approximate model is
\begin{equation}\label{twolevel}
E_1-E_0 = \frac{2}{1-s^2}   [ ( x_{1,2} - s x_{2,2} ) ( x_{1,2} -
s x_{1,1}) ]^{1/2},
\end{equation}
where $E_0$ and $E_1$ are the qubit levels at the anticrossing.
Here, $s=\langle y_1 | y_2 \rangle$, and the matrix elements (for
$N_g=0$) are
\begin{equation}
\begin{split}
x_{i,j} = &  \sum_{N,M} \frac{1}{2} E_{J,M}( C^{i}_{N} C^{j}_{N+M}
+
C^{j}_{N} C^{i}_{N+M} ) \\
& + 4 E_c \sum_{N} N^2 C^{i}_{N} C^{j}_{N} .
\end{split}
\end{equation}
Knowledge of $C^{m}_{N}$ allows us to determine the gap for any
flux dependence of $E_{J,M}$. For the approximate results in
Fig.~\ref{gap} $C^{m}_{N}$ are found numerically but approximate
methods can also be applied. The anticrossing gap given by
Eq.~(\ref{twolevel}) is computed at the flux satisfying the
equality $x_{1,1}=x_{2,2}$, which gives the position of the
anticrossing point. This analysis indicates that since the
effective widths of the two potential wells forming the DW
Josephson potential (Sec.~\ref{wells}) are generally unequal, the
two potential minima centered at $\varphi_0=0$ and $\pi$ have to
be slightly detuned at the anticrossing.

Finally, we investigate the qubit Hamiltonian
[Eq.~(\ref{qubitham})] for $N_g\ne0$ and show the coherence times
of the parity protected qubit in Fig.~\ref{Ng}. Compared to
$N_g=0$, [Fig.~\ref{DWregime}(d) main text] now away from the
anticrossing point we have $\langle \psi_0 |
\partial H/ \partial N_g | \psi_1 \rangle \ne 0$ leading to
a finite relaxation time $T^{N_g}_1$. The same reasoning applies
for $T^{p}_1$ at any magnetic flux. The parity protection,
$\Gamma^{N_{g}}_1\approx0$, is again valid at the anticrossing
regardless of $N_g$. The characteristic peak in $T^{N_g}_1$ which
is observed for all $N_g$ values in Fig.~\ref{Ng} is the result of
this protection. When $N_g\ne0$ dephasing due to charge induced
noise becomes stronger but again the shortest time scale of the
qubit is $T^{\Phi}_{2}$ (flux induced dephasing). At the
anticrossing point and $N_g=0.45$ the time $T^{\Phi}_{2}$
decreases by nearly an order of magnitude compared with $N_g=0$.
This decrease is related to the fact that the anticrossing becomes
sharper with $N_g$ so $D_{\Phi\Phi}$ in Eq.~(\ref{T2}) increases.
The basic conclusion is that $N_g=0$ is the optimum offset charge
value for the parity-protected qubit.

\bibliography{biblio}

\end{document}